\documentclass[aps,groupedaddress,floatfix,nofootinbib]{revtex4}

\input epsf

\begin{document}

\bigskip

\title{    
       Strings at Finite Temperature:  \\
       Wilson Lines, Free Energies, and the Thermal Landscape}

\author{Keith R. Dienes$^{1,2,3}$\footnote{E-mail address:  {\tt dienes@physics.arizona.edu}},
        Michael Lennek$^{3}$\footnote{E-mail address:  {\tt mlennek@gmail.com}},
        Menika Sharma$^{3}$\footnote{E-mail address:  {\tt msharma@physics.arizona.edu}}}
\affiliation{
     $^1$ Physics Division, National Science Foundation, Arlington, VA  22230  USA\\
     $^2$ Department of Physics, University of Maryland, College Park, MD  20742  USA\\
     $^3$ Department of Physics, University of Arizona, Tucson, AZ  85721  USA}


\begin{abstract}
  According to the standard prescriptions, zero-temperature string theories 
   can be extended to finite temperature by compactifying their time directions
    on a so-called ``thermal circle'' and implementing certain orbifold twists.
    However, the existence of a topologically non-trivial thermal circle
    leaves open the possibility that a gauge flux can pierce this circle ---
   {\it i.e.}\/, that a non-trivial Wilson line (or equivalently a non-zero chemical potential)
    might be involved in the finite-temperature extension.  In this paper, we
   concentrate on the zero-temperature heterotic and Type~I strings in ten dimensions, 
   and survey the possible Wilson lines which might be introduced in their
   finite-temperature extensions.
     We find a rich structure of possible thermal string theories,
   some of which even have non-traditional Hagedorn temperatures,
   and we demonstrate that these new thermal string theories
    can be interpreted as extrema of a continuous thermal free-energy
   ``landscape''.  Our analysis also uncovers a unique finite-temperature extension
    of the heterotic $SO(32)$ and $E_8\times E_8$ strings which involves a non-trivial Wilson line, 
    but which --- like the traditional finite-temperature
     extension without Wilson lines --- is metastable in this thermal landscape.
\end{abstract}
 

\maketitle

\newcommand{\newc}{\newcommand}
\newc{\gsim}{\lower.7ex\hbox{$\;\stackrel{\textstyle>}{\sim}\;$}}
\newc{\lsim}{\lower.7ex\hbox{$\;\stackrel{\textstyle<}{\sim}\;$}}

\def\beq{\begin{equation}}
\def\eeq{\end{equation}}
\def\beqn{\begin{eqnarray}}
\def\eeqn{\end{eqnarray}}
\def\half{{\textstyle{1\over 2}}}
\def\quarter{{\textstyle{1\over 4}}}

\def\calO{{\cal O}}
\def\calE{{\cal E}}
\def\calP{{\cal P}}
\def\calT{{\cal T}}
\def\calM{{\cal M}}
\def\calF{{\cal F}}
\def\calS{{\cal S}}
\def\calY{{\cal Y}}
\def\calV{{\cal V}}
\def\calZ{{\cal Z}}
\def\calN{{\cal N}}
\def\ibar{{\overline{\imath}}}
\def\chibar{{\overline{\chi}}}
\def\ttwo{{\vartheta_2}}
\def\tthree{{\vartheta_3}}
\def\tfour{{\vartheta_4}}
\def\ttwob{{\overline{\vartheta}_2}}
\def\tthreeb{{\overline{\vartheta}_3}}
\def\tfourb{{\overline{\vartheta}_4}}
\def\Mob{{M\"obius}}

\def\qbar{{\overline{q}}}
\def\mm{{\tilde m}}
\def\nn{{\tilde n}}
\def\rep#1{{\bf {#1}}}
\def\ie{{\it i.e.}\/}
\def\eg{{\it e.g.}\/}

\def\chihat{{\widehat{\chi}}}
\def\nbar{{\overline{n}}}
\def\calP{{\cal P}}
\def\calW{{\cal W}}
\def\taut{{\tau_2}}
\def\tildeq{{\tilde{q}}}
\def\mathhalf{{\frac{1}{2}}}


\hyphenation{su-per-sym-met-ric non-su-per-sym-met-ric}
\hyphenation{space-time-super-sym-met-ric}
\hyphenation{mod-u-lar mod-u-lar--in-var-i-ant}


\def\inbar{\,\vrule height1.5ex width.4pt depth0pt}

\def\IC{\relax\hbox{$\inbar\kern-.3em{\rm C}$}}
\def\IQ{\relax\hbox{$\inbar\kern-.3em{\rm Q}$}}
\def\IR{\relax{\rm I\kern-.18em R}}
 \font\cmss=cmss10 \font\cmsss=cmss10 at 7pt
\def\IZ{\relax\ifmmode\mathchoice
 {\hbox{\cmss Z\kern-.4em Z}}{\hbox{\cmss Z\kern-.4em Z}}
 {\lower.9pt\hbox{\cmsss Z\kern-.4em Z}}
 {\lower1.2pt\hbox{\cmsss Z\kern-.4em Z}}\else{\cmss Z\kern-.4em Z}\fi}
\def\bp{{\bf p}}
\def\bx{{\bf x}}

\long\def\@caption#1[#2]#3{\par\addcontentsline{\csname
  ext@#1\endcsname}{#1}{\protect\numberline{\csname
  the#1\endcsname}{\ignorespaces #2}}\begingroup
    \small
    \@parboxrestore
    \@makecaption{\csname fnum@#1\endcsname}{\ignorespaces #3}\par
  \endgroup}
\catcode`@=12

\input epsf


\section{Introduction and motivation}
\setcounter{footnote}{0}

One of the most profound observations in theoretical physics is the relationship
between finite-temperature quantum theories and zero-temperature quantum theories
which are compactified on a circle.  
Indeed, the fundamental idea behind this so-called ``temperature/radius correspondence''
is that the free-energy density of a theory at finite temperature $T$ can be 
reformulated as the vacuum-energy density of the same theory at zero temperature,
but with the Euclidean time dimension compactified on a circle
of radius $R=(2\pi T)^{-1}$.
This connection between temperature and geometry
is a deep one, stretching from quantum mechanics and quantum field theory
all the way into string theory.

This extension to string theory is truly remarkable, given that
the geometric compactification of string theory gives rise to numerous features which
do not, at first sight, have immediate thermodynamic analogues or interpretations.
For example, upon spacetime compactification, closed strings
accrue not only infinite towers of Kaluza-Klein ``momentum'' states
but also infinite towers of winding states.
While the Kaluza-Kelin momentum states are easily interpreted in a thermal context
as the Matsubara modes corresponding to the original zero-temperature states,
it is not {\it a priori}\/ clear what thermal interpretation might be ascribed to these winding states.
Likewise, as a more general (but not unrelated) issue,
closed-string one-loop vacuum energies generally exhibit additional symmetries
such as modular invariance which transcend field-theoretic expectations.
While the emergence of modular invariance is
clearly understood for zero-temperature geometric compactifications,
the need for modular invariance is perhaps less obvious from the
thermal perspective in which one would simply write down a Boltzmann
sum corresponding to each string state which survives the GSO projections.

Both of these issues tended to dominate the earliest discussions of
string  thermodynamics in the mid-1980's.  Historically, they were first flashpoints
which seemed to show apparent conflicts between the thermal and geometric
approaches which had otherwise been consistent in quantum field theory.
However, it is now well understood 
that there are ultimately no conflicts between these two approaches~\cite{Polbook,Pol86}.  
Indeed, modular invariance emerges naturally upon relating the
integral of the Boltzmann sum over the ``strip'' in the complex $\tau$-plane
to the integral of the partition function over the fundamental domain of the modular
group~\cite{McClainRoth}.
Likewise, thermal windings emerge naturally as a consequence of modular
invariance and can be viewed as artifacts arising from this mapping between
the strip and the modular-group fundamental domain.  

There is, however, one additional feature which can generically arise when a
theory experiences a geometric compactification:  because of the topologically non-trivial
nature of the compactification, it is possible for a non-zero gauge flux 
to pierce the compactification circle.  In other words, the compactification
might involve a non-trivial Wilson line.  Viewed from the thermodynamic perspective,
this corresponds to nothing more than the introduction of a chemical potential.
However, as we shall see, this is ultimately a rather unusual chemical potential:  it is not only imaginary
but also temperature-dependent.  
Such chemical potentials have occasionally played a role in studies 
of finite-temperature field theory (particularly finite-temperature QCD~\cite{ft}).
However, with only a few exceptions, 
such chemical potentials (and the Wilson lines to which they correspond)
have not historically played a significant role in
discussions of finite-temperature string theory.

At first glance, it might seem reasonable to hope (or simply postulate) that Wilson
lines should play no role in discussions of finite-temperature string theory.
However, Wilson lines play such a critical role in determining the allowed possibilities
for self-consistent geometric compactifications of string theory that it is almost 
inevitable that they should play a significant role in finite-temperature string theories as well.
Indeed, the temperature/radius correspondence essentially guarantees this.
Thus, it is natural to expect that theories with
non-trivial Wilson lines will be an integral part of the full landscape of possibilities for string
theories at finite temperature --- \ie, that they will be part of the full ``thermal string landscape''.

A heuristic argument can be invoked in order to illustrate the connection
that might be expected between Wilson lines and string theories at finite temperature.   
As we know, thermal effects treat
bosons and fermions differently and thereby necessarily
break whatever spacetime supersymmetry might have existed at zero temperature.  
However, in string theory there are tight self-consistency constraints which relate the presence
or absence of spacetime supersymmetry to the breaking of the corresponding gauge symmetry,
and these connections hold even at zero temperature.
For example, the $E_8\times E_8$ heterotic string in ten dimensions is necessarily supersymmetric,
and it is inconsistent to break this supersymmetry without simultaneously introducing 
a non-trivial Wilson line (or in this context, a gauge-sensitive orbifold twist)
which also breaks the $E_8\times E_8$ gauge group. 
Indeed, the two are required together.
Even for the $SO(32)$ gauge group, a similar situation arises:
although there exist two $SO(32)$ heterotic strings, one supersymmetric and the other non-supersymmetric,
the $\IZ_2$ orbifold which relates them to each other is not simply given by the SUSY-breaking action $(-1)^F$, where
$F$ is the spacetime fermion number.  Rather, the required
orbifold which twists the supersymmetric $SO(32)$ heterotic string to become the non-supersymmetric
$SO(32)$ heterotic string is given by $(-1)^F W$ where $W$ is a special non-zero Wilson line which
acts non-trivially on the gauge degrees of freedom.
This example will be discussed further in Sect.~IV.~
Indeed, such a Wilson line is needed even though we are not breaking the $SO(32)$ gauge symmetry
in passing from our supersymmetric original theory to our final non-supersymmetric theory.
Such examples indicate the deep role that Wilson lines play in zero-temperature string theory, and which they
might therefore be expected to play in a finite-temperature context as well.

In this paper, we shall undertake a systematic examination of the role that such
Wilson lines might play in string thermodynamics.  We shall 
concentrate on the zero-temperature heterotic and Type~I strings in ten dimensions, 
and survey the possible Wilson lines which might be introduced in their
finite-temperature extensions.  As we shall see, this gives rise to 
a rich structure of possible thermal string theories, and we shall 
demonstrate that these new thermal string theories can be 
interpreted as extrema of a continuous thermal free-energy ``landscape''.  
In fact, some of these new thermal theories even have non-traditional Hagedorn temperatures,
an observation which we shall discuss (and explain) in some detail. 
Our analysis will also uncover a unique finite-temperature extension
of the heterotic $SO(32)$ and $E_8\times E_8$ strings which involves a non-trivial Wilson line, 
but which --- like the traditional finite-temperature
extension without Wilson lines --- is metastable in this thermal landscape.
Such new theories might therefore play an important role in describing the correct
thermal vacuum of our universe.

This paper is organized as follows.
In order to set the stage for our subsequent analysis,
in Sect.~II 
we provide general comments concerning
string theories at finite temperature 
and in Sect.~III we discuss the possible
role that Wilson lines can play in such theories.  
We also discuss the equivalence between such thermal
Wilson lines and temperature-dependent chemical potentials.
In Sect.~IV, we then survey the specific Wilson lines that may
self-consistently be introduced when constructing our thermal theories,
concentrating on the two supersymmetric heterotic strings in ten dimensions
as well as the supersymmetric Type~I string in ten dimensions.
In Sect.~V, we demonstrate that non-trivial Wilson lines can also affect the
Hagedorn temperatures experienced by these strings, and show
how such shifts in the Hagedorn temperature can be reconciled with 
the asymptotic densities of the zero-temperature bosonic 
and fermionic string states.
Then, in Sect.~VI, we extend our discussion in order to consider
continuous thermal Wilson-line ``landscapes'' for both heterotic and
Type~I strings.  It is here that we discuss which Wilson lines lead
to ``stable'' and/or ``metastable'' theories.
Finally, in Sect.~VII, we conclude with some general
comments and discussion.
An Appendix summarizes the notation and conventions that we shall be using
throughout this paper.

\section{Strings at finite temperature}
\setcounter{footnote}{0}

We begin by discussing the manner in which a given zero-temperature
string model can be extended to finite temperature.
This will also serve to establish our conventions and notation.
Because of its central role in determining the thermodynamic properties
of the corresponding finite-temperature string theory, we shall focus on the
calculation of the one-loop string thermal partition function $Z_{\rm string}(\tau,T)$.
The situation is slightly different for closed and open strings, so we shall
discuss each of these in turn.

\subsection{Closed strings}

In order to begin our discussion of closed strings at finite temperature,
we begin by reviewing 
the case of a one-loop partition function 
for a closed string at zero temperature.
Our discussion will be as general as possible, and will 
therefore apply to all closed strings, be they bosonic strings, Type~II superstrings,
or heterotic strings.
For closed strings, the one-loop partition function is defined as 
\beq
      Z_{\rm model}(\tau)~\equiv~ {\rm Tr} ~ (-1)^F~\overline{q}^{H_R}\, q^{H_L}
\label{PF}
\eeq
where the trace is over the complete Fock
space of states in the theory, weighted by a spacetime statistics factor $(-1)^F$.
Here $q\equiv \exp(2\pi i \tau)$ where $\tau$ is the one-loop (torus) modular parameter, and $(H_R,H_L)$
denote the worldsheet energies for the right- and left-moving worldsheet
degrees of freedom, respectively.
Note that in general, $Z_{\rm model}$ is the quantity which appears
in the calculation of the one-loop cosmological
constant (vacuum-energy density) of the model:
\beq
         \Lambda^{(D)} ~\equiv~
     -\half \,{\cal M}^D\, \int_{\cal F} {d^2 \tau\over ({\rm Im}\, \tau)^2}
             Z_{\rm model}(\tau)~
\label{cosconstdef}
\eeq
where $D$ is the number of uncompactified spacetime dimensions,
where ${\cal M}\equiv M_{\rm string}/(2\pi)$ is the reduced string scale,
and where
\beq
   {\cal F}~\equiv ~\lbrace \tau:  ~|{\rm Re}\,\tau|\leq \half,~
 {\rm Im}\,\tau>0, ~|\tau|\geq 1\rbrace
\label{Fdef}
\eeq
is the fundamental domain
of the modular group.
Of course, the quantity in Eq.~(\ref{cosconstdef}) is divergent
for the compactified bosonic string
as a result of the physical bosonic-string tachyon.

Given the general form for the zero-temperature one-loop string partition function
in Eq.~(\ref{PF}), 
it is straightforward to construct its generalization to finite temperature.
As is well known in field theory, 
the free-energy density $F_{b,f}$ of a boson (fermion)
in $D$ spacetime
dimensions at temperature $T$ is nothing but 
the zero-temperature vacuum-energy density $\Lambda$
of a boson (fermion) in
$D$ spacetime dimensions, where the (Euclidean) timelike dimension is compactified
on a circle of radius $R\equiv 1/(2\pi T)$ about which the boson (fermion) is
taken to be periodic (anti-periodic).
We shall refer to this observation as the ``temperature/radius correspondence''.
This correspondence generally extends to string theory as well~\cite{Polbook,Pol86},
state by state in the string spectrum.
However, 
for closed strings there is an important extra ingredient:  we must include not only
the ``momentum'' Matsubara states
that arise from the compactification of the timelike direction, 
but also the ``winding'' Matsubara states that arise due to the closed nature
of the string.
Indeed, both types of states are necessary for the modular invariance of the
underlying theory at finite temperature.
As a result, a given zero-temperature string state 
will accrue not a single sum of Matsubara/Kaluza-Klein modes
at finite temperature,
but actually a {\it double sum}\/ consisting of the
Matsubara/Kaluza-Klein momentum modes 
as well as the Matsubara winding modes.

The final expressions for our finite-temperature
string partition functions $Z(\tau,T)$ must also be modular invariant,
satisfying the constraint $Z(\tau,T)=Z(\tau+1,T)=Z(-1/\tau,T)$.
Because our thermal theory necessarily includes 
two groups of momentum quantum numbers (namely those with $m\in\IZ$ as well as those with $m\in\IZ+1/2$)
which are treated separately (corresponding to spacetime bosons and fermions respectively), modular  
invariance turns out to imply that winding numbers $n\in \IZ$ which are even will likewise 
be treated separately from those that are odd.
As a result, the most general thermal string-theoretic partition function 
will take the form~\cite{Rohm,AlvOso,AtickWitten,KounnasRostand}
\beqn
        Z_{\rm string}(\tau,T)  &=&
           Z^{(1)}(\tau) ~ \calE_0(\tau,T) ~+~
           Z^{(2)}(\tau) ~ \calE_{1/2}(\tau,T) \nonumber\\
           && ~~+~  Z^{(3)}(\tau) ~ \calO_{0}(\tau,T) ~+~
           Z^{(4)}(\tau) ~ \calO_{1/2}(\tau,T) ~.
\label{EOmix}
\eeqn
Here $\calE_{0,1/2}$ and $\calO_{0,1/2}$ represent
the thermal portions of the partition function, namely the
double sums over appropriate combinations of thermal
momentum and winding modes~\cite{Rohm}.  Specifically, the $\calE_{0,1/2}$ functions
include the contributions from even winding numbers $n$
along with either integer or half-integer momenta $m$,
while the $\calO_{0,1/2}$ functions 
include the contributions from odd winding numbers $n$
with either integer or half-integer momenta $m$.
These functions are defined explicitly in the Appendix.
Likewise, the terms $Z^{(i)}$ ($i=1,...,4$) represent
the traces over those subsets of the zero-temperature
string states in Eq.~(\ref{PF})
which accrue the corresponding thermal modings at finite temperature.
For example, $Z^{(1)}$ represents a trace over those
string states in Eq.~(\ref{PF}) which accrue 
even thermal windings $n\in 2\IZ$ and integer thermal 
momenta $m\in \IZ$, and so forth.
Modular invariance for $Z_{\rm string}$ as a whole is
then achieved by demanding that each $Z^{(i)}$ 
transform exactly as does its corresponding $\calE/\calO$ function.

In the $T\to 0$ limit, it is easy to verify that
$\calO_0$ and $\calO_{1/2}$ each vanish while $\calE_0,\calE_{1/2} \to \calM/T$
with $\calM\equiv 1/\sqrt{\alpha'}$.
As a result, we find that
\beq
          Z_{\rm string}(T) ~\to~ {\calM \over T}\, \left[ Z^{(1)} + Z^{(2)}\right]~ 
             ~~~~~{\rm as}~~T\to 0~.
\label{preorigmodel}
\eeq
The divergent prefactor proportional to $1/T$ in Eq.~(\ref{preorigmodel}) is 
a mere rescaling factor which 
reflects the effective change of the dimensionality of the theory in the $T\to 0$ limit. 
Specifically, this is an expected dimensionless volume factor which emerges as the
spectrum of surviving Matsubara momentum states becomes continuous.  
However, we already know that $Z_{\rm model}$ in Eq.~(\ref{PF})  
is the partition function of the zero-temperature theory.
As a result, we can relate Eqs.~(\ref{PF}) and (\ref{EOmix})
by identifying
\beq
          Z_{\rm model} ~=~ Z^{(1)} + Z^{(2)}~.
\label{origmodel}
\eeq

We see, then, that the procedure for extending a given zero-temperature string
model to finite temperature is relatively straightforward.
Any zero-temperature string model is described by a partition function
$Z_{\rm model}$, the trace
over its Fock space.  The remaining task is then simply to determine 
which states within $Z_{\rm model}$
are to accrue integer modings around the thermal circle,
and which are to accrue half-integer modings.
Those that are to accrue integer modings become part of $Z^{(1)}$, while
those that are to accrue half-integer modings become part of $Z^{(2)}$.
In this way, we are essentially decomposing $Z_{\rm model}$ in 
Eq.~(\ref{origmodel}) into separate components $Z^{(1)}$ and $Z^{(2)}$.
Once this is done, modular invariance alone determines the unique resulting forms
for $Z^{(3)}$ and $Z^{(4)}$.
The final thermal partition function $Z_{\rm string}(\tau,T)$ is then given in Eq.~(\ref{EOmix}).
In complete analogy to Eq.~(\ref{cosconstdef}),
we can then proceed to define the $(D-1)$-dimensional vacuum-energy density
\beq
         \Lambda^{(D-1)} ~\equiv~
     -\half \, {\cal M}^{D-1}\, \int_{\cal F} {d^2 \tau\over \tau_2^2}
             Z_{\rm string}(\tau,T)~
\label{reducedlambda}
\eeq
(where $\tau_2\equiv {\rm Im}\/\tau$),
whereupon the corresponding $D$-dimensional free-energy density $F(T)$ is given by
\beq
                F(T)~=~  T \, \Lambda^{(D-1)}~.
\label{freeenergydef}
\eeq

As we see from this discussion, the only remaining critical question is to determine
how to decompose $Z_{\rm model}$ as in Eq.~(\ref{origmodel}) into the pieces 
$Z^{(1)}$ and $Z^{(2)}$  --- \ie, to determine 
which states within $Z_{\rm model}$ are to accrue integer thermal modings
(and thereby be included within $Z^{(1)}$), and which are to accrue 
half-integer modings 
(and thereby be included within $Z^{(2)}$).
However, this too is relatively simple.
In general, a given string model will give rise to states which are spacetime
bosons as well as states which are spacetime fermions.
In making this statement, we are identifying ``bosons'' and ``fermions''
on the basis of their spacetime Lorentz spins.
(By the spin-statistics theorem, this is equivalent to identifying
these states on the basis of their Bose-Einstein or Fermi-Dirac
quantizations.)
As a result, we can always decompose $Z_{\rm model}$ into separate
contributions from spacetime bosons and spacetime fermions: 
\beq
             Z_{\rm model} ~=~ Z_{\rm boson} + Z_{\rm fermion}~.
\label{bosfermdecomp}
\eeq
However, the temperature/radius correspondence instructs us that
bosons should be periodic around the thermal circle, and fermions
should be anti-periodic around the thermal circle. 
In the absence of any other effects, a field which is periodic
around the thermal circle will have integer momentum quantum
numbers $m\in \IZ$, while a field which is anti-periodic will
have half-integer momentum quantum numbers $m\in\IZ+1/2$. 
Thus, 
given the decomposition in Eq.~(\ref{bosfermdecomp}),
the standard approach which is taken in the string literature is to 
identify
\beq
     Z^{(1)}= Z_{\rm boson}~,~~~~ Z^{(2)}= Z_{\rm fermion}~.
\label{boltzmann}
\eeq 
This makes sense, since $Z^{(1)}$ corresponds to the $\calE_0$ sector which 
accrues integer thermal Matsubara modes $m\in \IZ$ while
$Z^{(2)}$ corresponds to the $\calE_{1/2}$ sector which 
accrues half-integer thermal Matsubara modes $m\in \IZ+1/2$.
Indeed, the choice in Eq.~(\ref{boltzmann}) is the unique choice which 
reproduces the standard Boltzmann sum for the states in the string
spectrum.

We can illustrate this procedure by explicitly writing down the 
standard thermal partition functions for the ten-dimensional supersymmetric
$SO(32)$ and $E_8\times E_8$
heterotic strings at finite temperature.
At zero temperature, both of these string theories have partition functions
given by
\beq
         Z_{\rm model} ~=~ Z^{(8)}_{\rm boson}~ (\overline{\chi}_V-\overline{\chi}_S)
         \, {\cal L}
\label{pfs}
\eeq
where $Z_{\rm boson}^{(8)}$ denotes the contribution from the eight worldsheet bosons
and where the contributions from the right-moving worldsheet
fermions are written in terms of the barred characters $\chibar_i$ of the
transverse $SO(8)$ Lorentz group.
These quantities are defined in the Appendix.
By contrast, ${\cal L}$ denotes the contributions from the left-moving (internal) worldsheet
degrees of freedom.  
Written in terms of products $\chi_i\chi_j$ of the unbarred
characters $\chi_i$ of the $SO(16)$ gauge group,
these left-moving contributions are given by
\beq
    {\cal L} ~=~ \cases{ 
    \chi_I^2 +  \chi_V^2 + \chi_S^2 + \chi_C^2 ~~~ & for $SO(32)$ \cr
         (\chi_I +  \chi_S)^2 & for $E_8\times E_8$~.}
\label{pfs2}
\eeq

States which are spacetime bosons or fermions contribute to the terms
in Eq.~(\ref{pfs})
which are proportional to $\chibar_V$ or $\chibar_S$, respectively.
The standard Boltzmann prescription in
 Eq.~(\ref{boltzmann})
therefore leads us to identify
\beq
    Z^{(1)} =               Z^{(8)}_{\rm boson} \, \chibar_V \, {\cal L} ~,~~~~
    Z^{(2)} =              -Z^{(8)}_{\rm boson} \, \chibar_S \, {\cal L}~,
\label{boltzII}
\eeq
whereupon modular invariance requires that
\beq
    Z^{(3)} =              -Z^{(8)}_{\rm boson} \, \chibar_C \, {\cal L} ~,~~~~
    Z^{(4)} =               Z^{(8)}_{\rm boson} \, \chibar_I \, {\cal L} ~.
\label{dictated}
\eeq
We therefore obtain the thermal partition functions
\beq
    Z(\tau,T) ~=~  Z^{(8)}_{\rm boson} \,\times \,\bigl\lbrace ~
            \chibar_V \, \calE_0 
         -  \chibar_S \, \calE_{1/2} 
         -  \chibar_C \, \calO_{0} 
         +  \chibar_I \, \calO_{1/2} ~ \bigr\rbrace~ {\cal L}~.
\label{fake}
\eeq
This is indeed the standard result in the string literature~\cite{AtickWitten}.

\subsection{Type~I strings}

We now turn to the case of Type~I strings.
Such strings, of course, have both closed and open sectors. 
Because Type~I strings are unoriented, their one-loop vacuum
vacuum energies receive four separate contributions:  those from
the closed-string sectors have the topologies of a torus
and a Klein bottle, while those from the open-string sectors
have the topologies of a cylinder and a M\"obius strip.   
We therefore must consider four separate partition functions:  
$Z_{\rm T}$, $Z_{\rm K}$, $Z_{\rm C}$, and $Z_{\rm M}$.

At zero temperature, both $Z_{\rm T}$ and $Z_{\rm K}$ are traces over
the closed-string states in the theory:
\beq
      Z_{\rm T} (\tau)~\equiv~ \half\, {\rm Tr} ~(-1)^F~\overline{q}^{H_R}\, q^{H_L}~,~~~~
      Z_{\rm K} (\tau)~\equiv~ \half\, {\rm Tr} ~\Omega ~ (-1)^F~\overline{q}^{H_R}\, q^{H_L}~
\label{torklein}
\eeq
where $\Omega$ is the orientation-reversing operator
which 
exchanges left-moving and right-moving worldsheet degrees of freedom.
Thus, taken together, the sum $Z_{\rm T} + Z_{\rm K}$ represents a single trace over those
closed-string states which are invariant under $\Omega$, as appropriate for an unoriented string.
Note that because of the presence of the orientifold operator $\Omega$ within $Z_{\rm K}$, the
Klein-bottle contribution $Z_{\rm K}$
can ultimately be represented as a power series in terms of a single variable $q\equiv \exp(2\pi i\tau)$
where $\tau$ represents the modulus for double-cover of the torus, as given in Eq.~(\ref{tauCases}).
Likewise, corresponding to this are the traces over the open-string states in the theory:
\beq
      Z_{\rm C} (\tau)~\equiv~ \half\, {\rm Tr} ~(-1)^F~ q^{H}~,~~~~
      Z_{\rm M} (\tau)~\equiv~ \half\, {\rm Tr} ~\Omega ~ (-1)^F~  q^{H}~,
\label{cylmob}
\eeq
where $H$ is the open-string worldsheet energy and where $q\equiv \exp(2\pi i\tau)$ with $\tau$
defined in Eq.~(\ref{tauCases}).   The presence of $\Omega$ within $Z_{\rm M}$ guarantees
that $Z_{\rm C}+Z_{\rm M}$ represents a single trace over an orientifold-invariant set 
of open-string states.

Extending these contributions to finite temperatures is also straightforward.
The extension of
the torus contribution $Z_{\rm T}$ to finite temperatures proceeds exactly as discussed 
above for closed strings, 
ultimately leading to an expression of the same form as in 
Eq.~(\ref{EOmix}), with four different thermal sub-contributions 
$\lbrace Z^{(1)}_{{\rm T}},
Z^{(2)}_{{\rm T}},
Z^{(3)}_{{\rm T}},
Z^{(4)}_{{\rm T}}\rbrace$.
The corresponding finite-temperature Klein-bottle contributions can be derived from the 
finite-temperature torus contribution
by implementing the orientifold projection in the finite-temperature trace,
ultimately leading to an expression which can be recast in the form
\beq
         Z_{\rm K}(\tau,T) ~=~ Z^{(1)}_{\rm K}(\tau) \,\calE(\tau,T) + Z^{(2)}_{\rm K}(\tau) \,\calE'(\tau,T)
\eeq
where the thermal functions ${\cal E}$ and ${\cal E}'$ are defined in Eq.~(\ref{EEprime})
and serve as the open-string analogues of the closed-string thermal $\calE_{0,1/2}$ functions. 
Likewise, the open-string sector extends to finite temperatures in complete analogy with 
the closed-string sector, by associating certain states with ${\cal E}$ and others with ${\cal E}'$:
\beqn
         Z_{\rm C}(\tau,T) &=& Z^{(1)}_{\rm C}(\tau) \,\calE(\tau,T) + Z^{(2)}_{\rm C}(\tau) \,\calE'(\tau,T)\nonumber\\
         Z_{\rm M}(\tau,T) &=& Z^{(1)}_{\rm M}(\tau) \,\calE(\tau,T) + Z^{(2)}_{\rm M}(\tau)\,\calE'(\tau,T)~.
\eeqn
Note that ${\cal E}(T)$ and ${\cal E}'(T)$ become equal as $T\to 0$.
It therefore follows that
$Z^{(1)}_{\rm X} + Z^{(2)}_{\rm X} = Z_{\rm X}$ for ${\rm X}\in \lbrace {\rm K,C,M}\rbrace$.

Once these four partition functions are determined,
the corresponding free-energy density is easily calculated.
The contribution from the torus to the free-energy density is given by
\beq
            F_{\rm T}(T)~=~  -\half \,T \,\calM^{9} \, \int_\calF
           {d^2\tau\over \tau_2^2}\, Z_{\rm T}(\tau,T)~,
\label{un}
\eeq
in complete analogy with Eqs.~(\ref{reducedlambda}) and (\ref{freeenergydef}).
By contrast, the remaining contributions to the free-energy density
are each given by
\beq
    F_{\rm X}(T) ~=~ -\half \,T \,\calM^{9} \, \int_0^\infty {d\taut\over \tau_2^2} \, Z_{\rm X}(\taut,T)~~~~~
 {\rm where}~~ {\rm X}\in \lbrace {\rm K}, {\rm C}, {\rm M}\rbrace~.
\label{deux}
\eeq
The total free-energy density of the thermal string model is then given by
\beq
          F(T) ~=~  F_{\rm T}(T) +  F_{\rm K}(T) +  F_{\rm C}(T)+  F_{\rm M}(T)~.
\label{trois}
\eeq

Thus, just as for closed strings,
we see that the art of extending a given zero-temperature Type~I string theory to finite
temperatures ultimately boils down to choosing the manner in which the zero-temperature
partition functions $Z_{\rm T}$ and $Z_{\rm C}$
are to be decomposed into the separate thermal contributions
$Z_{\rm T}^{(1,2)}$ 
and
$Z_{\rm C}^{(1,2)}$ respectively.
Once these choices are made, the rest follows uniquely:  modular invariance 
dictates $Z_{\rm T}^{(3,4)}$,
and orientifold projections determine 
$Z_{\rm K}^{(1,2)}$ and
$Z_{\rm M}^{(1,2)}$.
Moreover, just as for closed strings, it turns out that 
the traditional Boltzmann sum is reproduced 
in the finite-temperature theory by making the particular choices
for $Z_{\rm T}^{(1,2)}$ and $Z_{\rm C}^{(1,2)}$ such that
the spacetime bosonic (fermionic) states within 
$Z_{\rm T}^{(1,2)}$ and $Z_{\rm C}^{(1,2)}$ are associated with
$\calE_{0}$ ($\calE_{1/2}$) and $\calE$ ($\calE'$) respectively.  

To illustrate this procedure, let us consider
the single self-consistent zero-temperature ten-dimensional Type~I string model which is both
supersymmetric and anomaly-free:  this is the $SO(32)$ Type~I 
string~\cite{Polbook}.
Note that this string can be realized as
the orientifold projection of the ten-dimensional
zero-temperature Type~IIB superstring, whose partition function 
is given by
\beq
      Z_{\rm IIB} ~=~ Z_{\rm boson}^{(8)} ~
           (\chibar_V-\chibar_S)\,(\chi_V-\chi_S) ~.
\label{TypeIIBPartition}
\eeq
Here $Z_{\rm boson}^{(8)}$ denotes the contribution from the eight worldsheet coordinate bosons,
just as for the heterotic strings discussed above, and
the contributions from the left-moving (right-moving) worldsheet fermions 
are written in terms of the holomorphic (anti-holomorphic) characters $\chi_{V,S,C}$ ($\chibar_{V,S,C}$)
of the transverse $SO(8)$ Lorentz group.
Implementing the orientifold projection is relatively straightforward, and
leads to the Type~I contributions
\beqn
 {\rm torus}:  ~~~~~&      Z_{\rm T}(\tau) &=~ \phantom{+}\,\half\, \phantom{N^2\,} Z_{\rm boson}^{(8)}\,  (\chibar_V -\chibar_S)\,(\chi_V-\chi_S) \nonumber\\
 {\rm Klein}:  ~~~~~&      Z_{\rm K}(\taut) &=~ \phantom{+}\,\half\, \phantom{N^2\,} Z_{\rm boson}^{\prime (8)}\, (\chi_V -\chi_S ) \nonumber\\
 {\rm cylinder}:  ~~~~~&   Z_{\rm C}(\taut) &=~ \phantom{+}\,\half \, N^2 \, Z_{\rm boson}^{\prime (8)} \,(\chi_V -\chi_S ) \nonumber\\
 {\rm Mobius}:  ~~~~~&   Z_{\rm M}(\taut) &=~ -{\hskip -0.04 truein}\half\, N\phantom{^2} {\hskip 0.02 truein}\hat{Z}_{\rm boson}^{\prime (8)} \, (\chihat_V -\chihat_S )~,
\label{TypeIfourcontribs}
\eeqn
where we have used the notation and conventions defined in the Appendix.
Tadpole anomaly cancellation ultimately requires that we take $N=32$, thereby leading to
the $SO(32)$ gauge group.  
Note that while the cylinder contribution scales as $N^2$ [representing the sum of the dimensionalities
of the symmetric and anti-symmetric tensor representations of $SO(32)$], 
the \Mob\ contribution scales only as $N$ (representing their difference).

Given the results for the zero-temperature $SO(32)$ Type~I theory in Eq.~(\ref{TypeIfourcontribs}),
it is straightforward to construct their finite-temperature extension.
Within the torus contribution in Eq.~(\ref{TypeIfourcontribs}), we recognize that the states which are spacetime
bosons are those which contribute to 
$\chibar_V\chi_V+\chibar_S\chi_S$,
while those that are spacetime fermions contribute to  
$\chibar_V\chi_S+\chibar_S\chi_V$.
Following the standard Boltzmann description, we thus identify
\beqn
         Z_{\rm T}^{(1)}  &=& \phantom{-}\half \, Z_{\rm boson}^{(8)} \,(\chibar_V\chi_V+\chibar_S\chi_S)\nonumber\\
         Z_{\rm T}^{(2)}  &=& -          \half \, Z_{\rm boson}^{(8)} \,(\chibar_V\chi_S+\chibar_S\chi_V)~.
\eeqn
Similar reasoning for the cylinder contribution in Eq.~(\ref{TypeIfourcontribs}) also leads
us to identify 
\beqn
     Z_{\rm C}^{(1)} &=& \phantom{-} \half \, N^2 \, Z_{\rm boson}^{\prime (8)} \, \chi_V \nonumber\\
     Z_{\rm C}^{(2)} &=&  -          \half \, N^2 \, Z_{\rm boson}^{\prime (8)} \, \chi_S~.
\eeqn
Given these choices, the remaining terms in the total thermal partition function are determined
through modular transformations and orientifold projections, leading to the final finite-temperature result
\beqn
 {\rm torus}:  ~~~~~&      Z_{\rm T}(\tau,T) 
         &=~ \phantom{+}\,\, \half\, \phantom{N^2\,} Z_{\rm boson}^{(8)}\, \times \,\bigl\lbrace ~
    \phantom{+}
           \lbrack \chibar_V\chi_V + \chibar_S\chi_S
        \rbrack
      \calE_0
      \nonumber\\
  &~& \phantom{=~ \,\,\phantom{-} \half\, \phantom{N^2\,} Z_{\rm boson}^{(8)}\, \times \,\bigl\lbrace }
   -
      \lbrack \chibar_V\chi_S + \chibar_S\chi_V
    \rbrack
    \calE_{1/2}
   \nonumber\\
  &~& \phantom{=~ \,\,\phantom{-} \half\, \phantom{N^2\,} Z_{\rm boson}^{(8)}\, \times \,\bigl\lbrace }
   +
     \lbrack \chibar_I\chi_I + \chibar_C\chi_C \rbrack
    \, \calO_0
       \nonumber\\
  &~& \phantom{=~ \,\,\phantom{-} \half\, \phantom{N^2\,} Z_{\rm boson}^{(8)}\, \times \,\bigl\lbrace }
   -
     \lbrack \chibar_I\chi_C + \chibar_C\chi_I \rbrack
    \, \calO_{1/2}
     ~~\bigr\rbrace~ 
               \nonumber\\
 {\rm Klein}:  ~~~~~&      Z_{\rm K}(\taut,T) &=~ \phantom{+} \,\,\half\, \phantom{N^2\,} Z_{\rm boson}^{\prime(8)}\, (\chi_V -\chi_S ) \, \calE~\nonumber\\
  {\rm cylinder}:  ~~~~~&  Z_{\rm C}(\taut,T) &=~ \phantom{+}\,\,\half \, N^2\, Z_{\rm boson}^{\prime(8)} \,
                               (\chi_V \,\calE -\chi_S\, \calE') \nonumber\\
 {\rm Mobius}:  ~~~~~&   Z_{\rm M}(\taut,T) &=~ -{\hskip -0.02 truein}\half\, N\phantom{^2}\, \widehat{Z}_{\rm boson}^{\prime (8)} \,
                               (\chihat_V \,\calE -\chihat_S \,\calE')~.
\label{TypeIthermal}
\eeqn
This, too, is the standard result in the string-thermodynamics literature.

\section{Wilson lines and imaginary chemical potentials}
\setcounter{footnote}{0}

As we have seen in the previous section, there is a simple procedure by
which a given zero-temperature string model can be extended to finite temperatures.
Indeed, because of constraints coming from modular invariance and/or orientifold
projections, we have relatively little choice in how this is done.  For 
closed strings, the only freedom we have is related to how our (torus) partition
function $Z_{\rm model}$ is decomposed into $Z^{(1)}$ and $Z^{(2)}$ --- \ie,
into the separate contributions that determine which of the zero-temperature states in
the theory are to receive integer modings $m\in\IZ$ around the thermal circle,
and which states are to receive half-integer modings $m\in \IZ+1/2$.
Likewise, for Type~I strings, we have an additional freedom which concerns
how the same choice is ultimately made for the open-string sectors.
However, once those choices are made, all of the resulting thermal 
properties of the theory are completely determined.

As discussed in the previous section, the standard prescription is to identify
those states which are spacetime bosons with 
integer modings $m\in \IZ$ around the
thermal circle, and those which are spacetime fermions with half-integer
modings $m\in\IZ+1/2$. 
Indeed, this is ultimately the unique choice
for which the resulting string partition functions correspond to the
standard field-theoretic Boltzmann sums for each string state (a fact
which is most directly evident after certain Poisson resummations are 
performed, essentially transforming our theory from the so-called ${\cal F}$-representation
we are using here to the so-called ${\cal S}$-representation in which the modular invariance
of the torus contributions is not manifest).

Given this observation, it might seem that there is therefore 
no choice in how our zero-temperature string theories are extended to
finite temperatures.
However, this is not entirely correct.
It is certainly true that the temperature/radius correspondence instructs us
to treat bosonic states as periodic around the thermal circle and fermionic
states as anti-periodic.
However, this does not necessarily imply that all bosonic states will correspond
to integer momentum modings $m\in\IZ$, or that 
all fermionic states will correspond to half-integer momentum modings $m\in\IZ+1/2$.
Indeed, in the presence of a non-trivial Wilson line,
this result can change.

In order to understand how this can happen,
let us first recall how the standard ``temperature/radius correspondence''
is derived (see, \eg, Ref.~\cite{DolanJackiw}).
As is well known, this correspondence is most directly formulated in quantum
field theory (as opposed to string theory) and ultimately rests upon 
the algebraic similarity between the free-energy density of a thermal 
theory and the vacuum-energy density of the zero-temperature theory in which
the Euclidean timelike direction is geometrically compactified on a circle.
This similarity can be demonstrated as follows.
Let us begin on the thermal side, and consider the thermal (grand-canonical) partition functions
corresponding to a single real $D$-dimensional bosonic field and a single $D$-dimensional
fermionic field of mass $m$:
\beq
      Z_{b,f}(T) ~=~ \prod_\bp \, (1 \mp  e^{-E_p/T})^{\mp 1}
\label{pfs3}
\eeq
with $E_\bp^2 \equiv \bp\cdot \bp + m^2$.
In Eq.~(\ref{pfs3}), the products are over all $(D-1)$-dimensional spatial momenta ${\bf p}$.
Given these thermal partition functions, the corresponding
$D$-dimensional free-energy densities are given by
\beq
         F_{b,f}(T) ~\equiv~ -T \log Z_{b,f}(T) ~=~ 
             \pm T \int {d^{D-1} \bp\over (2\pi)^{D-1}} \log(1\mp e^{-E_\bp/T})~.
\eeq
However, thanks to certain infinite-product representations for the hyperbolic trigonometric functions, 
it is an algebraic identity that
\beq
     {\rm log}(1\mp e^{-E/T}) ~=~
      \frac{1}{2} \sum_{n = -\infty}^{\infty}
        \, {\rm log} \left[ E^2 + 4\pi^2 (n+c_\mp)^2 T^2 \right] ~+~ ... 
\label{iden}
\eeq
where $c_- = 0$ and $c_+ =1/2$.
In writing Eq.~(\ref{iden}), we have followed standard practice and
dropped terms beyond the infinite products
as well as terms which compensate for the dimensionalities of the arguments of
the logarithms.
We therefore find that
\beq
     F_{b,f}(T) ~=~ \pm {T\over 2}\, \int {d^{D-1} \bp\over (2\pi)^{D-1}}
      \sum_{n = -\infty}^{\infty}
        \, {\rm log} \left[ E_\bp^2 + 4\pi^2 (n+c_\mp)^2 T^2 \right] ~+~ ...
\label{freeEs}
\eeq
On the zero-temperature side, by contrast, we can consider
the zero-point one-loop vacuum-energy density corresponding
to a single real quantum field of mass $m$ in $D$ uncompactified dimensions:
\beq
    \Lambda ~\equiv ~ \half \,(-1)^F \,
                \int {d^D p \over (2\pi)^D} \, \log(p^2+m^2)~.
\label{Lambdadef}
\eeq
Here $(-1)^F$ indicates the spacetime statistics of the quantum field
($=1$ for a bosonic field, $= -1$ for a fermionic field).
Moreover, if we imagine that the time dimension is compactified on a circle of radius $R$
(so that the integral over $p^0$ can be replaced by a discrete sum),
and if the quantum field in question is taken to be periodic (P) or anti-periodic (A) around this
compactification circle,
then Eq.~(\ref{Lambdadef}) takes the form
\beq
    \Lambda_{P,A} ~=~ \half \,
                {1\over 2\pi R}\, (-1)^F\, \int {d^{D-1} \bp \over (2\pi)^{D-1}} \,
           \sum_{n=-\infty}^\infty \, \log[\bp\cdot \bp +m^2+(n+c_{P,A})^2/R^2]~
\label{Lambdanext}
\eeq
where $c_{P}=0$, $c_A=1/2$.
Given these results,
it is now possible to make the ``temperature/radius correspondence'':
comparing Eq.~(\ref{freeEs}) with Eq.~(\ref{Lambdanext}),
we see that we can identify the free-energy density\/ $F_{b,f}$ of a boson (fermion)
in $D$ spacetime
dimensions at temperature $T$ with the zero-temperature vacuum-energy density $\Lambda_{P,A}$
of a boson (fermion) in
$D$ spacetime dimensions, where a (Euclidean) timelike dimension is compactified
on a circle of radius $R\equiv 1/(2\pi T)$ about which the boson (fermion) is
taken to be periodic (anti-periodic).

Given this derivation of the temperature/radius correspondence, it may at first
glance seem that the identification of bosons and fermions with integer and half-integer
modings around the thermal circle is sacrosanct.  
However, let us consider what happens when we repeat this derivation in the presence
of a non-trivial gauge field $A^\mu$ on the geometric (zero-temperature) side.
When we calculate the vacuum energies of our bosonic or fermionic quantum field
in the presence of a non-trivial gauge field $A^\mu$,
we must use the kinematic momenta $\Pi^\mu \equiv p^\mu - \vec \lambda\cdot \vec A^\mu$ where $\vec \lambda$
is the charge (expressed as a vector in root space) of the field in question.
Of course, if the field $A^\mu$ is pure-gauge (\ie, with vanishing corresponding field strength) and our spacetime
geometry is trivial, then this change in momenta from $p^\mu$ to $\Pi^\mu$ will have no physical
effect.  {\it However, if we are compactifying on a circle, there is always the possibility that
our compactification encloses a gauge-field flux.}\/
As in the Aharonov-Bohm effect,
this then has the potential to introduce a non-trivial change in modings
for fields around this circle, even if the gauge field $A^\mu$ is pure-gauge at all
points along the compactification circle.
Indeed, such a flat (pure-gauge) background for the gauge field $A^\mu$ is nothing but a Wilson line.

To be specific, let us first consider the situation in which our compactification circle
of radius $R$ completely encloses a $U(1)$ magnetic flux of magnitude $\Phi$
which is entirely contained within a radius $\rho<R$.
At all points along the compactification circle,
this then corresponds to a $U(1)$ gauge field $A^\mu$ whose only non-zero component
is the component $A^i= -\Phi/(2\pi R)$ along the compactified dimension.
Because of the non-trivial topology of the circle, we then find that the shift from
$p^\mu$ to $\Pi^\mu$ for a state with $U(1)$ charge $\lambda$
induces a corresponding shift in the corresponding
modings:\footnote{
   This discussion of the effects of Wilson lines is mostly field-theoretic.  For closed strings,
   however, there will also be an additional shift due to the possible appearance of
   a non-trivial winding number.  This will be discussed below,
   but we shall disregard these additional shifts here since
   since they play no essential role in the present discussion.}
\beq
          {n\over R}~\to~{n\over R} ~+~ {1\over 2\pi R} \, \lambda\Phi~.
\eeq
While this result holds for $U(1)$ gauge fields, it is easy to generalize this
to the gauge fields of any gauge group $G$.
For any gauge group $G$,
we can describe a corresponding gauge flux in terms of the parameters
$\Phi_i$ for each $i=1,...,r$, where $r$ is the rank of $G$.
Collectively, we can write $\vec\Phi$ as a vector in root space.
Likewise, the gauge charge of any given state can be described in terms of
its Cartan components $\lambda_i$ for $i=1,...,r$;  collectively, $\vec \lambda$ is nothing
but the weight of the state in root space.
We then find that the modings are shifted according to
\beq
          {n\over R}~\to~{n\over R} ~+~ {1\over 2\pi R} \, \vec \lambda\cdot \vec\Phi~.
\eeq
As a result, complex fields which are chosen to be periodic
(P) or anti-periodic (A) around the compactification circle will have vacuum energies given by
\beq
    \Lambda_{P,A} ~=~ {1\over 2\pi R}\, (-1)^F\, \int {d^3 \bp \over (2\pi)^3} \,
    \sum_{n=-\infty}^\infty \, \log\left[E_{\bf p}^2 +  {1\over R^2} \left(n+ c_{P,A} 
          + {1\over 2\pi} \, \vec \lambda\cdot \vec \Phi\right)^2 \right] ~
\label{Lambdanext2}
\eeq
where $E^2_{\bf p}\equiv {\bf p}\cdot{\bf p} + m^2$.
Note that in each case, the underlying periodicity properties of the field are unaffected.
Rather, it is the {\it manifestations of these periodicities in terms of the modings}\/ which are
affected by the appearance of the Wilson line.

This, then, explains how a non-trivial Wilson line can produce unexpected modings due
to the non-trivial compactification geometry.
However, we still wish to understand the appearance of such a Wilson line thermally.
What is the thermal analogue  of the non-trivial Wilson line?
Or, phrased somewhat differently, 
what effect on the thermal side can restore the temperature/radius correspondence
if a non-trivial Wilson line has been introduced on the geometric side?

It turns out that introducing a non-trivial Wilson line on the geometric side
corresponds to introducing a non-zero chemical potential on the thermal side.
In fact, this chemical potential will be imaginary.
To see this, let us reconsider the partition functions of complex
bosons and fermions in the presence of a non-zero chemical potential $\mu\equiv i\tilde \mu$
where $\tilde \mu\in\IR$.
In general, a complex bosonic field
will have a grand-canonical partition function given by
\beq
        Z_b(T)~=~ \prod_{\bf p}
          \left[ 1+ e^{-(E_{\bf p}-\mu)/T} + e^{-2(E_{\bf p}-\mu)/T} +...\right]
          \left[ 1+ e^{-(E_{\bf p}+\mu)/T} + e^{-2(E_{\bf p}+\mu)/T} +...\right]
\label{parfB}
\eeq
where the two factors in Eq.~(\ref{parfB}) correspond to particle and
anti-particle excitations respectively.
The corresponding free energy $F_b(T)\equiv -T \log Z_b$ then takes the form
\beqn
       F_b(T) &=& T \int {d^{D-1} {\bf p}\over (2\pi)^{D-1}} \left\lbrace
                \log[ 1-e^{-(E_{\bf p}-\mu)/T}]  + \log[ 1-e^{-(E_{\bf p}+\mu)/T}]
                         \right\rbrace\nonumber\\
       &=&  {T\over 2} \int {d^{D-1} {\bf p}\over (2\pi)^{D-1}} \sum_{n=-\infty}^\infty \left\lbrace
                \log[ (E_{\bf p}-\mu)^2 + 4\pi^2 n^2 T^2 ] +
                \log[ (E_{\bf p}+\mu)^2 + 4\pi^2 n^2 T^2 ]
                         \right\rbrace\nonumber\\
       &=&  {T\over 2} \int {d^{D-1} {\bf p}\over (2\pi)^{D-1}} \sum_{n=-\infty}^\infty
                \log[ (E_{\bf p}^2-\tilde\mu^2 + 4\pi^2 n^2 T^2 )^2 + 4 \tilde\mu^2 E_{\bf p}^2 ]
                         \nonumber\\
       &=&  {T\over 2} \int {d^{D-1} {\bf p}\over (2\pi)^{D-1}} \sum_{n=-\infty}^\infty
                \log[ E_{\bf p}^4 + 2E_{\bf p}^2 (4\pi^2 n^2 T^2 +\tilde\mu^2) +
                     (4\pi^2 n^2 T^2 -\tilde\mu^2)^2 ]
                         \nonumber\\
       &=&  {T\over 2} \int {d^{D-1} {\bf p}\over (2\pi)^{D-1}} \sum_{n=-\infty}^\infty
                \log[ E_{\bf p}^4 + 2E_{\bf p}^2 (2\pi n T+\tilde\mu)^2
                            + 2E_{\bf p}^2 (-2\pi n T+\tilde\mu)^2 \nonumber\\
                   &&~~~~~~~~~~~~~~~~~~~~~~~~~~~~~~~~
                    + (2\pi n T+\tilde\mu)^2 (-2\pi n T+\tilde\mu)^2]\nonumber\\
       &=&  {T\over 2} \int {d^{D-1} {\bf p}\over (2\pi)^{D-1}} \sum_{n=-\infty}^\infty \left\lbrace
                \log[ E_{\bf p}^2 + (2\pi n T + \tilde \mu)^2]
              + \log[ E_{\bf p}^2 + (-2\pi n T + \tilde \mu)^2]  \right\rbrace\nonumber\\
       &=&  T  \int {d^{D-1} {\bf p}\over (2\pi)^{D-1}} \sum_{n=-\infty}^\infty \log[ E_{\bf p}^2 + (2\pi n T + \tilde \mu)^2]~.
\label{manysteps}
\eeqn
In Eq.~(\ref{manysteps}), the second equality follows from the
algebraic identities in Eq.~(\ref{iden})
while the final equality results upon
exchanging $n\to -n$ in the second term.
Thus, comparing the result in Eq.~(\ref{manysteps}) with the result in
Eq.~(\ref{Lambdanext2}), we see that the free energy of a bosonic field at temperature $T$
is equal to the vacuum energy of a periodically-moded field on a circle of radius $R$, where
$R\equiv 1/(2\pi T)$ and where
\beq
           \tilde\mu ~=~ (\vec \lambda\cdot \vec \Phi) \,T~~~~~~~~\Longrightarrow~~~~~~~
           \mu ~=~ i\, (\vec \lambda\cdot \vec \Phi) \,T~.
\label{chempotential}
\eeq
A similar result holds for complex fermions and anti-periodic fields, with the same
chemical potential.
We thus conclude that the introduction of a non-trivial Wilson line on the geometric side corresponds to
the introduction of an imaginary, temperature-dependent chemical potential on the thermal side.
This result is well known in field theory~\cite{ft},
and has also recently been discussed in a string-theory context~\cite{kounn}.

Before concluding, we should remark that the above discussion has been somewhat field-theoretic.
Indeed, if we define the Wilson-line parameter
\beq
     \vec \ell ~\equiv~ {\vec \Phi\over 2\pi} ~=~ -{\vec A\over 2\pi T}~,
\eeq
then our primary result is that a non-trivial Wilson line $\vec\ell$  
induces a shift in the momentum quantum number of the form
\beq
    m ~\rightarrow~   m + \vec \lambda \cdot \vec\ell~
\label{momentumshift}
\eeq
for a state carrying charge $\vec \lambda$ with respect to the gauge
field constituting the Wilson line.
This result is certainly true in quantum field
theory, and also holds by extension for open-string states.
However, closed-string states can carry not only momentum quantum numbers $m$ but
also winding numbers $n$ which parametrize their windings around the thermal circle.
This is important, because
in the presence of a non-zero winding mode $n$,
a non-trivial Wilson line shifts not only the momentum $m$ but also the charge vector $\vec \lambda$
of a given state, so that Eq.~(\ref{momentumshift}) is generalized to~\cite{HetGauge}
\beq
     \cases{
         m  ~\to~    m + \vec \lambda \cdot \vec\ell - n \vec \ell\cdot \vec\ell /2  & ~\cr
         n  ~\to~    n  & ~ \cr
        \vec \lambda ~\to ~ \vec \lambda  - n \vec{\ell} ~. & ~\cr}
\label{shiftNumbersHet}
\eeq
It is clear that Eq.~(\ref{shiftNumbersHet}) reduces to Eq.~(\ref{momentumshift}) for $n=0$.

\section{Surveying possible Wilson lines}
\setcounter{footnote}{0}

We have already seen in Sect.~II that the manner in which a zero-temperature
string theory is extended to finite temperature depends on the choice
as to which zero-temperature states are to be associated with integer
momenta $m\in \IZ$ around the thermal circle, and which are to be associated
with half-integer momenta $m\in \IZ+1/2$.
Once this decision is made, the thermal properties
of the resulting theory are completely fixed.
Moreover, we have seen in Sect.~III that there is considerable freedom
in making this choice, depending on whether (and which) Wilson lines
might be present.  Indeed, in principle, each choice of Wilson line
leads to an entirely different thermal theory.  While all of these
thermal theories necessarily reduce back to the starting zero-temperature theory
as $T\to 0$, they each represent different possible finite-temperature
extensions of that theory which correspond to different possible chemical
potentials which might be introduced into their corresponding Boltzmann sums. 
Indeed, viewed from this perspective, we see that the traditional Boltzmann
choices merely correspond to one special case:  that without a Wilson line,
for which the corresponding chemical potential vanishes.

As an example, let us consider the supersymmetric $SO(32)$ heterotic string.
At zero temperature, the partition function of this theory is 
\beq
         Z_{\rm model} ~=~ Z^{(8)}_{\rm boson}~ (\overline{\chi}_V-\overline{\chi}_S)
         \, \left( \chi_I^2 +  \chi_V^2 + \chi_S^2 + \chi_C^2 \right)~,
\label{start}
\eeq
and without a Wilson line we would normally decompose this into the
separate thermal contributions $Z^{(1)}$ and $Z^{(2)}$
by making the associations
\beqn
    Z^{(1)} &=&\phantom{-}Z^{(8)}_{\rm boson} \, \chibar_V \, 
             \left( \chi_I^2 +  \chi_V^2 + \chi_S^2 + \chi_C^2 \right)~\nonumber\\
    Z^{(2)} &=& -Z^{(8)}_{\rm boson} \, \chibar_S  
         \, \left( \chi_I^2 +  \chi_V^2 + \chi_S^2 + \chi_C^2 \right)~.
\label{nowilson}
\eeqn
Indeed, this is precisely the decomposition discussed in Sect.~II, which leads
to the standard Boltzmann sum.
However, there are in principle other ways in which the zero-temperature
partition function in Eq.~(\ref{start}) might be meaningfully decomposed.
For example, let us consider an alternate decomposition of the form
\beqn
    Z^{(1)} &=& Z^{(8)}_{\rm boson} \, 
     \left\lbrack \chibar_V \,(\chi_I^2 + \chi_V^2)  ~-~  \chibar_S \,(\chi_S^2 + \chi_C^2) \right\rbrack\nonumber\\  
    Z^{(2)} &=& Z^{(8)}_{\rm boson} \, 
     \left\lbrack \chibar_V \,(\chi_S^2+\chi_C^2)  ~-~  \chibar_S \,(\chi_I^2+\chi_V^2) \right\rbrack~.
\label{alternate}
\eeqn
Unlike the standard Boltzmann decomposition, this 
alternate decomposition 
treats spacetime bosonic and fermionic states 
in ways which are also dependent on their corresponding gauge quantum numbers.
Specifically, while vectorial representations of the $SO(32)$ gauge group
are treated as expected, with spacetime bosons having integer momentum modings $m\in \IZ$
and spacetime fermions having half-integer momentum modings $m\in \IZ+1/2$,
the spinorial representations of the left-moving $SO(32)$ gauge group have
the opposite behavior, with 
spacetime bosons associated with modings $m\in \IZ+1/2$
and spacetime fermions associated with modings $m\in \IZ$.
[Note, in this connection, that the $SO(16)\times SO(16)$ character 
combination $\chi_S^2+\chi_C^2$ is nothing but
the $SO(32)$ character $\chi_S$, and likewise 
$\chi_I^2+\chi_V^2$ is nothing but $\chi_I$.]
However, as we have seen in Sect.~III, such ``wrong'' modings can be easily understood
as the effects of a non-trivial Wilson line.
Indeed, looking at Eq.~(\ref{shiftNumbersHet}), 
we see that 
the results in Eq.~(\ref{alternate})
are obtained directly
if our Wilson line $\vec\ell$ is chosen such that
$\vec \lambda\cdot \vec\ell = 1/2$~(mod~$1$) for states in spinorial
representations of $SO(32)$, while
$\vec \lambda\cdot \vec\ell = 0$~(mod~$1$) for states in vectorial representations of $SO(32)$.
Given that
$\lambda^i\in \IZ$ for vectorial representations of $SO(32)$ and
$\lambda^i\in \IZ+1/2$ for spinorial representations of $SO(32)$,
we see that a simple choice such as $\vec\ell=(1,0,...,0)$ can easily accomplish this.

However, at this stage, we have no knowledge as to whether or not such a Wilson
line represents a legitimate choice for the $SO(32)$ heterotic string.
For example, we have no idea whether such a Wilson-line choice is compatible
with a worldsheet interpretation in which the possible choices
of Wilson lines are tightly constrained by numerous string self-consistency constraints.
Moreover, along the same lines, we do not know what other Wilson lines might 
also be available.

In order to explore all of the potential possibilities, we shall therefore 
proceed to survey the set of all possible  
Wilson lines which might be self-consistently introduced when attempting
to extend a given zero-temperature string theory to finite temperatures. 
As we shall see, however,
the situation is somewhat different for closed strings and
Type~II strings.  We shall therefore consider these two cases separately.

\subsection{Closed strings}

In general, there are two classes of closed strings which
are supersymmetric and hence perturbatively stable:
Type~II superstrings and heterotic strings.
In ten dimensions, however, the Type~II superstrings lack gauge symmetries;
thus no possible Wilson lines can exist in their 
extensions to finite temperatures.
For this reason, when discussing closed strings,
we shall concentrate on the ten-dimensional
supersymmetric $SO(32)$ and $E_8\times E_8$ heterotic strings.
Note, however, that in lower dimensions, 
all of the closed strings 
will accrue additional gauge symmetries as a result
of compactification ---  indeed, this holds for Type~II strings
as well as heterotic.  Thus, in lower dimensions,
the sets of allowed Wilson lines in each case
are likely to be much more complex than we are considering here.

In general, the temperature/radius correspondence provides us with
a powerful tool to help determine the allowed Wilson lines
that may be introduced when forming our thermal theory:
we simply replace the temperature $T$ with $1/(2\pi R)$ and 
consider the corresponding problem of introducing a Wilson line
into the geometric compactification of our original zero-temperature
theory.
For example, if we are seeking the set of allowed Wilson lines that
can be introduced into the construction of the finite-temperature
ten-dimensional $SO(32)$ heterotic theory, we can instead investigate
the allowed Wilson lines that may be introduced upon compactifying
the zero-temperature $SO(32)$ theory to nine dimensions. 
In principle, the latter problem can be studied through any number of formalisms
having to do with the construction of self-consistent zero-temperature string models ---
such model-building formalisms are numerous and 
include various orbifold constructions, Narain lattice constructions,
constructions based on free worldsheet bosons and fermions, and so forth. 

However, for closed strings, it turns out that T-duality leads to a significant
simplification:  while the $R\to\infty$ (or $T\to 0$) limit reproduces our original
string model in the original $D$ spacetime dimensions, and while taking $0<R<\infty$
leads to a string model in $D-1$ spacetime dimensions, the formal $R\to 0$ (or $T\to\infty$) limit
actually yields a new string theory which is back in $D$ spacetime dimensions!       
Moreover, the structure of the finite-temperature string partition function
in Eq.~(\ref{EOmix}) guarantees that 
this new $D$-dimensional theory is nothing but a $\IZ_2$ orbifold of our original $D$-dimensional
theory;
indeed, while the original theory in the $R\to \infty$ limit has the partition
function $Z^{(1)}+Z^{(2)}$, the final theory in the $R\to 0$ limit has
the partition function $Z^{(1)}+Z^{(3)}$.
In some sense, the thermal theory in $(D-1)$ dimensions {\it interpolates}\/
between the original $D$-dimensional theory at $T=0$ and a different $D$-dimensional
theory as $T\to\infty$, these two $D$-dimensional theories being $\IZ_2$ orbifolds
of each other.
Thus, the allowed Wilson lines that may be introduced into
the finite-temperature extension of a given zero-temperature
closed string theory are in one-to-one correspondence\footnote{
    At a technical level, this correspondence is easy to understand.
    Starting from a given zero-temperature theory in $D$ dimensions, one may construct
    the corresponding thermal theory through a specific sequence
    of steps:  first, one compactifies the zero-temperature theory
    on a circle of radius $2R = 1/(\pi T)$, and then one orbifolds
    the resulting theory by the $\IZ_2$ action $(-1)^F {\cal T} W$ where
    $F$ is the spacetime fermion number,
    where ${\cal T}$ denotes a half-shift around the thermal circle,
     and where $W$ (the Wilson line) indicates
    an additional specific orbifold action which is sensitive to the gauge
    quantum numbers of each state.
    The resulting $(D-1)$-dimensional thermal theory then has the property 
    that the original $D$-dimensional theory is reproduced as $T\to 0$,
    and that a new $D$-dimensional theory emerges in the formal $T\to\infty$ limit.
    Moreover, it can also be shown that the new theory which emerges in the $T\to\infty$
    limit is a $\IZ_2$
    orbifold of the original theory, where the $\IZ_2$ orbifold
    in this case is nothing but $(-1)^F W$.  Thus, for each Wilson line
     $W$ which is involved in construction of the thermal theory in $(D-1)$-dimensions,
    there is a corresponding orbifold $(-1)^F W$ which directly relates the
    two ``endpoint'' $D$-dimensional theories to each other.  }
with the set of allowed
$\IZ_2$ orbifolds of that theory  --- \ie, the set of $\IZ_2$ orbifolds
which reproduce another self-consistent string theory in $D$ dimensions.

This correspondence provides us with exactly the tool we need, because the
complete set of self-consistent heterotic string theories in ten dimensions
is known.
Indeed, these have been classified
in Ref.~\cite{KLTclassification}, 
and it turns out that in addition to the 
supersymmetric $SO(32)$ and $E_8\times E_8$ heterotic theories,
there are only seven additional heterotic theories in ten dimensions.
These are the tachyon-free $SO(16)\times SO(16)$
string model~\cite{DH,SW}
as well as six tachyonic string models with
gauge groups $SO(32)$, $SO(8)\times SO(24)$, $U(16)$,
$SO(16)\times E_8$,
$(E_7)^2 \times SU(2)^2$,  and
$E_8$.

\begin{figure}[h!]
\centerline{
   \epsfxsize 6.5 truein \epsfbox {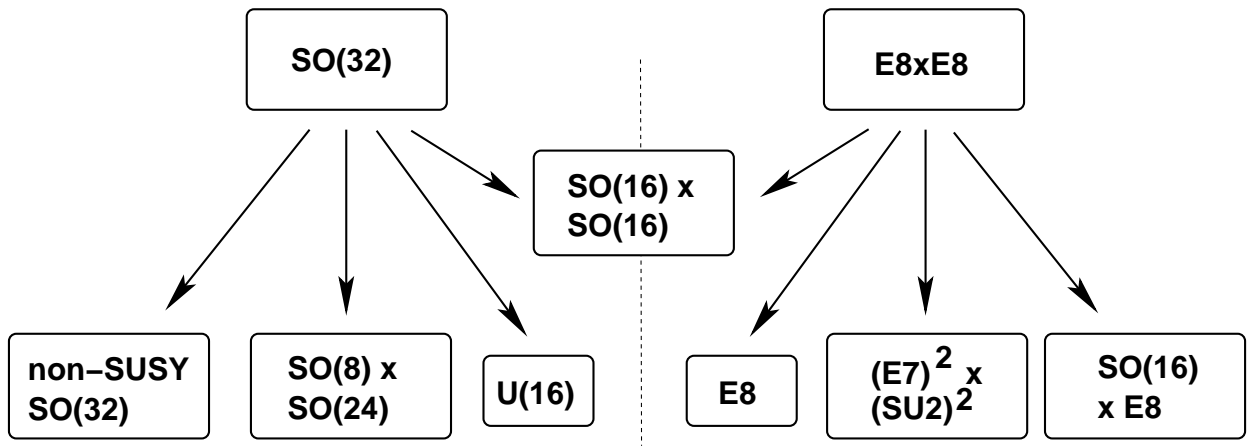}} 
\caption{
    Possible Wilson-line choices for the supersymmetric
     $SO(32)$ and $E_8\times E_8$ heterotic strings,  each corresponding
    to a $\IZ_2$ orbifold which breaks spacetime supersymmetry.
    Note that the $SO(16)\times SO(16)$ string is unique in that
    it can be realized as a $\IZ_2$ orbifold of either the $SO(32)$
    or $E_8\times E_8$ heterotic strings;  it is also the only
    non-supersymmetric heterotic string in ten dimensions which is
    tachyon-free.  By contrast, each of the remaining six non-supersymmetric 
     strings in ten dimensions has a physical tachyon with worldsheet
    energies $H_L=H_R= -1/2$.  }
\label{orbifoldfig}
\end{figure}

However, not all of these models can be realized as
$\IZ_2$ orbifolds of the 
original supersymmetric $SO(32)$ or $E_8\times E_8$ models.
Indeed, of the seven non-supersymmetric models listed above,
only four are $\IZ_2$ orbifolds of the supersymmetric $SO(32)$ string;
likewise, only four are $\IZ_2$ orbifolds of the $E_8\times E_8$ string.
These $\IZ_2$ orbifold relations are shown in Fig.~\ref{orbifoldfig}.

It is important to note that there also exists a non-trivial $\IZ_2$ orbifold relation
which directly relates the supersymmetric $SO(32)$ and $E_8\times E_8$ strings to each other. 
However, it is easy to see that this orbifold must be excluded from consideration.
On the thermal side, we know that finite-temperature effects 
necessarily treat bosons and fermions differently
and will therefore 
necessarily break whatever spacetime supersymmetry might have existed at
zero temperature.  This implies that we must restrict our attention
to those $\IZ_2$ orbifolds which project out
whatever gravitino might have existed in our original $D$-dimensional model.   
The $\IZ_2$ orbifold relating the supersymmetric $SO(32)$ and $E_8\times E_8$ strings to each
other does not have this property.
Likewise, there also exists a non-trivial $\IZ_2$ orbifold [specifically 
$(-1)^F$] which maps the supersymmetric $SO(32)$ and $E_8\times E_8$ heterotic strings
to chirality-flipped versions of {\it themselves}\/.   This somewhat degenerate orbifold 
actually corresponds to the situation {\it without}\/ a Wilson line, and has thus 
already been implicitly considered in Eq.~(\ref{fake}).

Given these results, we see that there are only four non-trivial candidate
Wilson-line choices for the finite-temperature $SO(32)$ heterotic
string.  Likewise, there are only four candidate Wilson-line choices for
the finite-temperature $E_8\times E_8$ heterotic string.  
For each of these Wilson-line choices, we can then construct the
corresponding finite-temperature theory.

It is straightforward to write down the partition functions
of these finite-temperature theories, some of which have already appeared in 
various guises in previous work (see, \eg, Refs.~\cite{Rohm,IT,julie}). 
In each case, we shall follow the exact notations and conventions
established in the Appendix.
However, for convenience, we shall also establish one further
convention.  Although the anti-holomorphic (right-moving)
parts of these partition functions will always be expressed in terms of
the (barred) characters $\overline{\chi}_i$ of the 
transverse $SO(8)$ Lorentz group, 
it turns out that we can express the holomorphic (left-moving) parts of
each of these partition functions in terms of the
(unbarred) characters $\chi_i \chi_j$ associated with the
group $SO(16)\times SO(16)$.
Indeed, it turns out that such a rewriting is possible in each case regardless 
of the actual gauge group $G$ of the ten-dimensional model that is produced by
the $\IZ_2$ orbifold.
Of course, if $SO(16)\times SO(16)$ is a subgroup of $G$, then such a rewriting
is meaningful and the characters which appear in the resulting partition function 
correspond to the actual gauge-group representations
which appear in spectrum of the model.
By contrast, if $SO(16)\times SO(16)$ is not a subgroup of $G$,
then such a rewriting is merely an algebraic exercise;
the $SO(16)\times SO(16)$ characters then have no meaning beyond their $q$-expansions,
and can appear with non-integer coefficients.
In all cases, however, these expressions represent the true partition
functions of these thermal theories as far as their $q$-expansions are concerned.
We shall therefore follow these conventions in what follows.

Let us begin by considering the 
zero-temperature supersymmetric $SO(32)$ heterotic string, 
which has the partition function given in Eq.~(\ref{start}).
For this string,
our four possible finite-temperature extensions are then as follows.
In each case we shall label each of the possibilities according to the 
$T\to\infty$ model produced by the corresponding $\IZ_2$ orbifold.
The partition function of the thermal model associated 
with the $\IZ_2$ orbifold producing the non-supersymmetric 
$SO(32)$ heterotic string is given by
\beqn
    Z_{SO(32)} ~=~  Z^{(8)}_{\rm boson} \,\times \,\bigl\lbrace 
     &\phantom{+}&
     \lbrack \chibar_V \,(\chi_I^2 + \chi_V^2)  ~-~  \chibar_S \,(\chi_S^2 + \chi_C^2) 
              \rbrack ~ \calE_0 \nonumber\\
     &+&
       \lbrack \chibar_V \,(\chi_S^2+\chi_C^2)  ~-~  \chibar_S \,(\chi_I^2+\chi_V^2) 
       \rbrack ~ \calE_{1/2} \nonumber\\
     &+&
      \lbrack \chibar_I \,(\chi_I\chi_V+\chi_V\chi_I) ~-~ \chibar_C \,(\chi_S\chi_C+\chi_C\chi_S) 
        \rbrack ~ \calO_{0} \nonumber\\
     &+&
       \lbrack \chibar_I \,(\chi_S\chi_C+\chi_C\chi_S) ~-~ \chibar_C \,(\chi_I\chi_V+\chi_V\chi_I) 
         \rbrack ~\calO_{1/2}~~\bigr\rbrace~, 
\label{A1}
\eeqn
while the partition functions of the thermal models associated the $\IZ_2$ orbifolds that
produce the the $SO(8)\times SO(24)$,
$U_{16}$, and $SO(16)\times SO(16)$ models are respectively given
by
\beqn
     Z_{SO(8)\times SO(24)} ~=~  Z^{(8)}_{\rm boson} \,\times \,\bigl\lbrace 
    &\phantom{+}&
     \lbrack \chibar_V \,(\chi_I^2 + \frac{1}{4}\chi_V^2 + \frac{3}{4}\chi_S^2)  
           ~-~  \chibar_S \,(\frac{1}{4}\chi_S^2  + \frac{3}{4}\chi_V^2 + \chi_C^2) \rbrack ~\calE_0 
           \nonumber\\
   &+&
       \lbrack \chibar_V \,(\frac{1}{4}\chi_S^2  + \frac{3}{4}\chi_V^2 + \chi_C^2)  
           ~-~  \chibar_S \,(\chi_I^2 + \frac{1}{4}\chi_V^2 + \frac{3}{4}\chi_S^2) \rbrack ~\calE_{1/2}  
            \nonumber\\
   &+&
      \lbrack \chibar_I \,(\frac{1}{2}\chi_I\chi_V + \frac{3}{2}\chi_V\chi_C) 
           ~-~ \chibar_C \,(\frac{1}{2}\chi_S\chi_C + \frac{3}{2}\chi_I\chi_S) \rbrack ~\calO_0 
           \nonumber\\
   &+&
       \lbrack \chibar_I \,(\frac{1}{2}\chi_S\chi_C + \frac{3}{2}\chi_I\chi_S) 
           ~-~ \chibar_C \,(\frac{1}{2}\chi_I\chi_V + \frac{3}{2}\chi_V\chi_C) \rbrack ~\calO_{1/2} 
         ~~\bigr\rbrace~,
\label{A2}
\eeqn
\beqn
     Z_{U(16)} ~=~  Z^{(8)}_{\rm boson} \,\times \,\bigl\lbrace 
    &\phantom{+}&
     \lbrack \chibar_V \,(\chi_I^2 + \frac{1}{16}\chi_V^2 + \frac{15}{16}\chi_S^2)  
           ~-~  \chibar_S \,(\frac{1}{16}\chi_S^2  + \frac{15}{16}\chi_V^2 + \chi_C^2) \rbrack ~\calE_0 
           \nonumber\\
       &+&
       \lbrack \chibar_V \,(\frac{1}{16}\chi_S^2  + \frac{15}{16}\chi_V^2 + \chi_C^2)  
           ~-~  \chibar_S \,(\chi_I^2 + \frac{1}{16}\chi_V^2 + \frac{15}{16}\chi_S^2) \rbrack ~\calE_{1/2}  
           \nonumber\\
   &+&
      \lbrack \chibar_I \,(\frac{1}{8}\chi_I\chi_V + \frac{15}{8}\chi_V\chi_C) 
           ~-~ \chibar_C \,(\frac{1}{8}\chi_S\chi_C + \frac{15}{8}\chi_I\chi_S) \rbrack ~\calO_0 
           \nonumber\\
   &+&
       \lbrack \chibar_I \,(\frac{1}{8}\chi_S\chi_C + \frac{15}{8}\chi_I\chi_S) 
           ~-~ \chibar_C \,(\frac{1}{8}\chi_I\chi_V + \frac{15}{8}\chi_V\chi_C) \rbrack ~\calO_{1/2} 
            ~~\bigr\rbrace~, 
\label{A3}
\eeqn
and
\beqn
    Z_{SO(16)\times SO(16)}  ~=~  Z^{(8)}_{\rm boson} \,\times \,\bigl\lbrace 
    &\phantom{+}&
    \lbrack \chibar_V \,(\chi_I^2 + \chi_S^2)  ~-~  \chibar_S \,(\chi_V^2+\chi_C^2) \rbrack ~\calE_0 
           \nonumber\\
   &+&
     \lbrack \chibar_V \,(\chi_V^2 + \chi_C^2)  ~-~  \chibar_S \,(\chi_I^2+\chi_S^2) \rbrack~ \calE_{1/2} 
           \nonumber\\
   &+&
    \lbrack  \chibar_I \,(\chi_V\chi_C+\chi_C\chi_V) ~-~ \chibar_C \,(\chi_I\chi_S+\chi_S\chi_I) \rbrack ~\calO_0  
           \nonumber\\
   &+&
     \lbrack \chibar_I \,(\chi_I\chi_S+\chi_S\chi_I) ~-~ \chibar_C\,(\chi_V\chi_C+\chi_C\chi_V) \rbrack~ \calO_{1/2} 
       ~~\bigr\rbrace~.
\label{firstso16}
\eeqn
Note that as $T\to 0$, each of these expressions reduces to 
the partition function of the zero-temperature supersymmetric $SO(32)$ heterotic string
in Eq.~(\ref{start}), as required.

As is easy to verify, these four different thermal extensions of the
supersymmetric $SO(32)$ heterotic string correspond to the Wilson lines
\beqn
                   {\rm non{-}SUSY}~SO(32):&~~~~~ \vec\ell ~=& ((1)^n (0)^{16-n})  ~~~~~{\rm for}~n\in 2\IZ+1~
\nonumber\\
         SO(8)\times SO(24):&~~~~~ \vec\ell ~=& ((\half)^4 (0)^{12}) ~~{\rm or}~~  ((\textstyle{3\over 2})^4 (0)^{12})
\nonumber\\
        SO(16)\times SO(16):&~~~~~ \vec\ell ~=& ((\half)^8 (0)^8) ~~{\rm or}~~  ((\textstyle{3\over 2})^8 (0)^8)
\nonumber\\
                     U(16):&~~~~~ \vec\ell ~=& ((\ell)^{16}) ~~~~~{\rm for}~~\ell \in \lbrace 
               \textstyle{1\over 4}, \textstyle{3\over 4}, \textstyle{5\over 4}, \textstyle{7\over 4} \rbrace~.
\label{hetcases}
\eeqn
Indeed, because our original supersymmetric $SO(32)$ heterotic theory contains only vectorial and
spinorial representations of $SO(32)$, each of the individual components of the Wilson line $\vec\ell$
is defined only modulo~2.

A similar situation exists for the zero-temperature $E_8\times E_8$ heterotic string,
which has partition function
\beq
     Z^{(8)}_{\rm boson} \,
      (\chibar_V -\chibar_S) \,(\chi_I + \chi_S)^2~
\label{e8pf}
\eeq
The partition function of the thermal extension 
of this model associated with the $\IZ_2$ orbifold producing the
non-supersymmetric $SO(16)\times E_8$ model
is given by
\beqn
      Z_{SO(16)\times E_8} ~=~  Z^{(8)}_{\rm boson} \,\times
            \,\bigl\lbrace 
             &\phantom{+}& 
                \lbrack \chibar_V\,\chi_I ~-~ \chibar_S\,\chi_S \rbrack ~\calE_0   
             \nonumber\\
            &+&
                \lbrack
                \chibar_V\,\chi_S ~-~ \chibar_S\,\chi_I \rbrack ~\calE_{1/2}  
             \nonumber\\
            &+&
                \lbrack
              \chibar_I\,\chi_V ~-~ \chibar_C\,\chi_C \rbrack ~\calO_0   
             \nonumber\\
            &+&
                \lbrack
               \chibar_I\,\chi_C  ~-~ \chibar_C\,\chi_V \rbrack ~\calO_{1/2}  
               ~~\bigr\rbrace~ \times~ (\chi_I+\chi_S)~,
\label{B1}
\eeqn
while the partition functions of the thermal models associated with the $(E_7)^2\times SU(2)^2$, $E_8$, 
and $SO(16)\times SO(16)$ orbifolds are respectively given by
\beqn
     Z_{(E_7)^2\times SU(2)^2} ~=~  Z^{(8)}_{\rm boson} \,\times \,\bigl\lbrace 
    &\phantom{+}&
                \lbrack \chibar_V \,(\chi_I^2 + \frac{1}{4}\chi_I \chi_S + \frac{3}{4}\chi_S^2)  
          ~-~  \chibar_S \,(\frac{1}{4}\chi_S^2 + \frac{7}{4}\chi_I \chi_S ) \rbrack ~\calE_0 
             \nonumber\\
   &+& \lbrack \chibar_V \,(\frac{1}{4}\chi_S^2 + \frac{7}{4}\chi_I \chi_S )  
          ~-~  \chibar_S \,(\chi_I^2 + \frac{1}{4}\chi_I \chi_S + \frac{3}{4}\chi_S^2) \rbrack ~\calE_{1/2}  
             \nonumber\\
   &+& \lbrack \chibar_I \,(\frac{1}{4}\chi_I\chi_V + \frac{7}{4}\chi_V\chi_S) 
          ~-~ \chibar_C \,(\frac{1}{4}\chi_S\chi_S + \frac{7}{4}\chi_I\chi_S) \rbrack ~\calO_0 
             \nonumber\\
   &+& \lbrack \chibar_I \,(\frac{1}{4}\chi_S\chi_S + \frac{7}{4}\chi_I\chi_S) 
          ~-~ \chibar_C \,(\frac{1}{4}\chi_I\chi_V + \frac{7}{4}\chi_V\chi_S) \rbrack ~\calO_{1/2} 
          ~~\bigr\rbrace~, 
\label{B2}
\eeqn
\beqn
     Z_{E_8} ~=~  Z^{(8)}_{\rm boson} \,\times \,\bigl\lbrace 
    &\phantom{+}&
     \lbrack \chibar_V \,(\chi_I^2 + \frac{1}{16}\chi_I \chi_S + \frac{15}{16}\chi_S^2)  
          ~-~  \chibar_S \,(\frac{1}{16}\chi_S^2 + \frac{31}{16}\chi_I \chi_S ) \rbrack ~\calE_0 
             \nonumber\\
   &+&
       \lbrack \chibar_V \,(\frac{1}{16}\chi_S^2 + \frac{31}{16}\chi_I \chi_S)  
          ~-~  \chibar_S \,(\chi_I^2 + \frac{1}{16}\chi_I \chi_S + \frac{15}{16}\chi_S^2) \rbrack ~\calE_{1/2}  
             \nonumber\\
   &+&
      \lbrack \chibar_I \,(\frac{1}{16}\chi_I\chi_V + \frac{31}{16}\chi_V\chi_S) 
          ~-~ \chibar_C \,(\frac{1}{16}\chi_S\chi_S + \frac{31}{16}\chi_I\chi_S) \rbrack ~\calO_0 
             \nonumber\\
   &+&
       \lbrack \chibar_I \,(\frac{1}{16}\chi_S\chi_S + \frac{31}{16}\chi_I\chi_S) 
          ~-~ \chibar_C \,(\frac{1}{16}\chi_I\chi_V + \frac{31}{16}\chi_V\chi_S) \rbrack ~\calO_{1/2} 
          ~~\bigr\rbrace~, 
\label{B3}
\eeqn
and
\beqn
    Z_{SO(16)\times SO(16)} ~=~  Z^{(8)}_{\rm boson} \,\times \,\bigl\lbrace 
    &\phantom{+}&
           \lbrack
          \chibar_V \,(\chi_I^2 + \chi_S^2)  ~-~  \chibar_S \,(\chi_I\chi_S + \chi_S\chi_I) \rbrack~ \calE_0   
             \nonumber\\
   &+&
         \lbrack
          \chibar_V \,(\chi_I\chi_S+\chi_S\chi_I)  ~-~  \chibar_S \,(\chi_I^2+\chi_S^2) \rbrack ~\calE_{1/2} 
             \nonumber\\
   &+&
           \lbrack
           \chibar_I \,(\chi_V\chi_C+\chi_C\chi_V) ~-~ \chibar_C \,(\chi_V^2 +\chi_C^2) \rbrack ~\calO_0   
             \nonumber\\
   &+&
             \lbrack
           \chibar_I \,(\chi_V^2+\chi_C^2) ~-~ \chibar_C \,(\chi_V\chi_C+\chi_C\chi_V) \rbrack ~\calO_{1/2}  
      ~~\bigr\rbrace~.
\label{secondso16}
\eeqn
Once again, using the identities listed in the Appendix, it is straightforward to
verify that each of these expressions reduces to Eq.~(\ref{e8pf}) as $T\to 0$.
Moreover, the expressions in Eqs.~(\ref{firstso16}) and (\ref{secondso16}) are actually
equal as the result of the further identity on $SO(16)$ characters given by
\beq
          \chi_I \chi_S + \chi_S \chi_I ~=~ \chi_V^2 + \chi_C^2~.
\label{identy}
\eeq
This is ultimately the identity which is responsible for the fact that 
the two expressions within Eq.~(\ref{pfs2})
are equal at the level of their $q$-expansions, \ie,
that the ten-dimensional supersymmetric $SO(32)$ and $E_8\times E_8$ heterotic strings 
have the same bosonic and fermionic state degeneracies at each mass level.

\begin{figure}[b!]
\centerline{
   \epsfxsize 3.5 truein \epsfbox {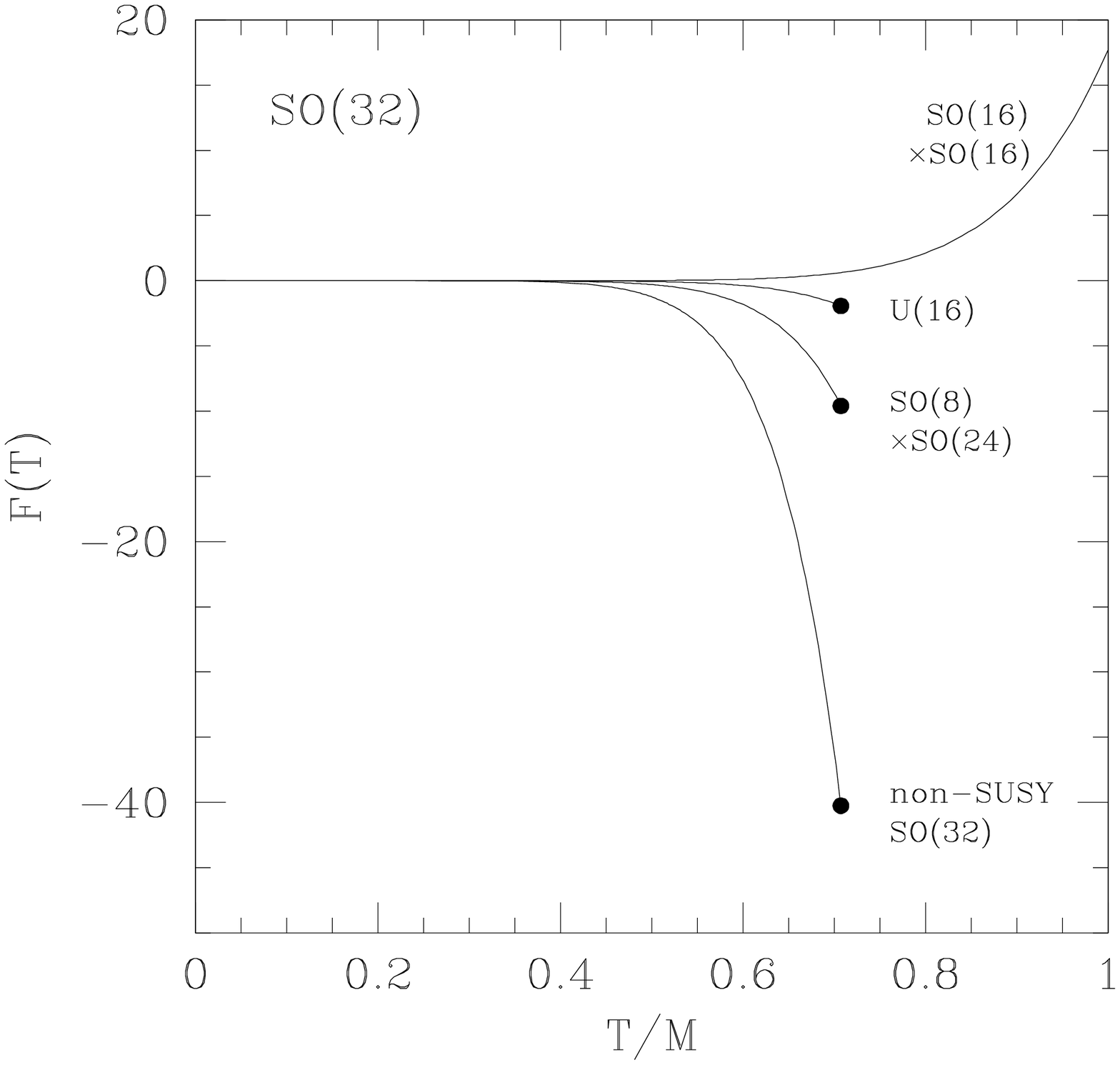} 
   \epsfxsize 3.5 truein \epsfbox {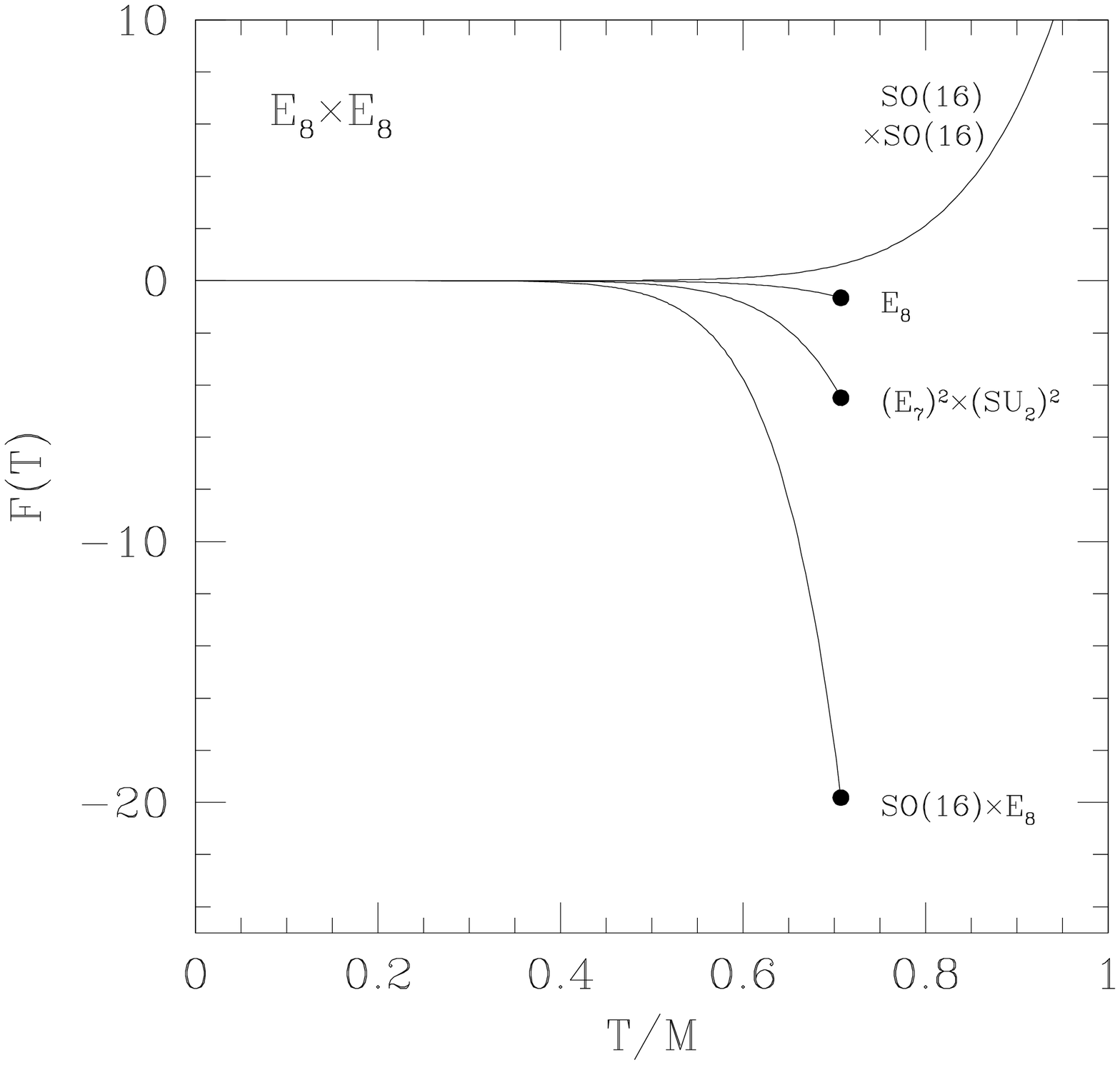} }
\caption{
    Free-energy densities $F(T)$ in units of $\half \calM^{10}=\half (M_{\rm string}/2\pi)^{10}$,
    plotted as functions of the normalized temperature $T/\calM$
    for the $SO(32)$ heterotic string (left plot) and
    $E_8\times E_8$ heterotic string (right plot).
    In each case, the free energies are shown for the four corresponding choices 
    of allowed non-trivial Wilson lines.
    We see that in general $F(T)\to 0$ as $T\to 0$, in accordance with the 
    spacetime supersymmetry which exists at zero temperature.  
    At non-zero temperatures, however, the spacetime supersymmetry is necessarily broken.
    Interestingly, we see that the non-trivial Wilson line 
    which leads to the smallest free-energy density
    in each case is the one which breaks the gauge group minimally:
    for the $SO(32)$ string, this is the Wilson line associated with the
    the non-supersymmetric $SO(32)$ orbifold, while for the $E\times E_8$ heterotic 
    string, this is the Wilson line associated with the
    $SO(16)\times E_8$ orbifold.
    With the sole exception of the Wilson line leading to the $SO(16)\times SO(16)$ heterotic
    string, each of the non-trivial Wilson-line choices in each
    case leads to a free energy which is negative for all $T>0$ and which diverges discontinuously
    at the critical temperature $T_H\equiv \calM /\sqrt{2}$ (indicated in 
    each case with a solid black dot).
    These divergences arise in each case due to the existence of   
    a thermal winding state which is massive for all $T<T_H$, massless at $T=T_H$,
    and tachyonic for all $T>T_H$, signalling a Hagedorn transition at $T=T_H$.} 
\label{Ffig}
\end{figure}

As an aside, it is interesting to note that all of these thermal functions can be written
in a common form parametrized by a single integer $\zeta$:
\beqn
       Z^{(8)}_{\rm boson} \,\times \,\biggl\lbrace 
    &\phantom{+}&
     \biggl\lbrack \chibar_V \,\left(\chi_I^2 + \frac{1}{\zeta}\chi_V^2 + \frac{\zeta-1}{\zeta}\chi_S^2\right)  -  
      \chibar_S \,\left(\chi_C^2 + \frac{1}{\zeta}\chi_S^2  + \frac{\zeta-1}{\zeta}\chi_V^2\right) \biggr\rbrack ~\calE_0 
            \nonumber\\
   &+& \biggl\lbrack \chibar_V \,\left(\chi_C^2 + \frac{1}{\zeta}\chi_S^2  + \frac{\zeta-1}{\zeta}\chi_V^2\right)  -  
     \chibar_S \,\left(\chi_I^2 + \frac{1}{\zeta}\chi_V^2 + \frac{\zeta-1}{\zeta}\chi_S^2\right) \biggr\rbrack ~\calE_{1/2}  
            \nonumber\\
   &+& \biggl\lbrack \chibar_I \,\left(\frac{1}{\zeta}\chi_I\chi_V + \frac{1}{\zeta}\chi_V\chi_I 
              +  \frac{\zeta-1}{\zeta}\chi_V\chi_C + \frac{\zeta-1}{\zeta}\chi_C\chi_V\right) 
                                       \nonumber\\ && \phantom{\lbrack}
                 -\chibar_C \,\left(\frac{1}{\zeta}\chi_S\chi_C 
            + \frac{1}{\zeta}\chi_C\chi_S +  \frac{\zeta-1}{\zeta}\chi_V^2 + \frac{\zeta-1}{\zeta}\chi_C^2\right) 
                       \biggr\rbrack ~\calO_0 
            \nonumber\\
   &+& \biggl\lbrack \chibar_I \,\left(\frac{1}{\zeta}\chi_S\chi_C + \frac{1}{\zeta}\chi_C\chi_S 
             + \frac{\zeta-1}{\zeta}\chi_V^2 + \frac{\zeta-1}{\zeta}\chi_C^2\right) 
                                       \nonumber\\ && \phantom{\lbrack}
            -\chibar_C \,\left(\frac{1}{\zeta}\chi_I\chi_V 
           + \frac{1}{\zeta}\chi_V\chi_I +  \frac{\zeta-1}{\zeta}\chi_V\chi_C + \frac{\zeta-1}{\zeta}\chi_C\chi_V\right) 
                         \biggr\rbrack ~\calO_{1/2} 
            ~\biggr\rbrace~.
\label{singleInterpolation}
\eeqn
In particular, the values $\zeta=\lbrace 1,2,4,8,16,32,\infty\rbrace$
correspond to the partition functions
in Eqs.~(\ref{A1}), (\ref{B1}), (\ref{A2}), (\ref{B2}), (\ref{A3}), 
(\ref{B3}), and (\ref{firstso16}) [or (\ref{secondso16})] 
respectively, where Eq.~(\ref{identy}) has been used wherever needed.

It is also instructive to examine the free-energy densities $F(T)$ associated with
each of these possible Wilson-line choices.
As we have seen, for each thermal partition function $Z(\tau,T)$ listed
above, the corresponding free-energy density $F(T)$ is given
by Eq.~(\ref{freeenergydef}).
Following this definition, 
we then obtain the results shown in Fig.~\ref{Ffig}.

We observe from Fig.~\ref{Ffig} that the non-trivial Wilson line 
which minimizes the free-energy density in each case is 
the one which breaks the gauge group minimally.
For the $SO(32)$ string, this is the Wilson line associated with  
the non-supersymmetric $SO(32)$ orbifold, while for the $E\times E_8$ heterotic 
string, this is the Wilson line associated with the non-supersymmetric
$SO(16)\times E_8$ orbifold.
It is tempting to say, therefore, that these particular non-trivial Wilson lines are somehow
``preferred'' in some dynamical sense over the others.  However, this assumption presupposes
the existence of a mechanism by which these Wilson lines can smoothly be deformed into each
other with finite energy cost.  
Given that these Wilson lines ultimately correspond to fluxes which are not only constrained
topologically but also presumably quantized, such Wilson-line-changing transitions would require
exotic physics (such as might occur on a full thermal landscape).  We shall discuss the structure
of such a landscape in Sect.~VI.
We also note that for both of our supersymmetric heterotic strings, there remains the traditional option
of constructing a thermal theory {\it without}\/ a non-trivial Wilson line. 
It turns out that the free energies corresponding to these choices are numerically almost identical (but 
ultimately slightly smaller) than those of the non-supersymmetric $SO(32)$ and $SO(16)\times E_8$ cases plotted in
Fig.~\ref{Ffig}.  These features will be discussed further in Sect.~VI.

We see, then, that have been able to construct four 
new thermal theories for the supersymmetric $SO(32)$ heterotic string 
as well as four new thermal theories for the supersymmetric 
$E_8\times E_8$ heterotic string.
Each of these theories has the novel feature that a non-trivial Wilson
line has been introduced when constructing the finite-temperature extension,
or equivalently that a non-trivial temperature-dependent chemical potential 
has been introduced into the Boltzmann sum.
Each of these theories reduces to the correct supersymmetric theory as $T\to 0$,
and moreover
each is modular invariant for all temperatures $T$.
Even more importantly, the temperature/radius correspondence guarantees that in each case,
the temperature variable $T$ --- like the radius variable $R$ to which it corresponds ---
is a bona-fide {\it modulus}\/ of the theory, able to be freely changed without disturbing
the worldsheet self-consistency of the string.
Despite the fact that the non-trivial Wilson lines we have introduced
in each case have led to certain unorthodox modings for our string states around
the thermal circle, 
none of these theories
violates any spin-statistics relations.
Indeed, the spin-statistics theorem relates the spacetime Lorentz spin
of a given quantum field 
to its thermal statistics, and the temperature/radius correspondence
relates such thermal statistics to the periodicity of such a field
around the thermal circle.  Indeed, it is only the relation between
this periodicity and the resulting algebraic moding which is altered
as a result of the non-trivial Wilson line.

\subsection{Type~I strings}
\setcounter{footnote}{0}

We now turn our attention to the corresponding situation for Type~I strings.
In ten dimensions, there is a single self-consistent Type~I string model which is both
supersymmetric and anomaly-free:  this is the $SO(32)$ Type~I string~\cite{Polbook}.
Our goal is therefore to survey the possible Wilson lines which can be 
introduced when formulating its thermal extension.

As discussed in Sect.~II, the 
ten-dimensional zero-temperature $SO(32)$ Type~I string has a partition
function given in Eq.~(\ref{TypeIfourcontribs}), and its extension to finite
temperature {\it without}\/ Wilson lines is given in Eq.~(\ref{TypeIthermal}). 
However, just as for the heterotic strings, we expect that new thermal possibilities
can be constructed when non-trivial Wilson lines are introduced~\cite{julie,ADS1998,AS2002}.

In general, as discussed in Sect.~II, there are two kinds of Wilson lines which
might be introduced for Type~I theories.
First, there are Wilson lines that might be introduced into the closed-string sectors
of such theories, much along the lines we have already discussed for the heterotic strings.  
However, the closed-string sectors of Type~I strings are essentially Type~II superstrings
(indeed, these are the strings from which the Type~I strings can be obtained by orientifolding),
and in ten dimensions the perturbative states of such Type~II strings do not carry gauge charges.
Thus, for the ten-dimensional Type~I string, it is not possible to introduce a non-trivial Wilson line
in the closed-string sector.  This guarantees that the results for $Z_{\rm T}(\tau,T)$ and $Z_{\rm K}(\tau,T)$
given in Eq.~(\ref{TypeIthermal}) will remain invariant regardless of what happens in the open-string
sector.

The question then boils down to determining 
the allowed Wilson lines that might be introduced in the {\it open-string}\/
sector of the ten-dimensional Type~I string.
Indeed, because
the states contributing to the cylinder and \Mob\ partition functions
carry $SO(32)$ gauge charges, their modings in Eq.~(\ref{TypeIthermal}) are potentially affected 
by the presence of an $SO(32)$ Wilson line.
Fortunately, thanks to the temperature/radius correspondence, this problem can be mapped
to the purely geometric issue of determining the allowed Wilson lines that can be introduced when compactifying
the Type~I string to nine dimensions on a circle --- indeed, the fact that we continually refer to ``Wilson lines''
and ``thermal circles'' already implicitly presupposes that this can be done!
It turns out that the allowed Wilson lines fall into two distinct classes.

The first class consists of Wilson lines of the form
\beq
                     \vec\ell ~=~ (\half,\half,\half,...,0,0,0,...)
\label{wilsonclass1}
\eeq
where the number of non-zero components is given by $n$, with $0\leq n<16$.
The $n=0$ special case corresponds to the case without a Wilson line, and in general
we shall define $n_1\equiv 2n$ and $n_2\equiv 32-n_1$.
For Wilson lines of this form, 
the \Mob\ contribution in Eq.~(\ref{TypeIthermal})
turns out to be independent of $n$, and thus 
remains the same as in Eq.~(\ref{TypeIthermal}) for all $n>0$:
\beq
 {\rm Mobius}:  ~~~~~~   Z_{\rm M}(\taut,T) ~=~ 
                         -\half\, \widehat{Z}_{\rm open}^{(8)} \, 
                               (n_1+n_2) \, (\chihat_V \,\calE -\chihat_S \,\calE')~.
\eeq
However, we find that the cylinder
contribution in Eq.~(\ref{TypeIthermal}) now takes the form
\beqn
 {\rm cylinder}:  ~~~~~Z_{\rm C}(\taut,T) ~=~ 
               \half \, Z_{\rm open}^{(8)} \,\times \,\bigl\lbrace 
          && \lbrack (n_1^2 + n_2^2) \chi_V  -  2 n_1 n_2  \chi_S \rbrack \, \calE \nonumber\\
         &-& \lbrack (n_1^2 + n_2^2) \chi_S  -  2 n_1 n_2  \chi_V \rbrack \, \calE' ~\bigr\rbrace~.
\label{newcylinder}
\eeqn
It is easy to demonstrate that as a result of the shift induced by this Wilson line,
the gauge group of the resulting model is broken to $SO(n_1)\times SO(n_2)$.
[In the T-dual picture, the choice of the Wilson line in Eq.~(\ref{wilsonclass1}) indicates that we have simply 
moved $n_1$ of the original 32 D8-branes in this theory to the opposite side of the thermal circle.]
Note, however, that the appearance of this Wilson line has also induced states with the ``wrong'' thermal modings
to appear in Eq.~(\ref{newcylinder}).  Specifically, we see from Eq.~(\ref{newcylinder}) that we
now have spacetime spinors accruing integer thermal momentum modes within $\calE$,
while we also have spacetime vectors accruing half-integer thermal modes within $\calE'$.

The second ``class'' of Wilson lines we shall consider consists of a single Wilson line
of the form
\beq
                     \vec\ell ~=~ (\quarter,\quarter,\quarter,...,\quarter)~.
\label{wilsonclass2}
\eeq
For this Wilson line, the cylinder and \Mob\ partition functions in Eq.~(\ref{TypeIthermal})
now take the form
\beqn
 {\rm cylinder}:  ~~~~~Z_{\rm C}(\taut,T)  &=&
               \phantom{-}\half\, Z_{\rm open}^{(8)} \,\times \,\bigl\lbrace 
          \phantom{+} \lbrack 2 n \nbar  \chi_V - (n^2 + \nbar^2) \chi_S  \rbrack \, \calE \nonumber\\
          && \phantom{-\half Z_{\rm open}^{(8)} \,\times \bigl\lbrace} 
        - \lbrack 2 n \nbar  \chi_S - (n^2 + \nbar^2) \chi_V  \rbrack \, \calE' ~\bigr\rbrace\nonumber\\
 {\rm Mobius}:  ~~~~~Z_{\rm M}(\taut,T)  &=&
               -\half \, \widehat Z_{\rm open}^{(8)} \,(n+\nbar)\,\bigl(
             -\chihat_S \,\calE + \chihat_V\,\calE' \bigr) ~
\label{newcylmob}
\eeqn
where $n=\nbar=16$.
In this case, the Wilson line has deformed the gauge group of our original $SO(32)$ theory to $U(16)$.
Note that the alternate Wilson line $\vec\ell=( \textstyle{3\over 4}, \textstyle{3\over 4}, \textstyle{3\over 4},...,
\textstyle{3\over 4})$
produces the same theory.
In either case, however, we once again observe that the 
Wilson line has induced states to appear in Eq.~(\ref{newcylmob}) with the ``wrong'' thermal modings.

It should be stressed that when discussing the possible ``allowed'' Wilson lines, we are not enforcing
the possible open-string NS-NS tadpole-anomaly constraints for all temperatures (as might normally be
done within a more general Type~I model-building framework).  
Indeed, only the $SO(16)\times SO(16)$ and $U(16)$ cases
outlined above satisfy these constraints and completely avoid NS-NS tadpole divergences at all temperatures;
in all other cases, these constraints are satisfied only for temperatures below the Hagedorn temperature.  
However, this approach
is justified in this context because we are not seeking to avoid the possible emergence of open-string tachyons.  
In fact, such tachyons and the divergences they induce are both desired
and expected, since these are precisely the features which ultimately trigger the Hagedorn transition 
for Type~I strings.

It should also be stressed that there are many different ways of
obtaining the models discussed in this section.  While one approach
involves compactifying the ten-dimensional supersymmetric Type~I string on
the thermal circle in the presence of various Wilson lines, it is also
possible to compactify the Type~II string directly on the thermal circle,
implementing the orientifold projection only after this compactification
is performed (see, \eg, Ref.~\cite{earlyOrientifold}).  The different
allowed choices for open-string sectors in this orientifold projection
then yield the models we have constructed here.  
Regardless of the approach taken, however, we see that there are only a
finite set of self-consistent possibilities which are available as
potential finite-temperature extensions of the zero-temperature Type~I
string.


\begin{figure}[t!]
\centerline{
   \epsfxsize 3.5 truein \epsfbox {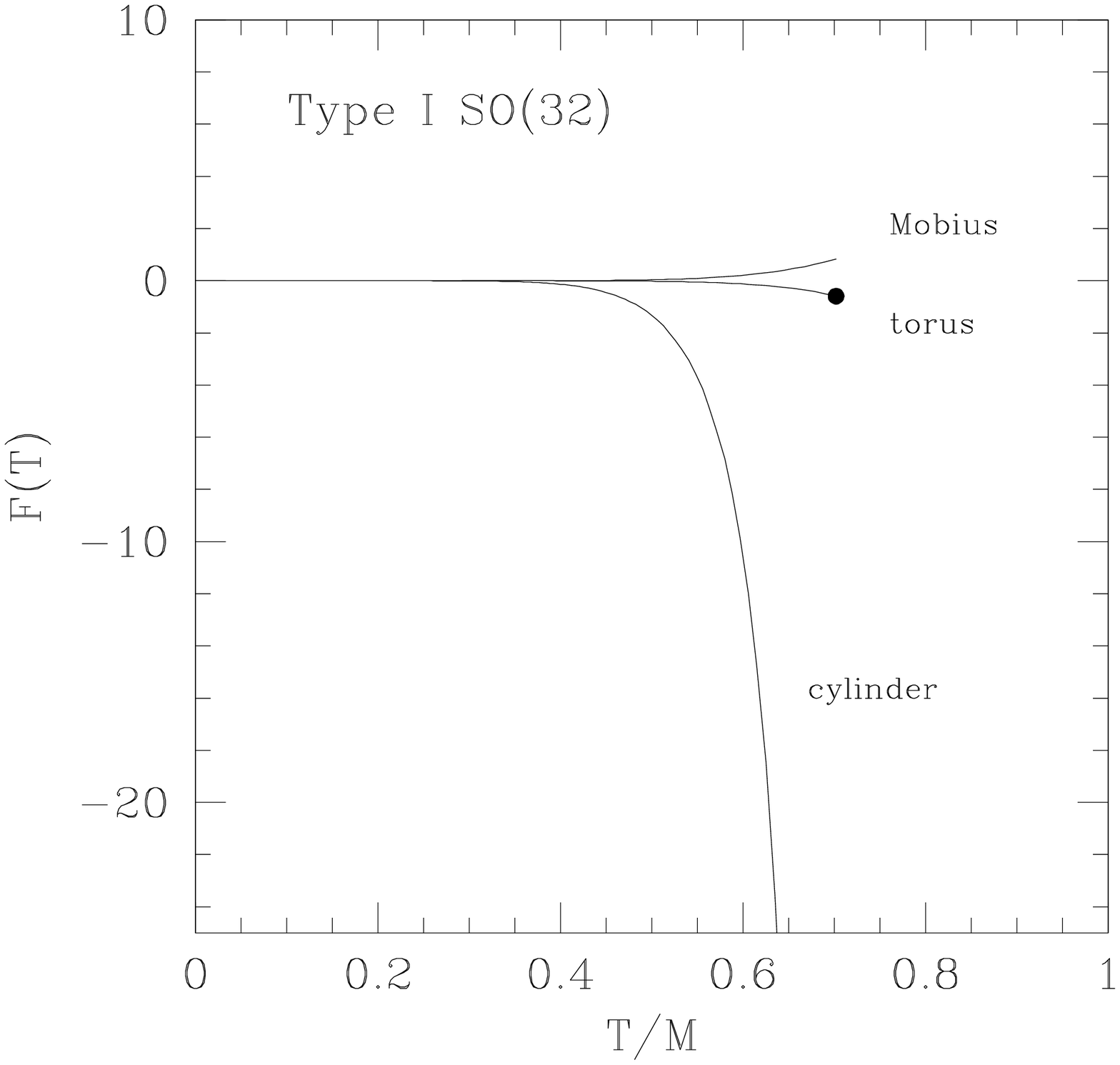}
   \epsfxsize 3.5 truein \epsfbox {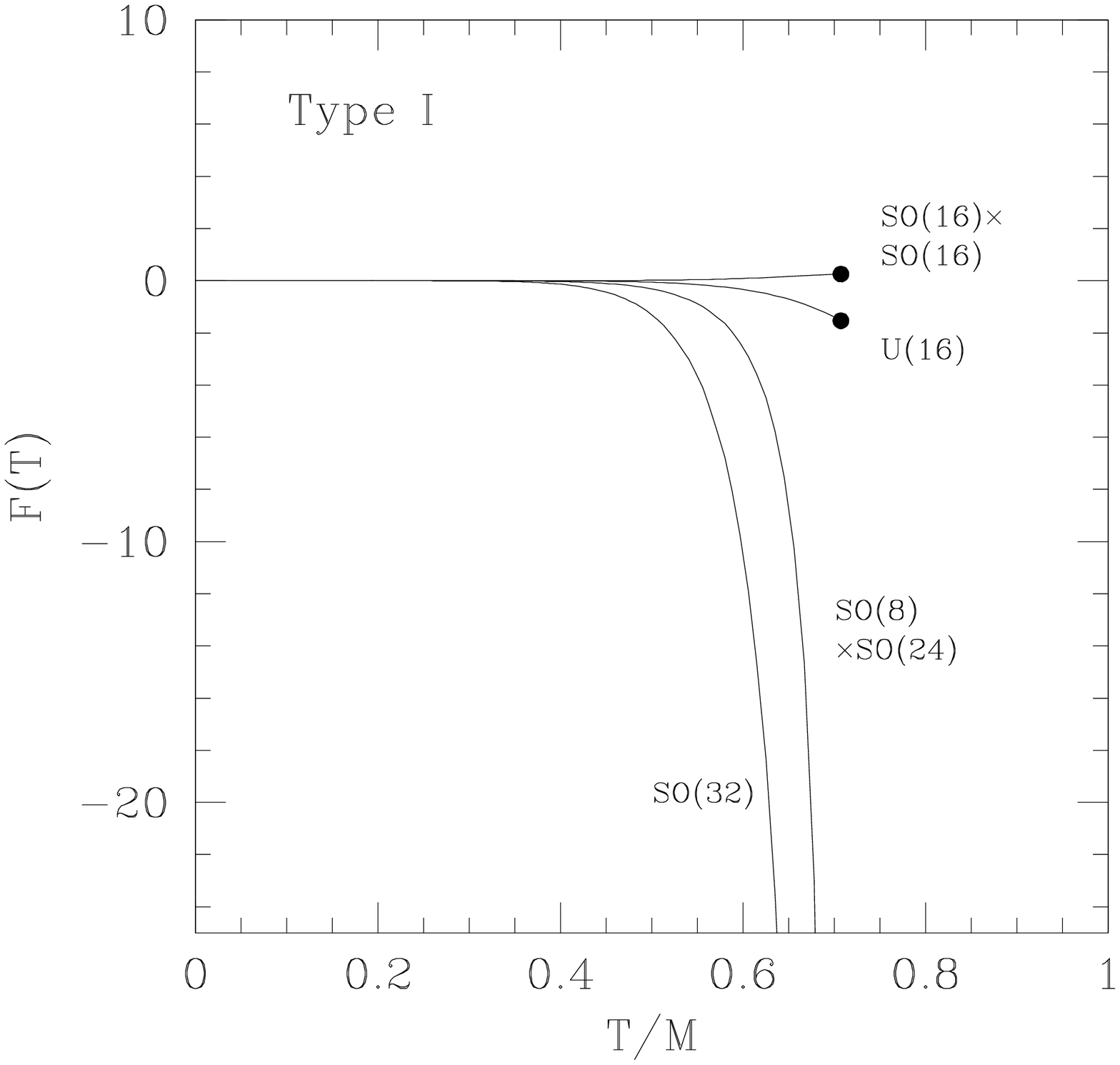}}
\caption{Free-energy densities $F(T)$ in units of $\half\calM^{10}=\half(M_{\rm string}/2 \pi)^{10}$, 
  plotted as functions of the normalized temperature $T/\calM$ for the zero-temperature 
   ten-dimensional $SO(32)$ Type~I string
  extended to finite temperature.
  (a) {\it Left plot:}\/  Individual torus, cylinder, and \Mob\ contributions to $F(T)$ 
  for the $SO(32)$ case without a Wilson line.   
  We see that in general each contribution vanishes as $T\to 0$, in accordance with the
  spacetime supersymmetry which exists at zero temperature;  likewise, the Klein-bottle
  contribution vanishes for all temperatures.
  Note that the \Mob\ contribution remains finite at the critical Hagedorn temperature
  $T_H\equiv \calM/\sqrt{2}$, while the torus contribution diverges discontinuously 
  at $T_H$ (as indicated with a solid dot) and the 
  cylinder contribution diverges continuously as $T\to T_H$.
  (b) {\it Right plot:}\/
  Total free-energy densities $F(T)$ corresponding to the 
  Wilson-line choices
  associated with the gauge groups $SO(32)$,
   $SO(8)\times SO(24)$, $SO(16)\times SO(16)$, and $U(16)$.
    Note that it is the $SO(32)$ case 
    which minimizes the free energy.
    Like the analogous case of the ten-dimensional $SO(32)$ heterotic string shown
    in Fig.~\protect\ref{Ffig}, this is also the choice which preserves 
    the zero-temperature gauge symmetry.  
    However, unlike the case of the heterotic string, we see that 
    the corresponding free energy in the Type~I case 
    actually grows without bound as the $T\to T_H$, a feature
    which suggests that the Hagedorn temperature is actually a limiting temperature
    for the Type~I string rather than the location of a phase transition.}
\label{type1plot}
\end{figure}

Given the set of Wilson lines outlined above, we can now
examine their corresponding free energies $F(T)$.
As discussed in Sect.~II, 
for Type~I string models the corresponding free-energy density receives separate contributions  
from the torus, the Klein bottle, the cylinder, and the \Mob\ amplitudes;  
these are shown in Eqs.~(\ref{un}), (\ref{deux}), and (\ref{trois}).
In particular, several particular Wilson lines will interest us, such as those which
yield the gauge groups we have considered for heterotic strings:
\beqn
         {\rm non-SUSY}~SO(32):&~~~~~ \vec\ell ~=& ((0)^{16}) \nonumber\\
         SO(8)\times SO(24):&~~~~~ \vec\ell ~=& ((\half)^4 (0)^{12})
\nonumber\\
        SO(16)\times SO(16):&~~~~~ \vec\ell ~=& ((\half)^8 (0)^8)
\nonumber\\
                     U(16):&~~~~~ \vec\ell ~=& ((\quarter)^{16})~~{\rm or}~~((\textstyle{3\over 4})^{16})~.
\label{type1cases}
\eeqn
The results are shown in Fig.~\ref{type1plot}.
Note that several of these results have also appeared in a different context in Ref.~\cite{julie}.

It is straightforward to understand the general features shown in Fig.~\ref{type1plot}.
First, we recall from the above discussion that 
all four of these possible finite-temperature extensions share the same torus and Klein-bottle contributions
to the free-energy density:
as shown in Fig.~\ref{type1plot}(a),
the torus contribution is relatively small and negative 
for $T>0$, remaining finite until
it diverges discontinuously at the critical temperature $T/\calM=1/\sqrt{2}$, while the Klein-bottle contribution
actually vanishes as a result of the identity $\chi_V=\chi_S$ which holds for the characters 
of the transverse $SO(8)$ Lorentz group.
Thus, as expected, it is the cylinder and \Mob\ contributions which are responsible for the relative differences
between these different Wilson-line choices.

Let us first consider the Wilson-line cases following from the choice in Eq.~(\ref{wilsonclass1}).
As indicated above, these cases all share the same \Mob\ contribution as well.  Like the torus
contribution, the \Mob\ contribution is also relatively small for $T>0$;  however, unlike
the torus contribution, we see from Fig.~\ref{type1plot}(a)
that it is positive rather than negative and does not diverge, even at
$T/\calM=1/\sqrt{2}$.
Indeed, if these were the only three contributions in the $SO$ Wilson-line cases,  
we would obtain the total curve which is shown in Fig.~\ref{type1plot}(b) for $SO(16)\times SO(16)$.
In this case, the discontinuous divergence at $T=\calM/\sqrt{2}$ is solely due to the 
 {\it closed-string}\/ tachyon coming from the torus amplitude in Fig.~\ref{type1plot}(a).
However, for all other cases with $n_1\not= n_2$, we also have a relatively huge cylinder contribution which is 
negative for $T>0$, with $F_{\rm C}(T)\to \infty$ smoothly as $T\to \calM/\sqrt{2}$.
In fact, recalling the $SO(8)$ character identity $\chi_V=\chi_S$, 
we see from Eq.~(\ref{newcylinder}) that the overall general magnitude of 
this contribution is proportional to 
\beq
          n_1^2 + n_2^2 - 2 n_1 n_2 ~=~
        (n_1-n_2)^2 ~=~ (32-n)^2~.
\eeq
Thus the $n=0$ case (\ie, the case with vanishing Wilson line) 
has the most negative cylinder contribution and correspondingly the most negative total free-energy density
from amongst the choices with gauge groups $SO(n_1)\times SO(n_2)$.

Of course, it still remains possible that the $U(16)$ case in Eq.~(\ref{wilsonclass2}) might yield
a free-energy density which is even more negative.
However, since $2n\nbar = (n^2+\nbar^2)$ when
$n=\nbar$, we see that the cylinder contribution
in Eq.~(\ref{newcylmob}) actually vanishes for this case.
The \Mob\ contribution remains small but switches sign, becoming negative, but  
it continues to remain finite, even at $T=\calM/\sqrt{2}$.
[Indeed, as discussed above, only in the $U(16)$ and $SO(16)\times SO(16)$ cases 
are all open-string tachyons avoided.]
As a result, the free-energy density in the $U(16)$ case remains small, diverging discontinuously
at $T=\calM/\sqrt{2}$ only because of the {\it closed-string}\/ tachyon in the torus amplitude.

Comparing Fig.~\ref{type1plot} for the Type~I string with the analogous plot for 
the $SO(32)$ heterotic string in Fig.~\ref{Ffig},
we see certain superficial similarities.
For example, the set of permitted gauge groups is similar in each case,
and moreover Wilson-line choice in each case that leads  
to the minimum free energy is the one that breaks the gauge group minimally.
However, we stress that despite such superficial similarities, there remains one 
fundamental distinction between the heterotic and Type~I cases:
except for the cases involving the particular $SO(16)\times SO(16)$ and $U(16)$ Wilson lines,
the Type~I cases lead to free-energy densities which actually diverge smoothly
as the Hagedorn temperature $T_H=\calM/\sqrt{2}$ is approached, \ie,  
$F(T)\to -\infty$ as $T\to T_H$, while the corresponding heterotic free-energy densities
actually remain finite as $T\to T_H$.
This feature is already well known in the string-thermodynamics literature:  it is a direct
result of the open-string tachyon at $T_H=\calM/\sqrt{2}$,
and suggests that the Hagedorn temperature is actually a limiting temperature for Type~I strings
rather than the location of a phase transition.

\section{Wilson lines and the Hagedorn temperature}

As we have seen, one of the most prominent aspects of thermal string theories is the existence of a Hagedorn 
transition at which the string free-energy density diverges.
However, the introduction of a non-trivial Wilson line can actually change the
temperature at which this transition takes place.
This is particularly true for heterotic strings, and
we have already seen evidence of this fact in Fig.~\ref{Ffig}:  
the free-energy densities corresponding to the different possible Wilson-line choices
all diverge at critical temperatures which differ from
the Hagedorn temperature $T_H=(2-\sqrt{2})\calM$ normally associated with the heterotic string without Wilson lines.

In some sense, it is to be expected that the introduction of non-trivial Wilson lines 
can affect the resulting Hagedorn temperature, since we have seen that such Wilson lines
affect the thermal partition function as a whole.
However, we can equivalently associate the Hagedorn temperature with the asymptotic densities
of bosonic and fermionic states   
in the original zero-temperature theory, and the zero-temperature theory
is clearly independent of the introduction of a non-trivial thermal Wilson line.
Our goal in this section is to explain these different perspectives, and to show how they can ultimately
be reconciled with each other in the presence of a non-trivial Wilson line.
For concreteness, we shall focus on 
the case of the $SO(32)$ and $E_8\times E_8$
heterotic strings,
and examine the consequences of
moving from the standard thermal theories in Eq.~(\ref{fake}) 
which do not involve non-trivial Wilson lines
to the new thermal theories [such as those in Eqs.~(\ref{A1}) through (\ref{firstso16})
for $SO(32)$, and those in Eqs.~(\ref{B1}) through (\ref{secondso16}) for $E_8\times E_8$] which do.

\subsection{The Hagedorn transition:  UV versus IR}

We begin with several preliminary remarks concerning the Hagedorn
transition and its dual UV/IR nature.

The Hagedorn transition is 
one of the central hallmarks of string thermodynamics.
Originally encountered in the 1960's through studies of hadronic
resonances and the so-called ``statistical
bootstrap''~\cite{Hagedorn,Huang,cudell},
the Hagedorn transition is now understood to be a generic feature
of any theory exhibiting a density of states which rises exponentially as
a function of mass.
In string theory,
the number of states of a given total mass depends on the number of ways
in which that mass can be partitioned amongst
individual quantized mode contributions, leading to an exponentially rising
density of states~\cite{Polbook}.
Thus, string theories should exhibit a
Hagedorn transition~\cite{earlystringpapers,vortices,McClainRoth,AtickWitten,longstrings}.
Originally, it was assumed that the Hagedorn temperature is a limiting temperature
at which the internal energy of the system diverges.  However,
later studies demonstrated that for closed strings the internal energy actually remains finite at
this temperature.  This then suggests that the Hagedorn temperature is merely the
critical temperature corresponding to a first- or second-order phase transition.

There has been much speculation concerning possible interpretations of this phase
transition, including a breakdown of the string worldsheet into vortices~\cite{vortices}
or a transition to a single long-string phase~\cite{longstrings}.
It has also been speculated that there is a
dramatic loss of degrees of freedom at high temperatures~\cite{AtickWitten}.
Over the past two decades, studies of the Hagedorn transition have reached
across the entire spectrum of modern string-theory research,
including open strings and D-branes,
strings with non-trivial spacetime geometries (including AdS backgrounds and $pp$-waves),
strings in magnetic fields, ${\cal N}{=}4$ strings, tensionless
strings, non-critical strings, two-dimensional strings, 
little strings, matrix models, non-commutative theories,
as well as possible cosmological implications and implications for the brane world.
A brief selection of papers in many of these areas appears in
Refs.~\cite{KounnasRostand,general,Dbranes,geometries,ppwaves,magnetic,tensionless,noncritical,little,matrix,NCOS,cosmology,ridge,2Dhet}.

In general, determining the Hagedorn temperature associated with a given 
finite-temperature thermal partition function is relatively straightforward.
Given this thermal partition function,
the one-loop free-energy density $F(T)$ is given   
by the modular integral in Eq.~(\ref{freeenergydef}),
whereupon the full panoply 
of thermodynamic quantities such as
the internal
energy $U$, entropy $S$, and specific heat $c_V$ then follow
from the standard definitions 
         $U \equiv F - T {dF/dT}$,
         $S \equiv  -{dF/dT}$,
and         $c_V \equiv -T {d^2F/ dT^2}$.
In string theory, 
the Hagedorn transition is usually associated with a divergence or other discontinuity
in the free energy $F(T)$ as a function of temperature. 
It turns out that are only two ways in which such a divergence may arise within the expression
in Eq.~(\ref{freeenergydef}).

First, of course, is the possibility of a divergence or discontinuity 
due to the well-known exponential rise in the degeneracy of string states which
contribute to $Z_{\rm string}(\tau,T)$.  This may be considered an ultraviolet
(UV) divergence because it is triggered by the behavior of the extremely massive 
portion of the string spectrum.
However, it turns out that this rise in the state degeneracies 
ultimately does not cause $F(T)$ to diverge.
To understand why,
we may expand $Z(\tau,T)$ in the form $\sum_{MN} a_{MN} \qbar^M q^N$
where 
$q\equiv e^{2\pi i\tau}$, where
$(M,N)$ describe the right- and left-moving worldsheet energies (with thermal contributions
included), and where $a_{MN}$ describe the corresponding degeneracies of bosonic minus fermionic
states.
Although the degeneracies $a_{MN}$ indeed experience exponential growth  
of the generic form $a_{MN}\sim \exp\left( C_R \sqrt{M} + C_L\sqrt{N}\right)$ where
$C_{L,R}$ are positive coefficients,
the contribution of each such state to the modular integrand
in Eq.~(\ref{freeenergydef})
is suppressed according to $|\qbar^M q^N|\sim \exp[ - 2\pi \tau_2 (M+N)]$.
For all $\tau_2>0$ and sufficiently large $(M,N)$, this 
exponential suppression easily overwhelms the exponential rise in the degeneracy of states.
As a result, the integrand in Eq.~(\ref{freeenergydef}) remains convergent  
everywhere except as $\tau_2\to 0$.  However, this dangerous UV region is explicitly
excised from the fundamental domain $\calF$ in Eq.~(\ref{Fdef}).
Thus, we conclude that the expression in Eq.~(\ref{freeenergydef}) does not suffer 
from any UV divergences resulting from the exponential growth in 
the asymptotic degeneracies of states.

On the other hand, the expression in Eq.~(\ref{freeenergydef}) may experience a divergence 
due to on-shell states within $Z_{\rm string}(\tau,T)$ 
which may become massless or tachyonic 
at specific critical temperatures. 
For example, as the temperature increases,
there may exist a critical temperature $T_H$ at which
certain states which were massive for $T<T_H$ become massless at $T=T_H$
and ultimately tachyonic for $T>T_H$.
This can therefore be considered an infrared (IR) divergence.
Since such on-shell tachyons correspond to states with worldsheet energies
$M=N<0$, their contributions to the modular integral in Eq.~(\ref{freeenergydef})
 {\it grow}\/ as $(\qbar q)^N\sim \exp( + 4\pi \tau_2 |N|)$.  
The contributions from the (infrared) $\tau_2\to\infty$ region of the fundamental
domain then lead to a divergence for $F(T)$. 

Thus, a study of the Hagedorn transition in string theory essentially reduces 
to a study of the {\it tachyonic}\/ structure of $Z_{\rm string}(\tau,T)$ as a function
of temperature.
Before proceeding further, however, we caution that we have reached this conclusion only because
we have chosen to work in the so-called $\calF$-representation for $F(T)$ given
in Eq.~(\ref{freeenergydef}).
By contrast,
utilizing Poisson resummations and modular transformations~\cite{McClainRoth},
we can always rewrite
$F(T)$ as the integration of a different integrand $Z'_{\rm string}(\tau,T)$
over the strip
\beq
   {\cal S}~\equiv ~\lbrace \tau:  ~|{\rm Re}\,\tau|\leq \half, {\rm Im}\,\tau>0 \rbrace~.
\label{Sdef}
\eeq
In such an $\calS$-representation, the IR divergence as $\tau_2\to\infty$ 
is transformed into a UV divergence
as $\tau_2\to 0$.  This formulation thus has the advantage of relating the Hagedorn
transformation directly to a UV phenomenon such as the exponential rise in the degeneracy of states.
However, both formulations are mathematically equivalent;  indeed, modular invariance
provides a tight relation between the tachyonic structure of a given partition function
and the rate of exponential growth in its asymptotic degeneracy of states~\cite{HR,Kani,missusy,kutasov}.
In the following, therefore, we shall utilize the $\calF$-representation for 
$F(T)$ and focus on only the tachyonic structure of $Z_{\rm string}(\tau,T)$,
but we shall comment on the connection to the asymptotic degeneracy of states
in Sect.~V.C.

\subsection{Effect of Wilson lines on Hagedorn temperature}

So what then are the potential tachyonic states within $Z_{\rm string}(\tau,T)$,
and at what temperatures $T_H$ do they first arise?  Note that we are concerned with
states whose masses are temperature-dependent:  positive at
temperatures below a certain critical temperature, zero at the critical temperature,
and tachyonic at temperatures immediately above the critical temperature. 
The sudden appearance of such new ``thermally massless'' states at a critical temperature $T_H$ 
is the signal of the appearance of the long-range order normally associated
with a phase transition, and the fact that such states generally become tachyonic
immediately above $T_H$ reflects the instabilities which are also normally associated with a
phase transition.

As a result, in order to derive the Hagedorn temperature of a given theory,
it is sufficient to search for 
states within the thermal partition function $Z_{\rm string}(\tau,T)$
whose masses decrease as a function of temperature, reaching (and perhaps even crossing)
zero at a certain critical temperature.
We shall refer to such states as ``thermally massless'' at the critical temperature.
Since thermal effects always provide a positive contribution to the squared masses of 
any states, such states must intrinsically be tachyonic at zero temperature. 
In other words, for such thermally massless states,
masslessness is achieved at the critical temperature
$T_H$ as the result of a balance  between a
tachyonic non-thermal mass contribution 
(arising from the characters $\overline{\chi}_i \chi_j\chi_k$ within $Z_{\rm string}$)
and an additional positive temperature-dependent thermal mass contribution 
(arising from the thermal $\calE,\calO$ functions). 

We can quantify this mathematically as follows.
A given state with worldsheet energies $(H_R,H_L)$ will contribute
a term of the form $\qbar^{H_R}q^{H_L}$ to the
characters $\overline{\chi}_i \chi_j\chi_k$ within $Z_{\rm string}$.
Likewise, as evident from their definitions in the Appendix,
the thermal $\calE,\calO$ functions will make an additional,
thermal contribution to these energies which is given by  
\beq
        [\Delta H_R, \Delta H_L] ~=~ [ \quarter(m a -n/a)^2 , ~\quarter (m a + n/a)^2 ]
\label{thermalmasscontribution}
\eeq
where $(m,n)$ are respectively the momentum and winding quantum numbers
around the thermal circle and where
$a\equiv T/\calM=T/(2\pi M_{\rm string})$.
The conditions for thermal masslessness then become
\beq
      H_R + \quarter (ma-n/a)^2 ~=~0~,~~~~~~~
      H_L + \quarter (ma+n/a)^2 ~=~ 0~,
\label{masslessness}
\eeq
which together imply the useful relation $mn=H_R-H_L$.
Since the thermal contributions in Eq.~(\ref{thermalmasscontribution}) are
strictly non-negative (and are not zero, according to our assumption of {\it thermal}\/ masslessness),
we see that the possibility of obtaining a thermally massless state requires that 
either $H_L$ or $H_R$ (or both) must be negative, and neither can be positive. 
In other words, the zero-temperature state contributing within
the characters  
$\overline{\chi}_i \chi_j\chi_k$ within $Z_{\rm string}$ must 
be a tachyon which is either on-shell (if $H_R=H_L$) or off-shell (if $H_R\not= H_L$);  
this tachyonic mode is then ``dressed'' with specific
thermal contributions in order to become massless at the critical temperature $a_H$. 
Moreover, if our solution to Eq.~(\ref{masslessness}) has non-zero $n$, 
then such a state will be massive for all temperatures
below this critical temperature, as desired.  It will also usually 
be tachyonic for temperatures immediately above this critical temperature.

Given these observations,
our procedure for determining the Hagedorn temperature corresponding to 
a given thermal partition function   
$Z_{\rm string}(\tau,T)$ is then fairly straightforward.
First, we identify any
zero-temperature states which are tachyonic (either on- or off-shell)
contributing to the characters appearing
within $Z_{\rm string}(\tau,T)$.
For each such state, we then
attempt to solve the conditions in Eq.~(\ref{masslessness}), subject to the
constraints that $(m,n)$ are restricted to the values 
which are appropriate for the corresponding thermal function 
(\ie, $m\in \IZ$ or $\IZ+1/2$ and  $n\in 2\IZ$ or $2\IZ+1$). 
If such a solution exists and has non-zero $n$, then we have succeeded in identifying 
a massive state in the full thermal theory which will become massless at 
the corresponding critical temperature $a_H$.
This then signals a Hagedorn transition.  
In situations where multiple thermally massless states exist,
the Hagedorn temperature is identified as the lowest of the corresponding
critical temperatures, since the presumed existence of a phase transition at that temperature
invalidates any analysis based on $Z_{\rm string}$ at temperatures above it. 

Let us now calculate the Hagedorn temperatures corresponding to the
heterotic partition functions $Z_{\rm string}(\tau,T)$ in Sect.~IV.
We focus first on the standard heterotic results without Wilson lines,
as given in Eq.~(\ref{fake}).
For both the $SO(32)$ and $E_8\times E_8$ cases, we find that
the sector $\chibar_I \chi_I^2\calO_{1/2}$ is the sector
which is capable of providing thermally massless states
at the lowest possible temperature.
Indeed, solving the conditions for masslessness in Eq.~(\ref{masslessness}),
we see that the $(H_R,H_L)=(-1/2,-1)$ off-shell tachyon
within $\chibar_I\chi_I^2$ ---
dressed with the thermal excitations $(m,n)=\pm(1/2,1)$ within $\calO_{1/2}$ ---
becomes thermally massless at the critical temperature
$T_H= 2\calM/(2+\sqrt{2}) = (2-\sqrt{2})\calM$.
This, of course, is nothing but the traditional Hagedorn temperature associated with the
$SO(32)$ and $E_8\times E_8$ heterotic strings.

By contrast,  let us now examine the thermal partition functions for the $SO(32)$
string which are constructed using non-trivial Wilson lines.
For example, if we concentrate on the partition function in Eq.~(\ref{A1}),
we now find that the term 
$\chibar_I (\chi_I\chi_V+\chi_V\chi_I)\calO_0$  
is the one which 
gives rise to thermally massless level-matched states
at the lowest possible temperature.
Indeed, the $SO(16)\times SO(16)$ character 
$(\chi_I\chi_V+\chi_V\chi_I)$ gives rise to 
32 on-shell $(H_R,H_L)=(-1/2,-1/2)$ tachyons,
and these are nothing but the 32 tachyons of the
non-supersymmetric $SO(32)$ heterotic string which serves as the $T\to\infty$ endpoint
of the corresponding Wilson-line orbifold.
Moreover, we find that the $(m,n)=(0,\pm 1)$ thermal excitations
of these states are massless at $T_H=\calM/\sqrt{2}$,
massive below this temperature,
and tachyonic above it.
Indeed, there are no other tachyonic sectors within Eq.~(\ref{A1}) which could 
give rise to other phase transitions at lower temperatures.
Thus the Hagedorn temperature associated with Eq.~(\ref{A1}) is actually
given by $T_H=\calM/\sqrt{2}$, not $T_H= (2-\sqrt{2}) \calM$, and agrees
with the locations of the divergences indicated in Fig.~\ref{Ffig}. 
Remarkably, this new temperature is exactly the same as the Hagedorn temperature
of the Type~I and Type~II strings. 

The same is true for Eqs.~(\ref{A2}) and (\ref{A3}) as well:
each of these thermal partition functions corresponds to $T_H=\calM/\sqrt{2}$,
not  $T_H= (2-\sqrt{2}) \calM$.
This makes sense, since each of these Wilson lines corresponds 
to a non-supersymmetric heterotic model containing
on-shell tachyons with worldsheet energies $(H_R,H_L)=(-1/2,-1/2)$.
Indeed, the only exception is the partition function in Eq.~(\ref{firstso16}).
This too makes sense, since the Wilson line in this case corresponds
the $SO(16)\times SO(16)$ heterotic 
string model.  Although non-supersymmetric, this string model is
tachyon-free.
Indeed, for the partition function given in Eq.~(\ref{firstso16}), we find
that off-shell tachyons with $(H_R,H_L)=(-1/2,0)$ 
arise within the term
$\chibar_I (\chi_I\chi_S+\chi_S\chi_I)\calO_{1/2}$;
these, when dressed with the $(m,n)=\pm(1/2,-1)$ thermal excitations within $\calO_{1/2}$,
become massless at $T_H=\sqrt{2}\calM$.  This is the lowest temperature at which such
thermally massless states appear, which identifies this as the Hagedorn temperature
corresponding to the $SO(16)\times SO(16)$ Wilson line.

A similar situation exists for the possible thermal extensions of the $E_8\times E_8$ string.
Examining Eq.~(\ref{B1}), we see that only the sector
$\chibar_I \chi_V \chi_I \calO_{0}$
is capable of
giving rise to thermally massless level-matched states;
once again, these are the tachyons with energies $(H_R,H_L)=(-1/2,-1/2)$ within
$\chibar_I\chi_V\chi_I$,
dressed with the
$(m,n)=(0,\pm 1)$ thermal excitations within $\calO_0$.
These states are massless at $T_H=\calM/\sqrt{2}$, massive below this temperature,
and tachyonic above it.
Thus, we see that 
$T_H=\calM/\sqrt{2}$ emerges as the Hagedorn temperature following 
from Eq.~(\ref{B1}) as well.
Indeed, the same is also true for Eqs.~(\ref{B2}) and (\ref{B3}),
while we find that $T_H=\sqrt{2}\calM$ for Eq.~(\ref{secondso16}).
These results are precisely in one-to-one correspondence with those for the
$SO(32)$ string.

We conclude, then, that the existence of non-trivial Wilson 
lines in the formulation of finite-temperature heterotic strings 
has, in most cases, shifted the corresponding heterotic Hagedorn temperature
from $T_H=(2-\sqrt{2})\calM$ to $T_H=\calM/\sqrt{2}$. 
Remarkably, this is the same Hagedorn temperature as that associated
with Type~II strings.
It is easy to understand why this is the case.
Without a non-trivial Wilson line, 
our thermal heterotic theories are described by Eq.~(\ref{fake}), and 
the lowest mode contributing within $\chibar_I\chi_I^2$
is the (tachyonic) ground state of the heterotic theory, with non-level-matched
vacuum energies $(H_R,H_L)=(-1/2,-1)$.
However, as we have seen, turning on the Wilson lines leading to Eqs.~(\ref{A1}), (\ref{A2}),
and (\ref{A3}) in the $SO(32)$ case, or to Eqs.~(\ref{B1}), (\ref{B2}), and (\ref{B3}) in the
$E_8\times E_8$ case, 
effectively projects this non-level-matched state out of the finite-temperature theory
and leaves behind only the ``next-deepest'' tachyon with $(H_R,H_L)=(-1/2,-1/2)$
within $\chibar_I \chi_I\chi_V$.  Thus, with these Wilson lines turned on,
this new tachyon becomes the effective ground state of the theory.
However, this ``next-deepest'' tachyon has exactly the same worldsheet
energies $(H_R,H_L)=(-1/2,-1/2)$ as the ground state of the Type~II string.  
Thus it is not surprising
that the presence of the non-trivial Wilson line shifts the corresponding heterotic Hagedorn
temperature in such a way that it now matches the Type~II value.

\subsection{Reconciling the shifted Hagedorn temperature with the asymptotic
    degeneracies of states}

As discussed in Sect.~V.A,
our analysis of the Hagedorn temperature
has thus far been based on an analysis of the tachyonic
structure of our thermal partition functions.
Yet we know that there is a tight relation between the Hagedorn
temperature of a given theory and the exponential rate of growth
of its asymptotic degeneracies of bosonic and fermionic states.    
Specifically, if $g_M$ denotes the number of string states with mass $M$, 
then the thermal partition function is
given by $Z(T)=\sum  g_M e^{-M/T}$.  
However, if $g_M \sim e^{\alpha M}$ as $M\to\infty$, then
$Z(T)$ diverges for $T\geq T_H\equiv 1/\alpha$.
This appears to provide a firm link between
the Hagedorn temperature and the asymptotic degeneracy of states.
Of course, 
$\sum  g_M e^{-M/T}$  
is not a proper string-theoretic partition function.
However, even when we utilize a proper string-theoretic partition function $Z_{\rm string}(\tau,T)$
and calculate a proper string-theoretic amplitude as in Eq.~(\ref{Fdef}) 
in the $\calS$-representation,
the same basic argument continues to apply.

We are thus left with the critical question:
 {\it How can we justify a shifted Hagedorn temperature $T_H=\calM/\sqrt{2}$ for
heterotic strings, given that  
the zero-temperature bosonic and fermionic densities of heterotic states 
are apparently unchanged?}\/
This question is particularly urgent, given that the feature which is inducing
this shift in the Hagedorn temperature --- namely the introduction of a non-trivial
thermal Wilson line --- does not affect the zero-temperature theory in any way.
Specifically, an increase in the Hagedorn temperature of the heterotic string 
from the traditional value  
$T_H= 2\calM/(2+\sqrt{2})$ to
a new, higher value $T_H=\calM/\sqrt{2}$
would seem to require a corresponding decrease in the exponential
rate of growth of the asymptotic density of heterotic string states. 
In what sense can we understand such a decrease?

To answer this question, let us look 
again at the original partition function of the zero-temperature
ten-dimensional $SO(32)$ heterotic string
model in Eq.~(\ref{start}).  Recall that $\chibar_V$ and $\chibar_S$ 
indicate the transverse $SO(8)$ Lorentz spins of the different 
states which contribute in this theory.
As a result of spacetime supersymmetry,  
this partition function vanishes identically --- \ie, all of its level-degeneracy coefficients
are identically zero.  There is no exponential growth here at all.
But of course one does not examine the {\it total}\/ partition function   
in order to derive a Hagedorn temperature;  one instead looks at its separate bosonic and
fermionic contributions.  Ordinarily, these contributions would
be identified on the basis of the Lorentz spins of these states
as
\beq
         Z^{\rm (bosonic)}_{SO(32)} = Z^{(8)}_{\rm boson}~
                     \chibar_V \,(\chi_I^2 + \chi_V^2 + \chi_S^2 + \chi_C^2)~,~~~~~~~
         Z^{\rm (fermionic)}_{SO(32)} = -Z^{(8)}_{\rm boson}~
                     \chibar_S \,(\chi_I^2 + \chi_V^2 + \chi_S^2 + \chi_C^2)~,
\label{usualcomponents}
\eeq
and indeed
each of these expressions separately exhibits an exponential rise in the degeneracy
of states which is consistent with the traditional heterotic Hagedorn temperature.

But what do we really 
mean by ``bosonic'' and ``fermionic'' in this context?
For most purposes, we would identify 
states as ``bosonic'' or ``fermionic'' based on their Lorentz spins,
as above.
Moreover, by the spin-statistics theorem, this is equivalent to identifying
states as bosonic or fermionic based on their zero-temperature quantization statistics.  
However, {\it for the purposes of computing a Hagedorn temperature}\/, 
we should really be focused on a {\it thermodynamic}\/ definition of
``bosonic'' and ``fermionic'' wherein we
identify states as bosons or fermions on the basis of their Matsubara frequencies,
\ie, on the basis of their modings around the thermal circle.
Of course, under normal circumstances, all three of these identifications
are equivalent.  However, we have already seen in Sect.~III that this
chain of equivalences is modified in the presence of a non-trivial Wilson
line:  certain states which are ``bosonic'' 
in terms of their Lorentz
spins and zero-temperature quantization statistics can nevertheless
have half-integer modings $m\in\IZ+1/2$ around the thermal circle,
while other states which are ``fermionic''    
in terms of their Lorentz
spins and zero-temperature quantization statistics can nevertheless
have integer modings $m\in \IZ$ around the thermal circle.
Thus, {\it in the presence of  a non-trivial Wilson line,}\/
certain bosonic states can behave as fermions from a 
thermodynamic standpoint, and certain fermionic states can behave as bosons.

We emphasize that this is {\it not}\/ a violation of the spin-statistics theorem.
Indeed, the spin-statistics theorem is believed to hold without alteration in 
string theory, providing a connection between the
Lorentz spin of a state and its zero-temperature quantization statistics~\cite{Polbook}.
Rather, as discussed in Sect.~III, the effect of the Wilson line  
is to modify the {\it thermodynamic manifestation}\/ of these properties
as far as their Matsubara modings are concerned.
For issues pertaining to zero-temperature physics, these thermodynamic
manifestations may be of little consequence.
However, when we seek to understand the thermal properties of a theory,
these modifications are critical.

Therefore, if we seek to understand the spectra of bosonic and fermionic states
in the heterotic string {\it for thermodynamic purposes}\/, 
we should return to the partition function in Eq.~(\ref{start}) 
and separate this expression into 
individual contributions from bosonic and fermionic states
 {\it on the basis of their Matsubara modings around the thermal circle}\/.  
It is here where the Wilson line comes into play.

Let us begin by considering the case of the $SO(32)$ string.
Without a Wilson line, we know that bosonic states will correspond
to integer Matsubara modings $m\in \IZ$
and fermionic states will correspond to half-integer modings $m\in\IZ+1/2$.
This corresponds to the traditional identifications in Eq.~(\ref{usualcomponents}),
and these lead to the traditional Hagedorn temperature $T_H=(2-\sqrt{2})\calM$.

However, under the influence a non-trivial Wilson line,
these identifications can change.
For example, let us consider the case of the Wilson line corresponding
to Eq.~(\ref{A1}).
In this case,
the ``bosonic'' contributions to Eq.~(\ref{start})
must be identified as those which multiply the thermal sum $\calE_0$ in Eq.~(\ref{A1}),
while the ``fermionic'' contributions to Eq.~(\ref{start}) must be identified as those
which multiply the thermal sum $\calE_{1/2}$ in Eq.~(\ref{A1}).
In other words, we replace the identifications in 
Eq.~(\ref{usualcomponents}) with 
\beqn
         \widetilde Z^{\rm (bosonic)}_{SO(32)} &=& \phantom{-} Z^{(8)}_{\rm boson}~
               \left\lbrack
          \chibar_V \,(\chi_I^2 + \chi_V^2)  ~-~  \chibar_S \,(\chi_S^2 + \chi_C^2)\right\rbrack
                                                      ~,\nonumber\\
     \widetilde    Z^{\rm (fermionic)}_{SO(32)} &=& -Z^{(8)}_{\rm boson}~
     \left\lbrack \chibar_S \,(\chi_I^2+\chi_V^2)  ~-~ \chibar_V \,(\chi_S^2+\chi_C^2)\right\rbrack~.
\label{newcomponents}
\eeqn
In the presence of the non-trivial Wilson line,
it is therefore {\it these}\/ expressions which serve to define our 
separate bosonic and fermionic contributions to Eq.~(\ref{start}),
and indeed their sum
\beq
         \widetilde Z^{\rm (bosonic)}_{SO(32)} ~+~  \widetilde Z^{\rm (fermionic)}_{SO(32)} 
\eeq
correctly reproduces the expression in Eq.~(\ref{start}).

Given these results, we can now calculate the exponential rates of growth
in the degeneracies of the states contributing to $\widetilde Z^{\rm (bosonic)}_{SO(32)}$
 and $\widetilde Z^{\rm (fermionic)}_{SO(32)}$ 
in Eq.~(\ref{newcomponents}).
 We find that while each individual term within 
$\widetilde Z^{\rm (bosonic)}_{SO(32)}$
 and $\widetilde Z^{\rm (fermionic)}_{SO(32)}$ 
in Eq.~(\ref{newcomponents})
continues to exhibit the traditional
rate of growth associated with the traditional Hagedorn temperature for heterotic 
strings, {\it the minus signs within 
$\widetilde Z^{\rm (bosonic)}_{SO(32)}$
 and $\widetilde Z^{\rm (fermionic)}_{SO(32)}$
have the net effect of cancelling this dominant exponential behavior,
leaving behind only a smaller exponential rate of growth for 
the state degeneracies corresponding to
$\widetilde Z^{\rm (bosonic)}_{SO(32)}$
 and $\widetilde Z^{\rm (fermionic)}_{SO(32)}$.}
Moreover, as expected, this smaller exponential rate of growth precisely  
matches the rate of growth that corresponds to the new (increased) heterotic Hagedorn temperature
$T_H=\calM/\sqrt{2}$.
Similar results also hold the Wilson lines associated with Eqs.~(\ref{A2}) and (\ref{A3}).
 
It may seem strange that two terms, each exhibiting a dominant exponential growth rate,
can be subtracted and leave behind a sub-dominant exponential growth rate.
Yet this phenomenon is well known for modular functions such as these,
and has been shown to operate in other string-theoretic contexts~\cite{HR,Kani,missusy,intprojs}.
We emphasize that this subtraction is relevant only in the sense 
that it deforms the exponential growth rate when we count 
$\widetilde Z^{\rm (bosonic)}_{SO(32)}$
 and $\widetilde Z^{\rm (fermionic)}_{SO(32)}$ separately.
Each string state still continues to appear with positive unit weight in the string Fock space,
as always, and still contributes to the overall partition function with unit weight and an appropriate sign
(positive for spacetime bosons, negative for spacetime fermions).

Similar results hold for the $E_8\times E_8$ heterotic string.
Without a Wilson line, the usual identification of bosonic and fermionic states
is nothing but 
\beq
         Z^{\rm (bosonic)}_{E_8\times E_8} = Z^{(8)}_{\rm boson}~
                     \chibar_V \,(\chi_I+ \chi_S)^2~,~~~~~~~
         Z^{\rm (fermionic)}_{E_8\times E_8} = -Z^{(8)}_{\rm boson}~
                     \chibar_S \,(\chi_I+ \chi_S)^2 ~.
\label{usualcomponentse8}
\eeq
However, in the presence of the Wilson line associated with Eq.~(\ref{B1}), 
we find that our new thermal identification
of bosonic and fermionic states is given by
\beqn
         \widetilde Z^{\rm (bosonic)}_{E_8\times E_8} &=& \phantom{-} Z^{(8)}_{\rm boson}~
                (\chibar_V\,\chi_I ~-~ \chibar_S\,\chi_S)\,(\chi_I+\chi_S)\nonumber\\
         \widetilde Z^{\rm (fermionic)}_{E_8\times E_8} &=& - Z^{(8)}_{\rm boson}~
              (\chibar_S\,\chi_I ~-~ \chibar_V\,\chi_S)\,(\chi_I+\chi_S)~. 
\label{e8newbosferm}
\eeqn
Just as with the $SO(32)$ string, the minus signs within Eq.~(\ref{e8newbosferm}) lead to
state degeneracies which show a
reduced exponential growth --- one which is precisely 
in accordance with the new, increased heterotic Hagedorn temperature associated with
this Wilson line.
Similar results also hold the Wilson lines associated with Eqs.~(\ref{B2}) and (\ref{B3}).

This, then, is the essence of the manner in which the asymptotic density of
states is ultimately reconciled with the modified Hagedorn temperature for heterotic strings.
The presence of the non-trivial Wilson line ``deforms'' the thermal identification
of bosonic and fermionic states, 
trading states between the separate sets of bosonic and fermionic 
states in such a way that the net exponential rate of growth for the asymptotic
state degeneracy of each set is reduced.

There are also other ways to understand this result.
For example, one might argue  on general conformal-field-theory (CFT) grounds that
such a change in the Hagedorn temperature should not be possible.
After all, there exists a general result which relates the Hagedorn temperature
of a given closed-string theory to the central charges $(c_R,c_L)$ of its underlying
worldsheet CFT's:
\beq
       T_H ~=~   \left(  \sqrt{c_L\over 24} + \sqrt{c_R\over 24}
              \right)^{-1}\, \calM~.
\label{THag}
\eeq
For Type~II strings, we have $(c_R,c_L)=(12,12)$,
while for heterotic strings, we have 
$(c_R,c_L)=(12,24)$.
However, in deriving Eq.~(\ref{THag}), there is only place 
in which the central charges enter:  this is in
setting the ground-state energies $(H_R,H_L)=(-c_R/24,-c_L/24)$. 
Moreover, as we have seen in Sect.~V.B, the 
Wilson line has effectively projected the true heterotic ground
state with $(H_R,H_L)=(-1/2,-1)$ out of the spectrum, leaving
behind only the ``next-deepest'' tachyon with $(H_R,H_L)=(-1/2,-1/2)$
to serve as the effective ground state of the theory.   
Thus, in the presence of the Wilson line, the effective central charges
of the theory become $(c_R,c_L)=(12,12)$, just as for Type~II strings.

We see, then, that the introduction of a non-trivial Wilson line
induces both a shift in the vacuum energy of the effective ground state
and a shift in the asymptotic rates of growth for the
state degeneracies.
These shifts, of course, are flip sides of the same coin, deeply related
to each other through modular transformations.
Indeed, these are nothing but equivalent UV/IR descriptions of the same
phenomenon, all arising due to the existence of the non-trivial Wilson
line.
Under Poisson resummation,
a half-integer shift in the moding of a given set of string states
around the thermal circle translates 
into an overall $\IZ_2$ phase (\ie, a minus sign)
in front of the corresponding character in the partition function.
Thus the Wilson line, which shifts the apparent thermal modings of
certain states in the theory, necessarily induces a corresponding
change in the asymptotic state degeneracies and a corresponding
shift in the Hagedorn temperature of the theory.

We thus conclude that the introduction of a non-trivial Wilson line
has the potential to change the Hagedorn temperature of the
resulting thermal theory, and in many cases actually shifts this
temperature from its traditional heterotic value to a new value
which is the same as that associated with Type~I and Type~II strings.
Indeed,
although the heterotic string would na\"\i vely appear to have a slightly lower
Hagedorn temperature than the Type~II string
due to its non-level-matched ground state, 
we see that the introduction of a non-trivial Wilson line has the potential
to eliminate this discrepancy.
Such Wilson lines
deform the effective worldsheet central charges of the heterotic theory
as far as its thermal properties are concerned, 
and lead to new, effective ground states for the theory as
well as modified rates of exponential growth for the corresponding
bosonic and fermionic densities of states.  Both effects then alter
Hagedorn temperature of the heterotic string, and potentially 
bring it into agreement with the Type~I and Type~II value.

\subsection{The ``spectrum'' of Hagedorn temperatures:  A general classification}

As we have seen in Sect.~V.B, the thermal heterotic theories without Wilson lines
have the Hagedorn temperature 
$T_H= (2-\sqrt{2})\calM$;  indeed, this is the case because the relevant
partition functions in Eq.~(\ref{fake}) 
each contain a term of the form $\chibar_I \chi_I^2 \calO_{1/2}$.
Indeed, such a term encapsulates the contribution of 
the $(H_R,H_L)=(-1/2,-1)$ tachyon, and this is the state which, when dressed with
the thermal excitations $(m,n)=\pm(1/2,1)$ within $\calO_{1/2}$,
is massive at low temperatures but becomes massless at $T_H= (2-\sqrt{2}) \calM$.

By contrast, we found that this state is projected out of the thermal 
partition function when non-trivial Wilson lines are introduced. 
For example,  
for the Wilson lines leading to Eqs.~(\ref{A1}) and (\ref{B1}), we found
that the ``next-deepest'' remaining tachyons have 
worldsheet energies $(H_R,H_L)=(-1/2,-1/2)$;
indeed, these are the tachyons which contribute to the 
terms $\chibar_I \chi_V \chi_I \calO_{0}$
which appear within Eqs.~(\ref{A1}) and (\ref{B1}).
These tachyons, when dressed with the
$(m,n)=(0,\pm 1)$ thermal excitations within $\calO_0$,
are massless at $T_H=\calM/\sqrt{2}$, massive below this temperature,
and tachyonic above it.
Thus, $T_H=\calM/\sqrt{2}$ emerges as the Hagedorn temperature corresponding
to these choices of Wilson lines.
We found that this result also holds for the Wilson lines leading to Eqs.~(\ref{A2}), (\ref{A3}),
(\ref{B2}), and (\ref{B3}).   By contrast, we found that the Hagedorn temperature is
$T_H=\sqrt{2}\calM$ for the Wilson lines
leading to Eqs.~(\ref{firstso16}) and (\ref{secondso16}).
Indeed, in these cases, even the $(H_R,H_L)=(-1/2,-1/2)$ tachyons are projected
out of the spectrum, so that the lowest possible remaining tachyons have
$(H_R,H_L)=(-1/2,0)$ and contribute to the $\calO_{1/2}$ sector.

Given these results, it is natural to wonder whether other Hagedorn temperatures
might also be possible.  Indeed, one might even wonder whether there exist Wilson lines
for which the Hagedorn transition might be avoided completely! 
Of course, neither of these options is possible for the supersymmetric
heterotic strings in ten dimensions, 
for we have presented a complete classification of all self-consistent Wilson lines
in such cases, and our results are quoted above.
However, in lower dimensions, this might no longer be the case.   
In fact, in lower dimensions, even the Type~II superstrings can have non-trivial
Wilson lines which are associated with the gauge symmetries that emerge upon
compactification.  Thus, these questions become relevant for Type~II strings as
well as heterotic. 

Towards this end, we shall now provide a general classification all of the Hagedorn
temperatures which can ever be realized for closed strings with non-trivial Wilson
lines.
Our analysis will apply to all closed strings, both Type~II and heterotic.

For concreteness, we shall restrict our attention to theories built from only $\IZ_2$ orbifolds,
so that $H_{L,R}$ are quantized in half-integer values.  
Given the heterotic constraints $H_L\geq -1$ and $H_R\geq -1/2$ (which also subsume
the Type~II constraints $H_{L,R}\geq -1/2$), 
we then find that there are only eight different terms which could possibly
appear in $Z_{\rm string}(\tau,T)$ and trigger a Hagedorn transition.
These are listed in Table~\ref{Hagedornia}, along with their corresponding
thermal excitations and Hagedorn temperatures [obtained by solving Eq.~(\ref{masslessness})].
It is interesting to note
the mathematical fact that these terms come in ``dual'' pairs under 
which $T_H/\calM \to 2\calM/T_H$ and
$(\calE_0,\calE_{1/2},\calO_0,\calO_{1/2})\to (\calE_0,\calO_0,\calE_{1/2},O_{1/2})$.
Roughly speaking, this duality corresponds to exchanging 
the direction of the corresponding ``interpolating'' thermal partition functions,
exchanging the $T=0$ and $T\to \infty$ endpoints.
Although this ``thermal duality'' phenomenon
has played a significant role
in other work~\cite{McClainRoth,AlvOso,AtickWitten,Polbook,dualityus,shyamoli}), 
it will not be critical for the following discussion.  We can 
therefore view the emergence of this duality within Table~\ref{Hagedornia}
as a mere mathematical curiosity.

\begin{table}[ht]
\begin{center}
\begin{tabular}{||c||c|c|c||c|c||}
         \hline
         \hline
       ~ & $H_R$  & $H_L$ &  ~Thermal Function~ & ~Thermal Modes $(m,n)$~ &  $T_H/\calM$ \\
       \hline
         \hline
     A &  $-1/2$   & $-1$    &   $\calO_{1/2}$   &  $\pm (1/2, 1)$   &  $2-\sqrt{2}$ \\
       ~& ~ & ~       & ~                 &          ~        &  ~(also $2+\sqrt{2}$)~ \\
    \hline
      B & ~~$-1/2$~~  & ~~$-1/2$~~  &   $\calE_{0}$     &  $(0, n)$, ~$n\in 2\IZ$     &  $|n|/\sqrt{2}$  \\
      ~& ~  & ~           &   ~               &  $(m, 0)$, ~$m\in \IZ$      &  $\sqrt{2}/|m|$  \\
    \hline
      C & $-1/2$  & $-1/2$  &   $\calO_{0}$     &  $(0, n)$, ~$n\in 2\IZ+1$   &  $|n|/\sqrt{2}$  \\
      D & $-1/2$  & $-1/2$  &   $\calE_{1/2}$   &  ~~$(m, 0)$, ~$m\in \IZ+1/2$~~  &  $\sqrt{2}/|m|$  \\
    \hline
      E & $0$     & $-1/2$  &   $\calO_{1/2}$   &  $\pm (1/2, 1)$            &  $\sqrt{2}$  \\
    \hline
      F & $-1/2$  & $0$     &   $\calO_{1/2}$   &  $\pm (1/2, -1)$           &  $\sqrt{2}$  \\
    \hline
      G & $0$     & $-1$    &   $\calO_0$       &  $\pm (1,1)$      &  $1$ \\
      ~H~ & $0$     & $-1$    &   ~~$\calE_{1/2}$~~   &  $\pm (1/2, 2)$   &  $2$ \\
         \hline
         \hline
\end{tabular}
\end{center}
\caption{Complete set of possible terms (labeled A through H)
         which can potentially trigger a Hagedorn transition for
         string models built with $\IZ_2$ orbifolds.  
         As discussed in the text, Case~A is responsible for the traditional
         heterotic Hagedorn transition, while Case~C with $n=1$ is responsible
         for the traditional Type~II Hagedorn transition as well as the ``shifted''
         heterotic Hagedorn transition.  Cases~B and D can only arise in theories which are 
         already tachyonic (and hence unstable) at zero temperature, while Case~H
         is guaranteed to arise for all heterotic strings which are supersymmetric at 
         zero temperature.  Observe that all of these possibilities
         come in ``dual'' pairs under which $T_H/\calM \to 2\calM/T_H$ and
         $(\calE_0,\calE_{1/2},\calO_0,\calO_{1/2})\to (\calE_0,\calO_0,\calE_{1/2},O_{1/2})$.
         Thus the two possibilities within Cases~A and B are dual to each other, 
         while Cases~C and G are dual to Cases~D and H respectively (and vice versa).
         By contrast, Cases~E and F are each self-dual.  Note that Cases~A, G, and H
         are unique to heterotic strings, while all other cases
         can in principle arise in both heterotic and Type~II strings. }
\label{Hagedornia}
\end{table}

As we have already seen, Case~A is responsible for the traditional
heterotic Hagedorn transition and leads to the lowest possible
Hagdorn temperature $T_H=(2-\sqrt{2})\calM$.
Likewise, 
Case~C with $n=1$ is responsible for the traditional Type~II Hagedorn 
transition as well as the shifted heterotic Hagedorn transitions
with $T_H=\calM/\sqrt{2}$.
Indeed, this case produces what is ultimately the ``next-lowest''
Hagedorn temperature, and as such it dominates 
[when present within $Z_{\rm string}(\tau,T)$] over
any other terms which may also simultaneously appear within $Z_{\rm string}(\tau,T)$. 
Indeed, in ten dimensions, our complete enumeration of all possible non-trivial Wilson
lines in the heterotic case has demonstrated that Case~C with $n=1$ arises in all
but $SO(16)\times SO(16)$ cases.
By contrast, the $SO(16)\times SO(16)$ Wilson lines are examples of Case~F.

Ultimately, the question of which of these terms ends up dominating for
a given string model in $D<10$ dimensions is likely to be addressable only 
on a case-by-case basis.
Nevertheless, 
it is easy to see that Cases~B and D can only arise for string models which
are already tachyonic (and hence unstable) at zero temperature;  this
follows from the fact that the solutions for their corresponding Hagedorn
temperatures, as shown in Table~\ref{Hagedornia}, 
always include the cases with $n=0$ or $m\to\infty$.  
This can also be seen by taking the direct $T\to 0$ limit of the terms 
in each of these cases.
Thus, Cases~B and D need not concern us further.

Given this situation, it is natural to wonder whether there are any
Wilson-line choices for which 
the Hagedorn transition is eliminated {\it completely}\/ --- 
\ie, string models in which 
no thermally massless states appear at {\it any}\/ temperature,
and in which {\it none}\/ of the remaining cases listed in Table~\ref{Hagedornia} arise.
However, we shall now prove that this cannot happen for 
any heterotic string which is supersymmetric at zero temperature,
regardless of its spacetime dimension.
In particular, we shall demonstrate that Case~H will {\it always}\/ arise
for such strings, giving rise to a Hagedorn transition at $T_H=2\calM$ if
no earlier Hagedorn transition has occurred at lower temperature.

Our argument is completely general since it is based on considerations of
the most generic massless states in the perturbative heterotic string: 
those associated with the gravity multiplet.
Recall that in the heterotic string, the graviton is realized
in the Neveu-Schwarz sector as
\beq
        \hbox{graviton:}~~~~~~~~~~~~
             g^{\mu\nu}~\subset ~ \tilde b_{-1/2}^\mu |0\rangle_R ~\otimes~ \alpha_{-1}^\nu |0\rangle_L~
\label{graviton}
\eeq
where $\tilde b_{-1/2}^\mu$ and $\alpha^\nu_{-1}$ are respectively the excitations
of the right-moving worldsheet Neveu-Schwarz
fermion $\tilde \psi^\mu$ and left-moving worldsheet coordinate boson $X^\nu$.
Since the Neveu-Schwarz heterotic-string ground state
has vacuum energies $(H_R,H_L)=(-1/2,-1)$, 
the states in Eq.~(\ref{graviton}) are both level-matched and massless, with
$(H_R,H_L)=(0,0)$.
These states include the spin-two graviton,
the spin-one antisymmetric tensor field, and the spin-zero dilaton.

In a similar vein, any model exhibiting spacetime supersymmetry
must also contain the gravitino state, realized in the Ramond sector of
the heterotic string as
\beq
        \hbox{gravitino:}~~~~~~~~~~~~
             \tilde g^{\alpha \nu} ~\subset ~
         \lbrace \tilde b_{0}\rbrace^\alpha  |0\rangle_R ~\otimes~ \alpha_{-1}^\nu |0\rangle_L~.
\label{gravitino}
\eeq
Here $\lbrace \tilde b_{0}\rbrace^\alpha$ schematically indicates the Ramond zero-mode
combinations which collectively give rise to the spacetime Lorentz spinor index $\alpha$,
as required for the spin-3/2 gravitino state.

Regardless of the particular GSO projections inherent in the particular
string model under consideration, we know that the graviton state
in Eq.~(\ref{graviton}) must always appear in the string spectrum.  Likewise, if the
model has spacetime supersymmetry, we know that the gravitino state in Eq.~(\ref{gravitino}) 
must exist as well.
However, it is then straightforward to show that this implies that certain additional
off-shell tachyons must also exist in the string spectrum.  Specifically, regardless
of the particular GSO projections, the off-shell spectrum will always contain
a spin-one ``proto-graviton'' state $\phi^\mu$ in the Neveu-Schwarz sector:
\beq
        \hbox{proto-graviton:}~~~~~~~~~~~~
             \phi^\mu ~\equiv~  \tilde b_{-1/2}^\mu |0\rangle_R ~\otimes~ ~|0\rangle_L~;
\label{protograviton}
\eeq
likewise, if the model is spacetime supersymmetric, the off-shell spectrum will always
contain a spin-1/2 ``proto-gravitino'' state $\psi^\alpha$ in the Ramond sector:
\beq
        \hbox{proto-gravitino:}~~~~~~~~~~~~
             \psi^\alpha ~\equiv~
         \lbrace \tilde b_{0}\rbrace^\alpha  |0\rangle_R ~\otimes~ ~ |0\rangle_L~.
\label{protogravitino}
\eeq
Note that these are the same states as the graviton/gravitino, except that in each
case the left-moving bosonic excitation is lacking.  However, it is important to realize
that {\it GSO projections are completely insensitive to
the presence or absence of excitations of the worldsheet coordinate
bosonic fields}\/.  This is indeed a general property of GSO projections.
Thus, since the graviton is always present in the on-shell
spectrum,
it then follows that the proto-graviton must also always be present in the
off-shell spectrum;
likewise, if the model is supersymmetric and the gravitino
is present in the on-shell spectrum, then the proto-gravitino must also
always be present in the off-shell spectrum.
Thus, we conclude that the proto-graviton and proto-gravitino are two off-shell tachyons
with worldsheet energies $(H_R,H_L)=(0,-1)$ which generically appear in all supersymmetric
heterotic string models. 

This does not, in and of itself, guarantee that these states will contribute
to the thermal partition function $Z_{\rm string}(\tau,T)$ 
within the specific ${\cal O}_{1/2}$ or $\calE_{1/2}$ sectors  
that Cases~G or H would require.
Fortunately, however, it is not too difficult to determine which sectors will contain
these states.
Like the graviton and gravitino states from which they are derived, these proto-graviton and
proto-gravitino states must exist in the zero-temperature theory
and thus must survive the zero-temperature limit.  This implies that  
these states must appear in the $\calE$ sectors, not the $\calO$ sectors.
Moreover, since neither of these states carries any gauge charges,
neither can be affected by the presence of a Wilson line.  As a result,
we know that the (bosonic) proto-graviton state must appear in the $\calE_{0}$
sector (which has integer modings around the thermal circle), 
while the (fermionic) proto-gravitino state must appear in the $\calE_{1/2}$
sector (which has half-integer modings).

Given these results, we conclude that while the proto-graviton state will never lead to
any of the cases in Table~\ref{Hagedornia}, the proto-gravitino state leads directly
to Case~H.~  Moreover, as we have argued on general grounds, 
this state is always present in any heterotic model which is supersymmetric at zero temperature.
As a result, we conclude that the proto-gravitino state  --- dressed with 
$(m,n)=\pm (1/2,2)$ thermal excitations ---
will always exist and trigger a Hagedorn-like transition
at temperature $T_H=2\cal M$ (provided no other phase transition has occurred at 
any lower temperature).
  
This transition is somewhat different from the typical Hagedorn transition, however.
In general, the total spacetime mass $M_{\rm tot}$ of a given $(H_R,H_L)$ state dressed with
$(m,n)$ thermal excitations varies with the temperature $T$ according to
\beq
           \alpha' M_{\rm tot}^2 = 2 \left[ H_R + \quarter (ma-n/a)^2 + H_L + \quarter(ma+n/a)^2 \right]~
\eeq
where $a\equiv T/\calM$.
However, for the proto-gravitino (Case~H), this becomes
\beq
        \alpha' M_{\rm tot}^2 ~=~
                   {a^2\over 4} ~+~ {4\over a^2} ~-~ 2~,
\label{massparabola}
\eeq
whereupon we see that
the thermal excitation of the proto-gravitino state
 {\it never becomes tachyonic}\/!  Indeed, this state is massive for all $a<2$,
and merely hits masslessness 
at $a=2$ before becoming massive again at higher temperatures.
Of course, this result is completely consistent with the fact that the proto-gravitino
state is fermionic, since the existence of a physical fermionic tachyon
at any temperature would violate Lorentz invariance.

However, given that this state never becomes tachyonic,
it is natural to wonder whether this state can ever give rise to a Hagedorn
transition.  Indeed, since no tachyon ever develops, the
free-energy density $F(T)$
will never diverge.
To study this issue, let us define the vacuum amplitude $\calV(T)\equiv F(T)/T$,
whereupon we  observe that the $(m,n)=(1/2,2)$ thermal excitation of the proto-gravitino state
makes a contribution to $\calV(T)$ given by
\beqn
    \calV(T)  &=& -\half \calM^{D-1}\,
              \int_\calF {d^2\tau\over \tau_2^2} \, \tau_2^{1-D/2} \sqrt{\tau_2}~
              ~{1\over q} ~ \left\lbrack
                \overline{q}^{(a/2-2/a)^2/4} q^{(a/2+2/a)^2/4} \right\rbrack ~+~...\nonumber\\
              &=& -\half \calM^{D-1}\,
              \int_\calF {d^2\tau\over \tau_2^2} \, \tau_2^{1-D/2} \sqrt{\tau_2}~
                ~e^{2\pi \tau_2}
            ~e^{-\pi \tau_2 (a^2/4 + 4/ a^2)}  ~+~...
\label{Vform}
\eeqn
where we have left the temperature $a\equiv T/\calM$ arbitrary.
Note that the leading $1/q$ factor in the first line of Eq.~(\ref{Vform})
represents the zero-temperature contribution
from the proto-gravitino, with $(H_R,H_L)=(0,-1)$,
while the remaining factor in brackets represents the thermal contribution
with $(m,n)=(1/2,2)$.
Likewise, we have carefully recorded all factors of
$\tau_2\equiv {\rm Im}\,\tau$:  
two factors of $\tau_2$ arise in the denominator from the modular-invariant
measure of integration,
$(1-D/2)$ factors arise in the numerator from the
zero-temperature partition function,
and an additional factor $\sqrt{\tau_2}$ arises in the numerator
from the definitions of the $\calE,\calO$ thermal sums.
However, at $a=2$, this expression reduces to
\beq
    \calV(T)\biggl|_{a=2}  ~=~ -\half \calM^{D-1}\,
               \int_\calF {d^2\tau\over \tau_2^2} \, \tau_2^{1-D/2} \sqrt{\tau_2} ~+~...
\eeq
and as $\tau_2\to\infty$, this contribution scales like
\beq
               \int^\infty {d\tau_2\over \tau_2^{(1+D)/2}}~.
\eeq
This contribution is therefore finite for all $D\geq 2$.
This, of course, agrees with our usual expectation that a massless state does
not lead to a divergent vacuum amplitude in two or more spacetime dimensions.

It is important to realize that even though $\calV(T)$ remains finite 
for all temperatures, a phase transition still occurs;
indeed the sudden appearance of a new massless state   
at a critical temperature signals the appearance of a new long-range
order that was not present previously.
Therefore, in order to elucidate the effects of this massless state, 
let us now investigate temperature derivatives of $\calV(T)$.
As evident from the second line of Eq.~(\ref{Vform}),
each temperature derivative $d/dT\sim d/da$
brings down an extra factor of $\tau_2$.  In general, this thereby  increases the
tendency towards divergence of our thermodynamic quantities.

Our results are as follows.
The contribution of this thermally excited proto-gravitino state
to the first derivative $d\calV/da$ is given by
\beq
    {d \calV\over da}  ~=~ \pi \calM^{D-1}\,
              \int_\calF {d^2\tau\over \tau_2^2} \, \tau_2^{1-D/2} \sqrt{\tau_2}~
               \tau_2 \left(  {a\over 4} - {4\over a^3} \right)\,
                ~e^{2\pi \tau_2}
            ~e^{-\pi \tau_2 (a^2/4 + 4/ a^2)}  ~+~...,
\label{Vformonederiv}
\eeq
but at the temperature $a=2$ we see that the factor in parentheses
within Eq.~(\ref{Vformonederiv}) actually vanishes:
\beq
    {d \calV\over da}\biggl|_{a=2}  ~=~ 0~.
\label{firstderiv}
\eeq
It turns out that this is a general property,
reflecting nothing more than the fact that the slope of the mass function
in Eq.~(\ref{massparabola}) vanishes at its minimum, as it must.
However, taking subsequent derivatives and evaluating at $a=2$, we find
the general pattern
\beq
            {d^p \calV\over da^p} \biggl|_{a=2} ~=~
  \calM^{D-1}\, \int_\calF {d^2\tau\over \tau_2^2} \, \tau_2^{1-D/2} \sqrt{\tau_2}
           ~ f_p(\tau_2) ~+~...
\eeq
where $f_p(\tau_2)$ for $p\geq 2$ is a rank-$r$ polynomial in $\tau_2$ of the form
\beq
              f_p(\tau_2)~=~ A_p \,\tau_2^r ~+~ B_p \, \tau_2^{r-1} ~+~ C_p \tau_2^{r-2}~...~,
\eeq
where
\beq
            r ~=~ \cases{ p/2 &  for $p$ even\cr
                          (p-1)/2 & for $p$ odd~,\cr}
\eeq
and where the leading coefficients $A_p$ are positive for
$p=1,2$~(mod 4) and negative for $p=0,3$~(mod 4), with alternating signs for the lower-order
coefficients $B_p$, $C_p$, {\it etc.}
Given these extra leading powers of $\tau_2$,
we thus find that as a result of the proto-gravitino state,
\beq
            {d^p \calV\over dT^p} ~~~~~\hbox{diverges for}~~~~~ \cases{
             D\leq p&  for $p$ odd \cr
             D\leq p+1 & for $p$ even~. \cr}
\eeq
Equivalently, in $D\geq 2$ spacetime dimensions, the proto-gravitino state results
in a divergence that first occurs for $d^p \calV/dT^p$,
where
\beq
            p ~=~ \cases{  D & for $D$ even \cr
                           D-1 & for $D$ odd~.\cr}
\label{order}
\eeq
This divergence then corresponds to a very weak, $p^{\rm th}$-order phase transition.
In particular, for $D=4$, this would be a fourth-order phase transition in which
$d^2 c_V/dT^2$ diverges, causing
$d c_V/dT$ to experience a discontinuity, the specific heat
$c_V$ itself to experience a kink, and the internal energy function
to have a discontinuous change in curvature.  Similar kinds of phase transitions
have also been discussed for two-dimensional heterotic strings in Ref.~\cite{2Dhet},
and for Type~I strings with non-trivial Wilson lines in Ref.~\cite{kounn}.
These results for heterotic strings were first discussed in Ref.~\cite{old}.

We stress that it is not merely the masslessness
of this thermally-enhanced proto-gravitino state that
results in this phase transition.  It is
the fact that this masslessness is achieved {\it thermally}\/,
with non-trivial thermal momentum and winding quanta,
that induces this phase transition.  By contrast, a regular massless state
such as the usual graviton or gravitino does not contribute
to any temperature derivatives of $\calV$.

Thus, we conclude that for supersymmetric heterotic strings,
it is never possible to completely evade a Hagedorn-like phase transition.
Indeed this result holds {\it regardless}\/ of the specific Wilson line chosen
when constructing the finite-temperature theory.
However, the phase transition associated with the proto-gravitino
state appears only at the relatively high temperature $T_H\equiv 2\calM$,
and thus will be completely irrelevant if tachyon-induced Hagedorn
transitions appear at lower temperatures.

\section{A global thermal ``landscape'':  ~~Stability and metastability for finite-temperature strings}
\setcounter{footnote}{0}

Finally, in order to obtain a more 
global sense of the thermodynamic relations between  different Wilson-line choices discussed in Sect.~IV ---
and also in order to perform a more detailed comparison between the Type~I and heterotic cases ---
we now enlarge our perspective and consider the general space 
of allowed Wilson lines for these thermal string theories.
Our goal is to understand the behavior of the corresponding free energies of these theories 
as the underlying Wilson lines are allowed to vary.
Note that in general, we could also consider the variation of a whole host of background fields
and other moduli (such as the dilaton and temperature, or equivalently the thermal compactification radius);  
such analyses appear, \eg, in Refs.~\cite{nonSUSYgauge,EBstrings}. 
Indeed, generic issues arising within this context are not only the dilaton-runaway problem but also
a {\it temperature-runaway}\/ problem (a stringy ``greenhouse'' effect!).
However, for the purposes of our discussion, it will be sufficient to restrict our attention to 
those flat background gauge fields with vanishing field strengths --- \ie, to the space of
allowed Wilson lines in these thermal theories.

Such an analysis is also important for another reason.  
As we have discussed, our approach to generating the finite-temperature extension of a given zero-temperature
string model is to treat these Wilson lines as free parameters that allow us to scan across all possible 
finite-temperature thermal partition functions,  and to attempt to identify 
which Wilson line might lead to a minimum of the free energy with respect to variations of the Wilson line.
Of course, in doing this we have tacitly been assuming 
that such a minimum is unique and is thus a global minimum.  However, this need not be the case:
such solutions might also correspond to {\it local}\/ minima.  
In other words, adopting a terminology that suggests the possibility of ``tunnelling'' transitions
between theories with different Wilson lines, 
we may refer to our preferred Wilson-line choice as leading to a thermal vacuum which is either 
  {\it thermodynamically stable}\/ or {\it thermodynamically metastable}\/ within this sixteen-dimensional space.
Resolving this issue therefore requires understanding something of the global structure of the free energy
as a function of the possible Wilson-line choices.

\subsection{The thermal $SO(32)$ Type~I landscape}

In this section, it will prove simpler to begin by considering the case of the Type~I string.
We have already seen in Sect.~IV that the $SO(32)$, $SO(8)\times SO(24)$, $SO(16)\times SO(16)$, and $U(16)$ cases
correspond respectively to Wilson-line parameters given in Eq.~(\ref{type1cases}).
However, we now shall enlarge our discussion by considering each of the sixteen components of $\vec\ell$
to be an independent general free parameter, and examine the free energy $F(T)$ as a 
function over the resulting sixteen-dimensional parameter space $\lbrace \ell_i\rbrace$, $i=1,...,16$. 

In general, it is relatively easy to calculate the general expressions
that describe the Type~I component partition functions as general
functions of 
the sixteen parameters $\lbrace \ell_i\rbrace$, $i=1,...,16$. 
As we have discussed in Sect.~IV,
since the closed-string states are neutral with respect to the $SO(32)$ gauge group,
their contributions to the total torus and Klein-bottle amplitudes
are insensitive to the appearance of the Wilson line.  As a consequence, the
results for $Z_{\rm T}(\tau,T)$ and $Z_{\rm K}(\tau,T)$ 
in Eq.~(\ref{TypeIthermal}) remain valid even when a Wilson line is turned on.
By contrast, as discussed above, the states contributing to the cylinder and \Mob\ partition functions
carry gauge charges and consist of an anti-symmetric tensor (the adjoint representation)
of the gauge group
as well as a (reducible) symmetric tensor of the gauge group.
If we denote by $\Lambda_S$ and $\Lambda_A$ the sets of gauge charges associated with these symmetric
and anti-symmetric
representations, respectively,
we then find using the results in Eq.~(\ref{momentumshift})
that an arbitrary Wilson line parametrized by $\vec\ell$
causes the thermal cylinder and \Mob\ partition functions in Eq.~(\ref{TypeIthermal})
to take the shifted forms~\cite{ADS1998, AHI2003}:
\beqn
 {\rm cylinder}:  ~~~&   Z_{\rm C}(\taut,T) &=~
               \half \, Z_{\rm open}^{(8)} \,
             \sum_{m\in\IZ} \left\lbrack
                 \sum_{\vec \lambda\in \Lambda_A}
              \left( \chi_V P_{m+ \vec \lambda\cdot \vec\ell} - \chi_S P_{m + \half + \vec \lambda\cdot\vec\ell} \right)
             +
                 \sum_{\vec \lambda\in \Lambda_S}
              \left( \chi_V P_{m+ \vec \lambda\cdot \vec\ell} - \chi_S P_{m + \half + \vec \lambda\cdot\vec\ell} \right)
               \right\rbrack\nonumber\\
 {\rm Mobius}:  ~~~&   Z_{\rm M}(\taut,T) &=~
                         \half\, \widehat{Z}_{\rm open}^{(8)} \,
             \sum_{m\in\IZ} \left\lbrack
                 \sum_{\vec \lambda\in \Lambda_A}
              \left( \chihat_V P_{m+ \vec \lambda\cdot \vec\ell} - \chihat_S P_{m + \half + \vec \lambda\cdot\vec\ell} \right)
             -
                 \sum_{\vec \lambda\in \Lambda_S}
              \left( \chihat_V P_{m+ \vec \lambda\cdot \vec\ell} - \chihat_S P_{m + \half + \vec \lambda\cdot\vec\ell} \right)
               \right\rbrack~.\nonumber\\
\label{TypeIfourcontribs34shifted}
\eeqn
It is straightforward to check that in the special case with $\vec\ell=0$, these expressions recombine
to reproduce the results in Eq.~(\ref{TypeIthermal}),
with the $\half N(N+1)$-dimensional symmetric representation of $SO(32)$ and the
$\half N(N-1)$-dimensional anti-symmetric representation of $SO(32)$
adding and subtracting to produce the overall multiplicities $N^2$ and $N$ respectively.

Given the general expressions given in Eqs.~(\ref{TypeIfourcontribs34shifted}),
we can now examine the corresponding free-energy thermal ``landscape''. 
Performing this analysis is relatively straightforward.
Setting to zero the first derivatives of these partition functions with respect to the 16 parameters $\ell_i$ 
gives us the critical points of this theory ---  note that this condition also ensures the consistency 
of the string vacuum in question by ensuring that all one-loop one-point functions vanish.
Whether the extremum in question is a local maximum, minimum, 
or saddle point (or potentially even lying along a flat direction)
can then be determined by examining the Hessian matrix of second derivatives.

Our results are not unexpected.  
As already anticipated, our thermodynamically preferred $SO(32)$ case 
corresponds to a local minimum which also turns out to be a global minimum.
Thus this solution is thermodynamically stable.
By contrast, each of the other cases listed in Eq.~(\ref{type1cases}) is either a saddle point or local maximum.
There are no metastable local (but not global) minima.

It proves instructive to consider a two-dimensional   
projection of this sixteen-dimensional parameter space.
One such projection which distinctly captures all of the cases in Eq.~(\ref{type1cases})
comes from restricting our attention to Wilson lines  
of the form
\beq
            \vec\ell ~=~ ((\ell)^n (0)^{16-n})~
\label{2Dsubspace}
\eeq
where $\ell$ and $n$ are taken to be our two free parameters, with $0\leq \ell<1$ and $0\leq n\leq 16$.
In the T-dual theory, 
this Wilson line corresponds to having $n$ D8-branes coincident at the point $2 \pi \ell$ on the thermal circle and 
the remaining $(16-n)$ D8-branes coincident with an orientifold fixed plane.  
In terms of these two parameters $(\ell,n)$,
the four cases in Eq.~(\ref{type1cases}) are given by
\beqn
                     {\rm non{-}SUSY}~ SO(32):&~~~~~ (\ell,n)~=& \cases{ 
                                (0,n) & for any $n$\cr
                                (\ell,0) & for any $\ell\in\IZ$\cr
                                (\half,16) & ~\cr }
\nonumber\\
         SO(8)\times SO(24):&~~~~~ (\ell,n)~=& (\half, 4)~{\rm and}~ (\half, 12)
\nonumber\\
        SO(16)\times SO(16):&~~~~~ (\ell,n)~=& (\half,8) 
\nonumber\\
                      U(16):&~~~~~ (\ell,n)~=& (\quarter, 16)~{\rm and}~ (\textstyle{3\over 4}, 16)~.
\label{type1casesln}
\eeqn
Although $n$ is restricted to an integer, we will allow $n$ to range continuously within the range $0\leq n\leq 16$.
Non-integer values of $n$ can be interpreted physically in the T-dual theory as
effectively capturing the dynamics of a configuration with a total of 16 branes (and 16 image branes), 
some of which may be located at points other than $0$ and $2\pi \ell$.  

\begin{figure}[b!]
\centerline{
   \epsfxsize 3.5 truein \epsfbox {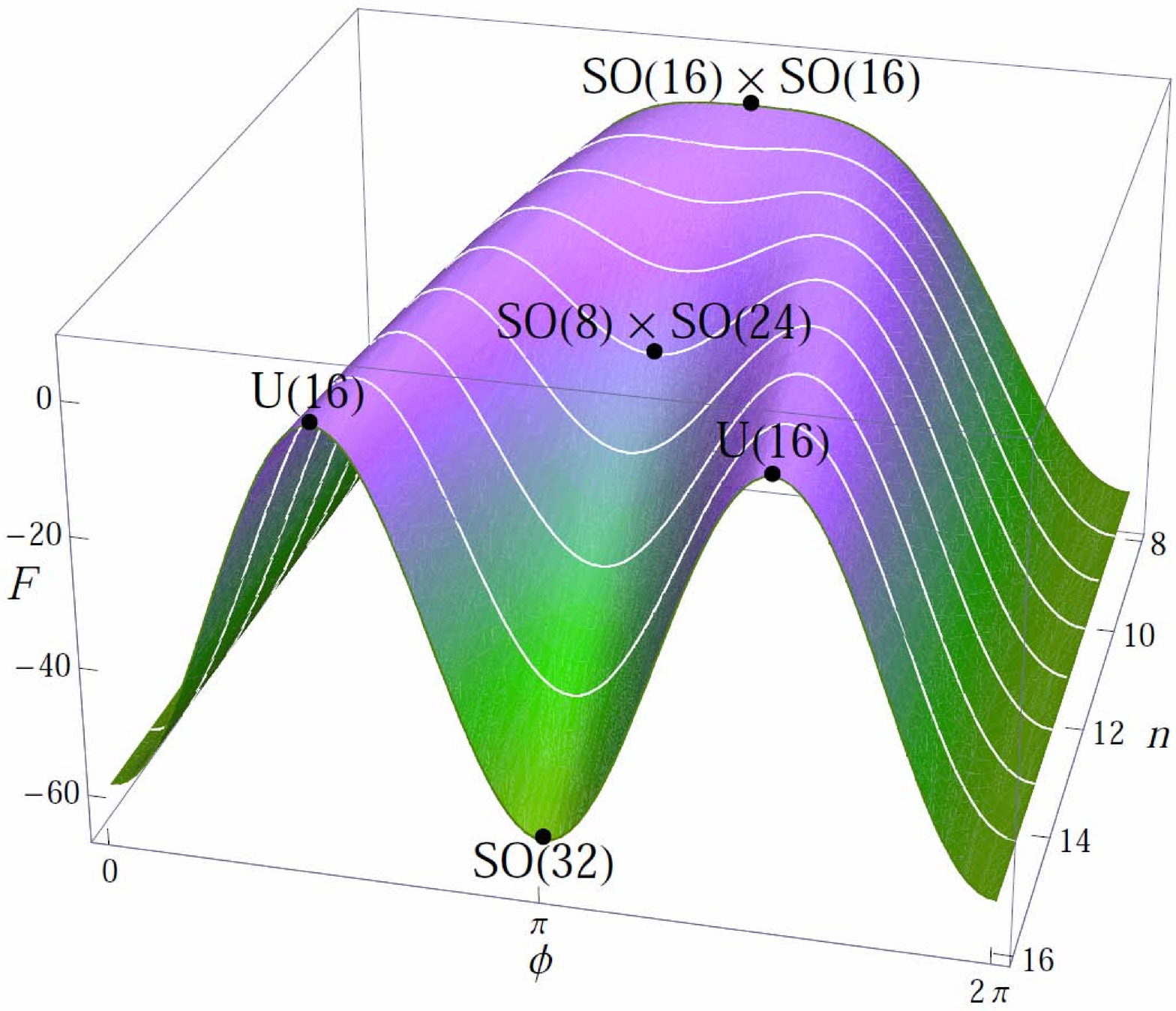}
   \epsfxsize 3.5 truein \epsfbox {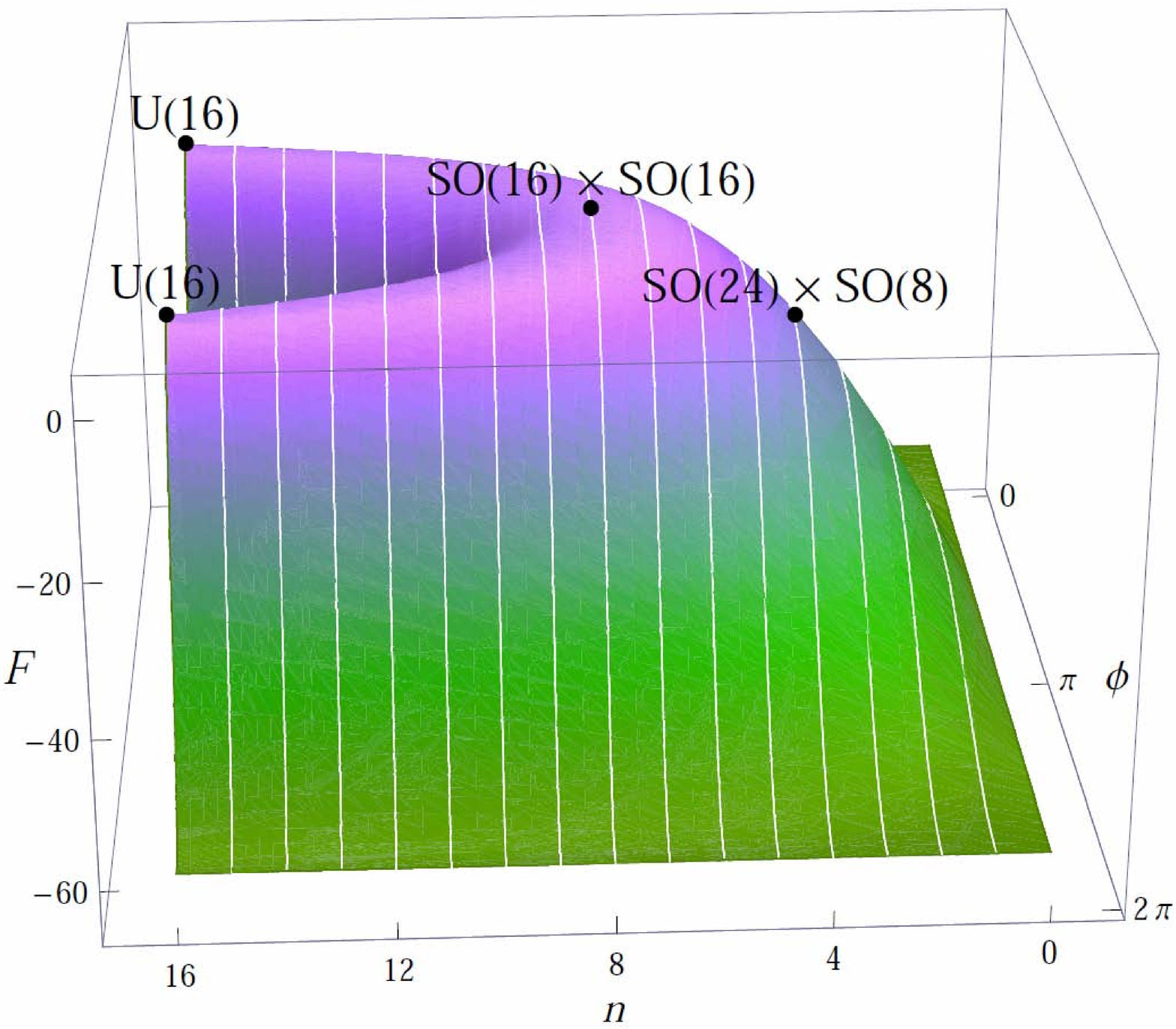} }
	\caption{Two views of the Type~I free-energy density $F(T)$ 
   in units of $\half\calM^{10}$, plotted as a function of 
   Wilson-line parameters $(\phi,n)$ for fixed reference temperature $T=2\calM/3$, 
     where $\phi\equiv 2\pi\ell$.
   The points corresponding to the specific cases listed in Eq.~(\ref{type1casesln}) 
   are also indicated.}
\label{fig:TypeIStability}
\end{figure}


In Fig.~\ref{fig:TypeIStability} we plot the total Type~I free-energy density $F(T)$
as a function of $n$ and $\phi\equiv 2 \pi \ell$. 
As we see from 
Fig.~\ref{fig:TypeIStability},
each of the cases we have examined in Eq.~(\ref{type1casesln})
appears as a critical point.  Moreover,
the non-SUSY $SO(32)$ theory appears as the global minimum.
It is worth noting that
the contour plot in Fig.~\ref{fig:TypeIStability} corresponds
to a fixed reference temperature $T=2\calM/3$.  As the temperature decreases, this contour becomes
increasingly flat,
becoming completely flat at $T=0$ (signalling the restoration of spacetime supersymmetry).  On the
other hand, as the temperature increases, the variations in this contour plot grow without bound,
ultimately diverging as $T\to \calM/\sqrt{2}$ (signalling the approach to the Hagedorn transition). 

It is also worth noting that this contour plot is periodic in the $\phi$ variable,
with a periodicity of magnitude $\Delta \phi=2\pi$ (or equivalently $\Delta \ell=1$).
This periodicity arises because the perturbative Type~I string contains states
in only those ``vectorial'' $SO(32)$ representations 
(corresponding to the adjoint representation, the symmetric tensor, end the singlet) 
whose $SO(32)$ charges $\vec \lambda$ have integer coefficients.
By contrast, if the perturbative Type~I string had contained states in spinorial representations
of $SO(32)$ (with charge vectors $\vec \lambda$ containing half-integer components), the periodicity
of the resulting free-energy contour plot would have been twice as large, with $\Delta\phi=4\pi$. 
In such a case, we could expand the list 
of Wilson lines in Eq.~(\ref{type1cases}), 
formally distinguishing two different classes of Wilson lines which
preserve the $SO(32)$ gauge group of the Type~I model:
\beqn
                     SO(32)_A :&~~~~~ \vec\ell ~=& ((1)^{n} (0)^{16-n})  ~~~{\rm for}~n\in 2\IZ~
\nonumber\\
                     SO(32)_B :&~~~~~ \vec\ell ~=& ((1)^{n} (0)^{16-n})  ~~~{\rm for}~n\in 2\IZ+1~.
\label{so32ab}
\eeqn
The case with vanishing Wilson line is thus of type~$SO(32)_A$.
Of course, since the perturbative $SO(32)$ Type~I string does not contain $SO(32)$ spinorial states,
both classes of Wilson lines lead to the same thermal Type~I model.

\subsection{The thermal $SO(32)$ heterotic landscape}

For the purpose of comparison, we now subject the ten-dimensional supersymmetric $SO(32)$ heterotic string 
to the same analysis.

The effects of Wilson lines on the thermal heterotic string are similar to the effects of
Wilson lines on the thermal Type~I string, with one notable exception:  
the fact that the heterotic string is closed
implies that the thermal string spectrum contains not only momentum modes but also winding modes
around the thermal circle.
The effects of Wilson lines therefore affect both of these quantum numbers, as indicated in 
Eq.~(\ref{shiftNumbersHet}).  

Given this, it is straightforward to generate the heterotic thermal partition functions which include
the effects of generalized Wilson lines.
To do this, we can begin with the 
thermal
partition function corresponding to the special case with $\vec\ell=0$: 
\beqn
    Z(\tau,T) ~=~  Z^{(8)}_{\rm boson} \,\times \,\bigl\lbrace ~
         \phantom{+} &  \chibar_V ~(\chi_I +  \chi_S) & \calE_0 \nonumber\\
         - &  \chibar_S ~(\chi_I +  \chi_S) &\calE_{1/2} \nonumber\\
         - &  \chibar_C ~(\chi_I +  \chi_S) &\calO_{0} \nonumber\\
         + &  \chibar_I ~(\chi_I +  \chi_S) &\calO_{1/2} ~ \bigr\rbrace~
\label{AWpartition}
\eeqn
where the holomorphic $\chi_i$ functions are the characters
associated with the $SO(32)$ gauge group.
We can then use the results in Eq.~(\ref{shiftNumbersHet})
in order to incorporate the effects of a general Wilson line $\vec\ell$.
To do this, we first recall from the Appendix that 
the $\calE$ and $\calO$ functions in Eq.~(\ref{AWpartition}) are given by double sums of the form
given in Eq.~(\ref{Zcircdef}),
where the thermal momentum and winding numbers
in the sum in Eq.~(\ref{Zcircdef}) are restricted
to the sets
\beqn
       \calE_0:~~~~~ \Lambda_{0,0} &\equiv & \lbrace  m\in\IZ,~n~{\rm even}\rbrace\nonumber\\
       \calE_{1/2}:~~~~~ \Lambda_{\half,0}  &\equiv & \lbrace  m\in\IZ+\half ,~n~{\rm even}\rbrace\nonumber\\
       \calO_0:~~~~~  \Lambda_{0,1} &\equiv & \lbrace  m\in\IZ,~n~{\rm odd}\rbrace\nonumber\\
       \calO_{1/2}:~~~~~ \Lambda_{\half,1} &\equiv & \lbrace  m\in\IZ+\half ,~n~{\rm odd}\rbrace~.
\label{EOfunctions2}
\eeqn
Likewise, we recognize that 
\beq
   \chi_I + \chi_S ~=~ {1\over \eta^{16}}\, \sum_{\vec \lambda\in \Lambda_{SO(32)}} q^{\vec \lambda\cdot \vec \lambda/2}
\eeq
where $\eta$ is the Dedekind eta-function 
and where the lattice of $SO(32)$ weights $\vec \lambda$ associated with the character sum $\chi_I+\chi_S$
is given by
\beq
\Lambda_{SO(32)} ~=~ \left\lbrace \lambda_i \in {\IZ},~~ \sum_{i=1}^{16} \lambda_i \in 2\IZ \right\rbrace 
   ~\oplus~ \left\lbrace \lambda_i \in {\IZ}+\half,~~ \sum_{i=1}^{16} \lambda_i \in 2\IZ \right\rbrace~. 
\label{WeightLatticeSO32}
\eeq
Given this, we then find that a general Wilson line $\vec\ell$ deforms the expression
in Eq.~(\ref{AWpartition}) to take the form 
\beqn
    Z[\tau,\vec\ell,T] ~=~  Z^{(8)}_{\rm boson} \,\times \,\bigl\lbrace ~
         \phantom{+}&  \chibar_V~&\Xi[\vec\ell,0,0]\nonumber\\
         - &~\chibar_S~&\Xi[\vec\ell,1/2,0]\nonumber\\
         - &~\chibar_C~&\Xi[\vec\ell,0,1]\nonumber\\
         + &~\chibar_I~&\Xi[\vec\ell,1/2,1]~\bigr\rbrace~
\label{WilsonLinePartition}
\eeqn
where
\beq
   \Xi[\vec\ell,r,s] ~\equiv~ {\sqrt{\tau_2} \over \eta^{16}}\, \sum_{\vec \lambda \in \Lambda_{SO(32)}}
        \sum_{m,n\in \Lambda_{r,s}}
       q^{(\vec \lambda  - n \vec{\ell})\cdot (\vec \lambda-n\vec \ell)/2} \, 
 \qbar^{[(m +\delta m)a - n/a]^2/4}\,
     q^{[(m +\delta m)a + n/a]^2/4}\, 
\label{Qseries}
\eeq
with $\delta m \equiv \vec \lambda\cdot \vec \ell - n \vec\ell\cdot \vec\ell/2$.
Indeed, the general expression in Eq.~(\ref{WilsonLinePartition}) 
is modular invariant for all values of $\vec{\ell}$
and successfully reproduces the partition
functions associated with each of the
thermal heterotic interpolations discussed in Sect.~IV for the specific Wilson-line choices
listed in Eq.~(\ref{hetcases}).
Note, in particular, that the $SO(32)$ heterotic string contains states transforming in 
spinorial representations of $SO(32)$ (\ie, representations which have charge components $\lambda_i\in \IZ+1/2$). 
As a result, the possible Wilson-line parameters 
$\ell_i$ must now be considered modulo 2 rather than modulo 1. 
  {\it As we shall discuss further below, this is an important distinction relative to the Type~I case.}

Despite the existence of the general expression in Eq.~(\ref{WilsonLinePartition}), it is important to bear in mind
that not all Wilson lines $\vec\ell$ correspond to self-consistent heterotic models. 
Indeed, as discussed in Sect.~IV, only the explicit choices listed in Eq.~(\ref{hetcases}) 
satisfy all necessary worldsheet constraints and lead to self-consistent heterotic models. 
We shall nevertheless consider the general unconstrained sixteen-dimensional parameter space of arbitrary Wilson-line
choices $\vec \ell$ in order to compare with the Type~I case. 

Given the general expression in Eq.~(\ref{WilsonLinePartition}), 
we can now examine 
the mathematical behavior of the corresponding free-energy density $F(T)$ as a 
function over the resulting sixteen-dimensional parameter space $\lbrace \ell_i\rbrace$, $i=1,...,16$. 
Unlike the Type~I case, however, we find that there are now {\it two}\/ distinct classes of minima:
\beqn
            SO(32)_A ~&\Longrightarrow&~  {\rm global ~(stable)~ minima} \nonumber\\
            SO(32)_B ~&\Longrightarrow&~  {\rm local ~(metastable)~ minima}~.
\eeqn
The free energies associated with these two classes of minima are nearly equal,
since these two classes of Wilson lines differ only in their treatment of the spinorial $SO(32)$ states,
and such states have large conformal dimensions $h_S=h_C=2$ 
and consequently do not appear until the second or third excited string mass level.
Their contributions to the overall free energy $F(T)$ are thus highly suppressed, and indeed their
free energies differ so minimally that the difference between their free-energy curves as a function
of temperature would not even be visible in Fig.~\ref{Ffig}(a)!

We have also verified that each of the other Wilson-line choices in Eq.~(\ref{hetcases}) corresponds to
either a saddle point or a local maximum in the full sixteen-dimensional parameter space $\lbrace \ell_i\rbrace$.
Thus, we see that the $SO(32)_A$ and $SO(32)_B$ choices are unique 
in that they are the only ones which correspond to minima in this space.
 
\begin{figure}[t!]
\centerline{
   \epsfxsize 6.0 truein \epsfbox {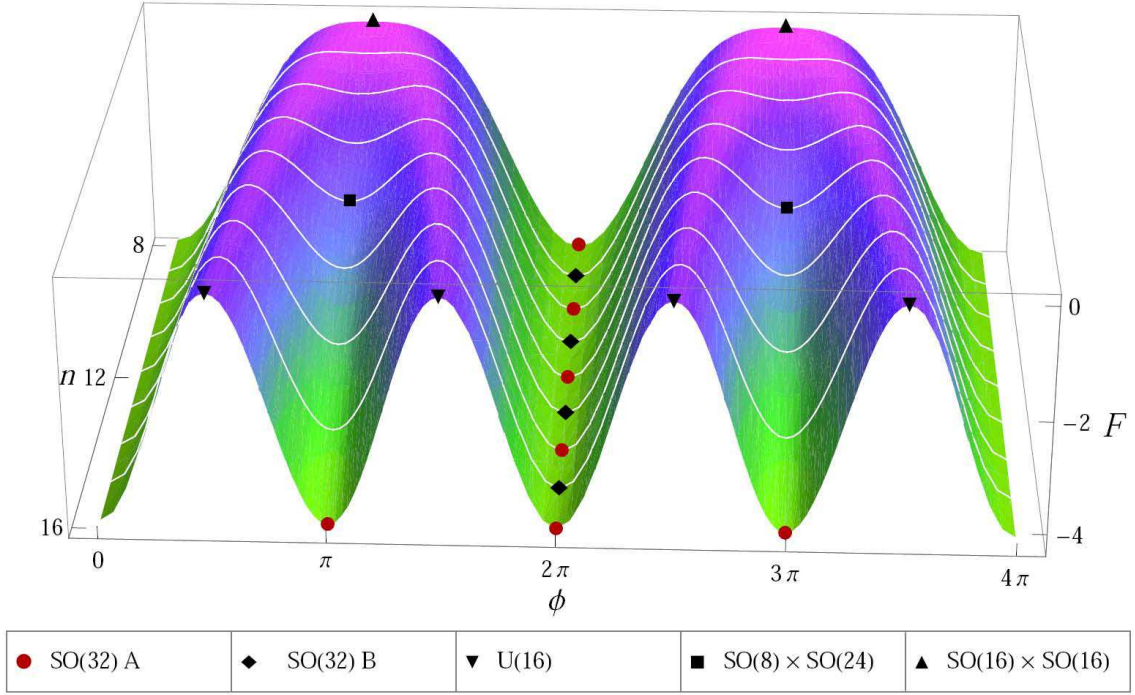}}
\caption{The heterotic free-energy density $F(T)$ 
in units of $\half\calM^{10}$, plotted as a function of 
Wilson-line parameters $(\phi\equiv 2\pi\ell,n)$ for $T=2{\cal M}/3$.
The points corresponding to the specific cases listed in Eq.~(\ref{hetcases}) 
   are also indicated, and these exactly match 
    the critical points of the Type~I theory illustrated in Fig.~\ref{fig:TypeIStability}.
     Unlike the perturbative Type~I case, however, this plot is periodic in $\phi$ with 
    period $\Delta \phi=4\pi$;  this is a direct consequence of the massive states
    which exist in the perturbative heterotic $SO(32)$ string and
    transform in spinorial representations of the gauge group. 
   As a result, the points along the central ``valley'' at $\phi=2\pi$ alternate 
    between $SO(32)_A$ and $SO(32)_B$ Wilson lines  for even and odd $n$ respectively. 
    Note that both classes of $SO(32)$ Wilson lines lead to local free-energy minima 
    in the full sixteen-dimensional $\lbrace \ell_i \rbrace$ parameter space.
    However, unlike the perturbative Type~I case, these two classes of Wilson 
    lines do not lead to the same physics.  Thus, we see that there exist two distinct
    heterotic analogues of the single Type~I $SO(32)$ thermal theory.} 
\label{hetplot}
\end{figure}

As in the Type~I case, it is also instructive to consider the heterotic 
free energy as a contour over the two-dimensional Wilson-line ``landscape''
$(\ell,n)$ parametrized in Eq.~(\ref{2Dsubspace}).
As discussed above, the perturbative heterotic $SO(32)$ string spectrum contains states transforming
in spinorial $SO(32)$ representations;
as a result, our Wilson lines must now be considered modulo $\Delta \phi=4\pi$ rather than $\Delta \phi=2\pi$, 
where $\phi=2\pi\ell$.
The resulting contour plot is shown in Fig.~\ref{hetplot}.

As we have seen, the free energies in the Type~I case are dominated by their cylinder contributions --- contributions which do not even exist in the heterotic case.  Likewise, the free energies in the heterotic case are completely described by modular-invariant expressions which include the contributions from not only momentum modes but also winding modes.
Nevertheless, we see that the qualititative shape of the heterotic contour in Fig.~\ref{hetplot} 
bears a striking similarity to the qualitative shape of the Type~I contour in Fig.~\ref{type1plot}.  
In both cases, the $SO(32)$ points indicate the minima of the contours, while all other critical points
in Eq.~(\ref{hetcases}) are saddle points and/or local maxima.

Of particular interest are the points along the central ``valley'' at $\phi=2\pi$.
These points alternate behaviors, corresponding to $SO(32)_A$ Wilson lines for even $n$ and
$SO(32)_B$ Wilson lines for odd $n$. 
For the Type~I string, of course, we saw that both cases lead to identical 
physics.  However, for the heterotic string, these choices lead
to different physics: 
the first choice leads to the standard thermal theory with
Hagedorn temperature $T= (2-\sqrt{2})\calM$, 
while the second corresponds to  
a thermal theory with a non-trivial Wilson line, one with Hagedorn temperature $T=\calM/\sqrt{2}$. 
As discussed in Sect.~II, the first option is the ``traditional'' choice in which 
essentially no non-trivial Wilson line is introduced;  indeed, this choice
reproduces the Boltzmann sum without a chemical potential.
By contrast,
as discussed in Sect.~IV,
the second
option involves a non-trivial Wilson line and thus introduces
a non-trivial chemical potential into the Boltzman sum.
Nevertheless, in this general ``landscape'' framework, we now see that all of these differences boil down to 
a single distinction:  choosing odd $n$ versus even $n$ along this central valley.
As we have seen, both options place us at local minima in the full sixteen-dimensional $\lbrace \ell_i\rbrace$
parameter space.

In some sense, this observation brings our discussion full circle.
On the surface, it might have seemed unexpected that there exists 
a non-trivial Wilson line which --- like the vanishing Wilson line --- leads
to a (meta)stable vacuum.
However, we now see that this is not a ``random'' Wilson line which has this property:
this is the unique Wilson line which
preserves the heterotic gauge group,  and it 
is also precisely the unique non-zero Wilson line which 
is paired with the vanishing Wilson line along the central valley.
Moreover, comparing the heterotic and Type~I thermal landscapes,
we see that this is the only possible non-trivial
Wilson line which yields a legitimate counterpart to the standard thermal theory on the Type~I side.
Indeed, in terms of matching the physics on the Type~I side,
we see from Fig.~\ref{hetplot} that both the $SO(32)_A$ and $SO(32)_B$ classes of Wilson lines 
are on equal footing and are in this sense equally compelling as thermal extensions of the
ten-dimensional supersymmetric $SO(32)$ heterotic string.
In fact, it is only due to the existence of $SO(32)$ spinorial states on the 
heterotic side that the two Wilson-line choices inherent in the $SO(32)_A$ and $SO(32)_B$ theories
lead to distinct heterotic physics.

Finally, before concluding, we remark that the existence of the general expression in Eq.~(\ref{WilsonLinePartition})
also allows us to deduce the corresponding heterotic Hagedorn temperature as a function of $\vec\ell$:
\beq
 T_H(\vec\ell)~=~ \frac{\sqrt{2} \,\calM}{\sqrt{3-\vec\ell\cdot\vec\ell + \sqrt{8-4\vec\ell\cdot\vec\ell}}}~
\label{HTWilson}
\eeq
where $\vec\ell\cdot\vec\ell$ is evaluated 
in the range $0\leq \vec\ell\cdot\vec\ell \leq 2$.
We thus find that
\eject
\beqn
           SO(32)_A:&~~~~ \vec\ell\cdot\vec\ell = 0~~~~\Longrightarrow&~~~~ T_H=   (2-\sqrt{2})\calM \nonumber\\
    SO(32)_B,~SO(8)\times SO(24),~U(16):&~~~~
           \vec\ell\cdot\vec\ell = 1~~~~\Longrightarrow&~~~~ T_H=   \calM/\sqrt{2}  \nonumber\\
SO(16)\times SO(16):&~~~~
           \vec\ell\cdot\vec\ell = 2~~~~\Longrightarrow&~~~~ T_H=   \sqrt{2} \calM ~,~~~~~~~~~~~~~~~~~~
\label{hagwilson}
\eeqn
in accordance with our previous expectations from Fig.~\ref{Ffig}(a).
Of course, the existence of a single expression such as that in Eq.~(\ref{HTWilson}) reflects the fact that
variations in $\vec\ell$ do not change which massive thermal mode in the general heterotic spectrum first 
becomes massless (and potentially also tachyonic) as a 
function of temperature, thereby triggering the Hagedorn transition.

\subsection{Two distinct thermal theories for heterotic strings?}

Given these results, it is perhaps time to take stock of where we stand.

On the Type~I side, we have seen that we can turn on a variety of Wilson lines.
Although all of these possibilities correspond to extrema of the corresponding 
free-energy densities $F(T)$, we have seen that
only the $SO(32)$ case without a Wilson line yields
a local or global minimum.
This case corresponds to the standard thermal extension of the Type~I string
that usually appears in the literature.

On the heterotic side, by contrast, the situation is more complicated.
For the supersymmetric $SO(32)$ heterotic string,
the standard case without a Wilson line [the so-called $SO(32)_A$ theory] continues to provide
a global minimum.  However,
although the set of self-consistent Wilson lines which may be introduced is more restricted than
for the Type~I string,
we find that there exists a unique non-trivial Wilson line for the $SO(32)$
heterotic string which also leads to a local minimum of the free-energy density:
this is the Wilson line leading to the so-called $SO(32)_B$ theory.

Although we have not examined the corresponding $E_8\times E_8$ thermal ``landscape'' in this paper,
we believe that a
similar situation also exists for the $E_8\times E_8$ heterotic string.
Once again, the standard case without a Wilson line [which we may call the $(E_8\times E_8)_A$ theory] 
continues to provide
a global minimum.  However,
here too there exists a unique non-trivial Wilson line 
which is completely analogous to that for the $SO(32)_B$ theory
and which is also likely to 
lead to a local minimum of the free-energy density:
this is the Wilson line which corresponds to the $SO(16)\times E_8$ orbifold in Eq.~(\ref{B1}).
We may therefore analogously refer to this as the $(E_8\times E_8)_B$ theory,
and we shall assume in what follows that it, like its $SO(32)_B$ counterpart, is metastable.

\begin{table}[t]
\begin{center}
\begin{tabular}{||c||c|c||}
         \hline
         \hline
      ~& $SO(32)_A$ theory & $SO(32)_B$ theory  \\
       \hline
       \hline
  ~&\multicolumn{2}{|c||}{~}\\
~~zero-temperature theory~~ &
    \multicolumn{2}{|c||}{$Z_{SO(32)} = Z_{\rm boson}^{(8)} \,(\chibar_V-\chibar_S)\,(\chi_I + \chi_S)$}\\ 
  ~&\multicolumn{2}{|c||}{~}\\
\hline
  ~&~&~\\
~~general Wilson line & ~~~~~$\vec\ell =\lbrace \ell_i\rbrace,~\ell_i\in\IZ,~\sum_{i=1}^{16} \ell_i={\rm even}~~~~~$  
                & ~~~~~$\vec\ell =\lbrace \ell_i\rbrace,~\ell_i\in\IZ,~\sum_{i=1}^{16} \ell_i={\rm odd}~~~~~$  \\
  ~&~&~\\
       \hline
  ~&~&~\\
~~sample Wilson line & $\vec\ell =( (0)^{16} )$  
                & $\vec\ell =( (1)(0)^{15} )$   \\
  ~&~&~\\
        \hline
  ~&~&~\\
  ~ &  $Z= Z_{\rm boson}^{(8)} \times \bigl\lbrace $~~~~~~~~~~~~~~~~   & 
                              $Z= Z_{\rm boson}^{(8)} \times \bigl\lbrace $~~~~~~~~~~~~~~~~   \\
        ~ & $~~~~\phantom{+}\chibar_V (\chi_I +\chi_S) \,\calE_0 $  &
          $~~~~\phantom{+}(\chibar_V \chi_I -\chibar_S \chi_S) \,\calE_0 $  \\
        ~~thermal partition function~~ & $~~~~~ - \chibar_S (\chi_I +\chi_S) \,\calE_{1/2} $  &
            $~~~~~+(\chibar_V \chi_S -\chibar_S \chi_I) \,\calE_{1/2} $  \\
        ~& $~~~~ - \chibar_C (\chi_I +\chi_S) \,\calO_{0} $  &
            $~~~~+(\chibar_I \chi_V -\chibar_C \chi_C) \,\calO_{0} $  \\
        ~& $~~~~~ ~~~~+ \chibar_I (\chi_I +\chi_S) \,\calO_{1/2} ~\bigr\rbrace$  &
            $~~~~~  ~~~~+(\chibar_I \chi_C -\chibar_C \chi_V) \,\calO_{1/2} ~\bigr\rbrace$  \\
    ~&~&~\\
      \hline
  ~&~&~\\
 stability & {globally stable} &  {locally stable} \\
  ~&~&~\\
   \hline
  ~&~&~\\
  ~~~Hagedorn temperature~~~ &
   $T_H=(2-\sqrt{2}){\cal M} $ 
    &  $T_H= {\cal M}/\sqrt{2} $  \\
  ~&~&~\\
         \hline
         \hline
\end{tabular}
\end{center}
\caption{Two possible thermal theories for the finite-temperature $SO(32)$
       heterotic string in ten dimensions.
       The $SO(32)_A$ theory is traditionally assumed in the
      string literature~\cite{AtickWitten}, while the $SO(32)_B$ theory involves a non-trivial
      Wilson line (or equivalently a non-trivial chemical potential).  
     All holomorphic characters correspond to the $SO(32)$ gauge group.
       As we have seen, both theories are equally compelling as heterotic analogues 
      of the thermal Type~I $SO(32)$ theory, and both 
       locally minimize the corresponding free-energy density and are thus
      locally stable within the thermal heterotic ``landscape''.
      Unlike the traditional $SO(32)_A$ thermal heterotic theory, the $SO(32)_B$ thermal heterotic theory
      more closely resembles the thermal Type~I and Type~II strings by sharing a common
      Hagedorn temperature and exhibiting a non-supersymmetric formal $T\to\infty$ limit.} 
\label{SO32table}
\end{table}

\begin{table}[ht]
\begin{center}
\begin{tabular}{||c||c|c||}
         \hline
         \hline
      ~& $(E_8\times E_8)_A$ theory & $(E_8\times E_8)_B$ theory  \\
       \hline
       \hline
  ~&\multicolumn{2}{|c||}{~}\\
~~zero-temperature theory~~ &
    \multicolumn{2}{|c||}{$Z_{SO(32)} = Z_{\rm boson}^{(8)} \,(\chibar_V-\chibar_S)\,(\chi_I + \chi_S)^2$}\\ 
  ~&\multicolumn{2}{|c||}{~}\\
\hline
  ~&~&~\\
~~general Wilson line~~ & ~~~~~$\vec\ell =\lbrace \ell_i\rbrace,~\ell_i\in\IZ,~$  
                        & ~~~~~$\vec\ell =\lbrace \ell_i\rbrace,~\ell_i\in\IZ,~$     \\
 ~& $~~~~~~\sum_{i=1}^{8}  \ell_i={\rm even}~,~
           \sum_{i=9}^{16} \ell_i={\rm even}~~~~~$  
  & $~~~~~~\sum_{i=1}^{8}  \ell_i={\rm odd}~,~
           \sum_{i=9}^{16} \ell_i={\rm even}~~~~~$       \\  
  ~&~&~\\
       \hline
  ~&~&~\\
~~sample Wilson line & $\vec\ell =( (0)^{16} )$  
                & $\vec\ell =( (1)(0)^{15} )$   \\
  ~&~&~\\
        \hline
  ~&~&~\\
  ~ &  $Z= Z_{\rm boson}^{(8)} \times \bigl\lbrace $~~~~~~~~~~~~~~~~   & 
                              $Z= Z_{\rm boson}^{(8)} \times \bigl\lbrace $~~~~~~~~~~~~~~~~   \\
        ~ & $~~~~\phantom{+}\chibar_V (\chi_I +\chi_S)^2 \,\calE_0 $  &
          $~~~~\phantom{+}(\chibar_V \chi_I -\chibar_S \chi_S) \,\calE_0 $  \\
        ~~thermal partition function~~ & $~~~~~ - \chibar_S (\chi_I +\chi_S)^2 \,\calE_{1/2} $  &
            $~~~~~+(\chibar_V \chi_S -\chibar_S \chi_I) \,\calE_{1/2} $  \\
        ~& $~~~~ - \chibar_C (\chi_I +\chi_S)^2 \,\calO_{0} $  &
            $~~~~+(\chibar_I \chi_V -\chibar_C \chi_C) \,\calO_{0} $  \\
        ~& $~~~~~ ~~~~+ \chibar_I (\chi_I +\chi_S)^2 \,\calO_{1/2} ~\bigr\rbrace$  &
            $~~~~~  ~~~~+(\chibar_I \chi_C -\chibar_C \chi_V) \,\calO_{1/2} ~\bigr\rbrace $  \\
     ~&~ & ~~~~~~~~~~~~~~~~~~~~$\times \, (\chi_I+\chi_S) $ \\
    ~&~&~\\
      \hline
  ~&~&~\\
 stability & {globally stable} &  {locally stable} \\
  ~&~&~\\
   \hline
  ~&~&~\\
  ~~~Hagedorn temperature~~~ &
   $T_H=(2-\sqrt{2}){\cal M} $ 
    &  $T_H= {\cal M}/\sqrt{2} $  \\
  ~&~&~\\
         \hline
         \hline
\end{tabular}
\end{center}
\caption{Two possible thermal theories for the finite-temperature $E_8\times E_8$
       heterotic string in ten dimensions.
       The $(E_8\times E_8)_A$ theory is traditionally assumed in the
      string literature~\cite{AtickWitten}, while the $(E_8\times E_8)_B$ theory involves a non-trivial
      Wilson line (or equivalently a non-trivial chemical potential).  
     Each holomorphic character corresponds to the $SO(16)$ gauge group.
      Unlike the traditional $(E_8\times E_8)_A$ thermal heterotic theory, the $(E_8\times E_8)_B$ 
       thermal heterotic theory
      more closely resembles the thermal Type~I and Type~II strings by sharing a common
      Hagedorn temperature and exhibiting a non-supersymmetric formal $T\to\infty$ limit.  Note
      the similarity between this table and Table~\ref{SO32table}:  essentially the $E_8\times E_8$
      partition functions in each case can be obtained from the corresponding $SO(32)$ functions 
      by viewing the left-moving characters as corresponding to $SO(16)$ rather than $SO(32)$ and
      multiplying by an additional modular-invariant factor $(\chi_I+\chi_S)$.
    This tight similarity between these two groups of theories 
    suggests that both the $(E_8\times E_8)_A$ and $(E_8\times E_8)_B$ theories
       locally minimize the corresponding free-energy density and are thus
      locally stable within the thermal heterotic ``landscape''.  }  
\label{E8table}
\end{table}

These two sets of theories are summarized in Tables~\ref{SO32table} and \ref{E8table},
and it is readily apparent that these theories share close similarities with each other.
In each case, the `A' theories correspond to thermal theories without Wilson lines,
while the `B' theories correspond to the Wilson line $\vec\ell = (1,0^{15})$.
Both options lead to thermal theories which are locally stable within the thermal
landscape, and indeed these are the {\it only}\/ theories which have this property.
Moreover, as remarked in Sect.~IV for the case of the $SO(32)$ string, 
the free-energy difference between the `A' theory
and the corresponding `B' theory is extremely small --- indeed,
such differences would not even be visible on the plots in Fig.~\ref{Ffig}.
Consequently, even though the `B' theories are technically only metastable,
there is very little dynamical ``force'' which would cause our universe
to flow from the `B' state to the corresponding `A' state.
(Phrased more precisely, an instanton analysis of the transitions from the `B' vacua
to the `A' vacua would lead to a very small decay width or equivalently an extremely long  
lifetime for the `B' theories.)  
As a result, it is quite possible that our universe, if somehow ``born'' in the 
`B' state, might reside there essentially forever.
However, as briefly mentioned in Sect.~IV, 
even this notion presupposes the existence of transitions between theories with different Wilson
lines --- something which is not at all obvious, given the quantized topological nature of 
the Wilson lines themselves.  

We are therefore faced with a situation in which both the `A' theories and the `B' theories may
be considered as legitimate ``ground'' states for our thermal heterotic strings.
Indeed, as we have explicitly seen in the case of the $SO(32)$ landscapes in Sect.~VI.B, 
both the $SO(32)_A$ and $SO(32)_B$ thermal heterotic theories have equal claims to be 
considered as the legitimate heterotic analogue of the $SO(32)$ Type~I thermal theory.

It is quite remarkable that the heterotic string gives rise to such a situation.
However, this phenomenon ultimately rests upon the existence of a unique Wilson line
which simultaneously has all of the properties needed in order to endow the resulting `B' theories 
with these critical features.
As we have seen, the Type~I string {\it a priori}\/ has a richer set of allowed Wilson lines,
yet none of these has the required properties.

Given the existence of these `B' theories, many questions naturally arise.
For example, it is well known that the zero-temperature supersymmetric $SO(32)$ Type~I  
and heterotic strings are related to each other under strong/weak coupling S-duality relations.
One naturally wonders, therefore, whether such S-duality relations extend to finite temperatures.
However, as we have seen, the ``landscape'' of the $SO(32)$ heterotic string at finite temperature
includes not only the $SO(32)_A$ theory but also the $SO(32)_B$ theory.
It would therefore be interesting to understand how this ``doubling'' phenomenon can be reconciled
with the existence of a unique thermal $SO(32)$ theory on the Type~I side~\cite{DS}.

Another important question concerns whether there might be some other way (\eg, through dynamical means,
or perhaps through self-consistency arguments) in order to develop a thermal ``vacuum selection'' criterion
and thereby assert that only one of these theories is preferred or allowed, and the other excluded.
This question will be examined in Refs.~\cite{D,DL}.
It is well known, for example, that the $SO(32)_A$ theory has a number of unexpected and
disturbing features, the least of which is the fact that the $T\to\infty$ limit of this theory
is again supersymmetric.  This is also true of the $(E_8\times E_8)_A$ theory.
This property is completely surprising, given our expectation that finite-temperature
effects should treat bosons and fermions differently, and is very different from what occurs for Type~I and Type~II 
thermal theories. Indeed, this behavior leads to several disturbing features (see, \eg, the discussion 
in Ref.~\cite{AtickWitten}).
By contrast, the $SO(32)_B$ theory is relatively natural from this point of view:  the $T\to\infty$ limit of this
theory lacks spacetime supersymmetry, but preserves the underlying $SO(32)$ gauge symmetry.
Even in the $E_8\times E_8$ case, the `B' theory has a non-supersymmetric $T\to\infty$ limit yet breaks the gauge symmetry as minimally as possible. 
Thus, if we could somehow argue that the `B' theories are the unique correct thermal heterotic theories,
we would then reach the remarkable conclusion that all string theories in ten dimensions, whether open or closed,
whether Type~I or Type~II or heterotic, actually have a unique Hagedorn temperature. 
In other words, we would be in the aesthetically pleasing situation 
in which we could assert a existence of a single Hagedorn temperature for string theory as a whole.
This issue is discussed further in Refs.~\cite{D,DL}.

\section{Discussion and conclusions}
\setcounter{footnote}{0}

In this paper, we investigated the consequences of introducing a non-trivial Wilson
line (or equivalently a non-trivial chemical potential) 
when formulating string theories at finite temperature.
We focused on the heterotic and Type~I strings in ten dimensions,
and surveyed the possible Wilson lines which might be introduced when extending
these strings to finite temperature.
We found a rich structure of resulting thermal string theories,
and showed that in the heterotic case, some of these new thermal theories 
even have Hagedorn temperatures 
which are shifted from their usual values.
Remarkably, these shifts in the Hagedorn temperature
are not in conflict with the densities of bosonic and fermionic states
which are exhibited by their zero-temperature counterparts.
We also demonstrated that our new thermal string theories
can be interpreted as extrema of a continuous thermal free-energy
``landscape''.  Finally, as part of this study, we also uncovered 
a pair of 
unique finite-temperature extensions
of the heterotic $SO(32)$ and $E_8\times E_8$ strings which involve a non-trivial Wilson line,
but which are nevertheless metastable in this thermal landscape.
As we have argued, these new thermal theories (the so-called `B' theories discussed in Sect.~VI.C)
represent {\it bona fide}\/ alternatives to the traditional thermal heterotic strings,
and may be viewed as equally legitimate candidate finite-temperature extensions
of the zero-temperature $SO(32)$ and $E_8\times E_8$ heterotic strings in ten dimensions.
Indeed, as we have seen, the analysis in this paper
also illustrates that the $SO(32)_A$ and $SO(32)_B$ heterotic theories are 
on equal footing as potential heterotic analogues of the thermal $SO(32)$ Type~I string.

Clearly, many outstanding questions remain.
For example, in this paper we have found that for each of the supersymmetric
heterotic strings in ten dimensions, there exists a unique non-trivial Wilson
line which leads to a metastable theory.
However, it would be interesting to understand more generally for which string theories
this will be the case.
Likewise, we have found that these new Wilson lines lead to 
an increased Hagedorn temperature.
It is therefore natural to wonder whether
there might exist situations in which metastable Wilson lines
manage to avoid the Hagedorn transition entirely.

It is also important to realize that in the Type~I case, our analysis
has essentially focused on those Wilson lines associated with the open-string sector.
In ten dimensions, this was a legitimate restriction, as one cannot introduce
Wilson lines in the closed-string sector of ten-dimensional Type~I strings.
However, it would clearly be of great interest to examine the situation in lower dimensions.
In lower dimensions, perturbative Type~II strings can have non-trivial gauge groups 
which are generated by compactification.  As a result, there can be non-trivial Wilson
lines that are associated with such Type~II compactifications, and therefore 
the potential Type~I orientifolds of such models can quickly become quite numerous.  In particular,
the set of candidate thermal extensions of a given Type~I string model might include models
which are distinct in terms of their closed-string sectors as well as their open-string sectors.

Of course, our analysis in this paper is subject to a number of important caveats.
First, we have been dealing with one-loop string vacuum amplitudes,
and likewise considering only the tree-level (non-interacting) particle spectrum.
Thus, we are neglecting all sorts of particle interactions.
Gravitational effects, in particular, can be expected to change the
spectrum quite dramatically, and have recently been argued to eliminate
the Hagedorn transition completely by deforming the resulting spectrum away from the expected
exponential rise in the degeneracy of states.
However, the purpose of this paper has been to show that even within the non-interacting
string theory which has been studied for more than two decades,
the introduction of non-trivial Wilson lines can have significant effects on the
resulting thermal theories.

This work has clearly focused on the thermal behavior of string theories
at temperatures {\it below}\/ the Hagedorn transition.
As such, it is not clear that these results will shed any light on
that feature which remains the most mysterious aspect of string thermodynamics:
the nature of the Hagedorn transition itself.  
However, the involvement of non-trivial Wilson lines may 
eventually have ramifications in this regard that are not yet apparent.

In this paper we have restricted our attention to the thermal properties
of {\it perturbative}\/ Type~I and heterotic strings, especially when
non-trivial Wilson lines are introduced into the mix.
However, it would clearly be of great interest to
consider the implications of these results at the non-perturbative level, once
D-branes and other non-perturbative structures are included.
In particular, it would be interesting to study the possible implications of these results
for the thermodynamics of D$p$-branes~\cite{D-BranesT,D-BranesPT} as well as  
for the cosmological applications of finite-temperature D-branes~\cite{D-branesC}.
In this connection, it would also be interesting to 
understand the thermal
consequences of our observations within the recent brane-world scenarios,
as these frameworks
involve a subtle dynamical interplay between 
bulk (closed-string) physics and
brane (open-string) physics.
Likewise, it would also be interesting to extend our results to non-flat backgrounds
in order to address important questions such as the thermodynamics of black holes,
the AdS/CFT correspondence, and so forth.

Continuing this line, it would also be interesting to understand
the implications of these results for the existence of strong/weak coupling 
duality relations at finite temperature.
It is certainly aesthetically pleasing that the heterotic and Type~I Hagedorn temperatures
can be brought into agreement through the introduction of non-trivial Wilson lines, and this 
immediately raises the question whether the zero-temperature heterotic/Type I dualities can be extended
to finite temperature. 
In particular, 
although the perturbative Type~I string lacks states transforming as $SO(32)$ spinors
--- which is ultimately the reason why the $SO(32)_A$ and $SO(32)_B$ theories are
equivalent on the Type~I side ---
such states do emerge non-perturbatively~\cite{NPTypeI}.
Thus we can expect that the non-perturbative Type~I theory will have a thermal behavior
which is even closer to that of the perturbative heterotic string,
and for which a non-trivial Wilson line might be examined as well.
These issues will be examined in more detail in Ref.~\cite{DS}.

But above all, perhaps the most pregnant issue is the existence of the 
so-called `B' theories, and the roles these theories may ultimately play in the general structure
of string theory at finite temperature.
Given that the `A' and the `B' theories appear to be equally compelling as finite-temperature
extensions of the traditional zero-temperature heterotic strings, it remains
to investigate whether there might be a thermal vacuum selection principle that favors one
over the other (see, \eg, Refs.~\cite{D,DL}).  Moreover,
if no such selection principle exists, it will be important to study
the different physics to which they each lead, and the possibility of phase transitions between them. 
These issues clearly warrant further study.

\section*{Acknowledgments}
\setcounter{footnote}{0}

We are happy to thank E.~Dudas for discussions.
This work was supported in part by the Department of Energy
under Grant~DE-FG02-04ER-41298, and by the National Science Foundation
through its employee IR/D program.
ML also wishes to acknowledge 
the Ecole Polytechnique in France 
and the Perimeter Institute and McMaster University in Canada,
where portions of this work were done.
In these latter institutions, ML was supported in part
by ANR grant ANR-05-BLAN-0079-02 (France), RTN contracts 
MRTN-CT-2004-005104 and MRTN-CT-2004-503369 (France), 
CNRS PICS~\#2530, \#3059, and \#3747 (France),
European Union Excellence Grant MEXT-CT-2003-509661,
and the Natural Sciences and Engineering Research Council (NSERC) of Canada.
The opinions and conclusions expressed herein are those of the authors, 
and do not represent either the Department of Energy or the National 
Science Foundation.


\appendix  

\section{~~Useful Trace Formulae}
\setcounter{footnote}{0}

In this Appendix, we collect the mathematical expressions which are used in this paper
for the traces over relevant string Fock spaces.
These results also serve to define our notations and conventions.

\subsection{Thermal Sums}

For any temperature $T$, we define the corresponding dimensionless
temperature $a\equiv 2\pi T/M_{\rm string} \equiv T/\calM$
where $\calM\equiv M_{\rm string}/(2\pi)=(2\pi\sqrt{\alpha'})^{-1}$.
We also define the associated thermal radius
$R\equiv  (2\pi T)^{-1}$.
A field compactified on a circle with this radius then accrues integer
Matsubara momentum and winding modes around this thermal circle,
resulting in left- and right-moving spacetime momenta of the forms
\beq
   p_R~=~ {1\over \sqrt{2\alpha'}} (ma-n/a)~,~~~~~~~~
   p_L~=~ {1\over \sqrt{2\alpha'}} (ma+n/a)~.
\eeq
Here $m$ and $n$ respectively represent the momentum 
and winding quantum numbers of the field in question.
The contribution to the partition function
from such thermal modes then takes the form of the double summation
\beq
     Z_{\rm circ}(\tau,T)~=~
        \sqrt{\tau_2}\, \sum_{m,n\in\IZ} \,
         \overline{q}^{\alpha' p_R^2/2} q^{\alpha' p_L^2/2}
        ~=~
     \sqrt{ \tau_2}\,
    \sum_{m,n\in\IZ} \,
      \overline{q}^{(ma-n/a)^2/4}  \,q^{(ma+n/a)^2/4}
\label{Zcircdef}
\eeq
where $q\equiv \exp(2\pi i\tau)$
and where $\tau_{1,2}$ respectively denote ${\rm Re}\,\tau$ and ${\rm Im}\,\tau$.
Note that $Z_{\rm circ}\to 1/a$ as $a\to 0$,
while  $Z_{\rm circ}\to a$ as $a\to \infty$.

The trace $Z_{\rm circ}$ is sufficient for 
compactifications on a thermal circle.
However, in this paper we are interested in
compactifications on $\IZ_2$ orbifolds
of the thermal circle.
Towards this end, we
introduce~\cite{Rohm} four new functions $\calE_{0,1/2}$ and $\calO_{0,1/2}$
which are the same as the summation in
$Z_{\rm circ}$ in Eq.~(\ref{Zcircdef}) except for the following
restrictions on their summation variables:
\beqn
       \calE_0 &=& \lbrace  m\in\IZ,~n~{\rm even}\rbrace\nonumber\\
       \calE_{1/2} &=& \lbrace  m\in\IZ+\half ,~n~{\rm even}\rbrace\nonumber\\
       \calO_0 &=& \lbrace  m\in\IZ,~n~{\rm odd}\rbrace\nonumber\\
       \calO_{1/2} &=& \lbrace  m\in\IZ+\half ,~n~{\rm odd}\rbrace~.
\label{EOfunctions}
\eeqn
Note that these functions
are to be distinguished from a related (and also often used)
set of functions with the same names in which the roles of $m$ and $n$ are exchanged.
Under the modular transformation $T:\tau\to\tau+1$,
the first three functions are invariant while
$\calO_{1/2}$ picks up a minus sign;
likewise, under $S:\tau\to-1/\tau$,
these functions mix according to
\beq
   \pmatrix{  \calE_0 \cr \calE_{1/2} \cr \calO_0  \cr  \calO_{1/2}  } (-1/\tau) ~=~
   \half \pmatrix{   1 & 1 & 1 & 1 \cr
                   1 & 1 & -1 & -1 \cr
                   1 & -1 & 1 & -1 \cr
                   1 & -1 & -1 & 1 \cr}
   \pmatrix{  \calE_0 \cr \calE_{1/2} \cr \calO_0  \cr  \calO_{1/2}  } (\tau) ~.
\label{Smixing}
\eeq
In the $a\to 0$ limit,
$\calO_0$ and $\calO_{1/2}$ each vanish while $\calE_0,\calE_{1/2} \to 1/a$;
by contrast, as $a\to\infty$,
$\calE_{1/2}$ and $\calO_{1/2}$ each vanish while $\calE_0,\calO_0 \to a/2$.
Clearly, $\calE_{0}+\calO_{0}= Z_{\rm circ}$.

The thermal ${\cal E}/{\cal O}$ functions are primarily of relevance for closed strings,
since such strings have both momentum and winding modes.
For open strings, by contrast, we instead define the thermal functions
\beq
      \calE \equiv \sum_{m\in \IZ} P_m~,~~~~~~~~\calE' \equiv \sum_{m\in \IZ} P_{m+1/2}~,
\label{EEprime}
\eeq
where $P_m\equiv \sqrt{\tau_2} \exp(-\pi^2 m^2 a^2 \tau_2)$, with $a\equiv T/\calM = 2\pi T/M_{\rm string}$.
Note that $\calE$ is the open-string analogue of $\calE_0$, while $\calE'$ is the open-string
analogue of $\calE_{1/2}$.  Indeed, the remaining closed-string 
functions $\calO_{0,1/2}$ do not have open-string analogues
because they both involve non-trivial winding modes, and winding modes do not survive the sorts of
direct orientifold projections that we implement in this paper in order to construct 
our thermal Type~I string models.

\subsection{$SO(2n)$ characters}

We begin by recalling the standard definitions
of the Dedekind $\eta$ and Jacobi $\vartheta_i$ functions:
\beqn
    \eta(q)  &\equiv&  q^{1/24}~ \displaystyle\prod_{n=1}^\infty ~(1-q^n)~=~
                \sum_{n=-\infty}^\infty ~(-1)^n\, q^{3(n-1/6)^2/2}\nonumber\\
    \vartheta_1(q)&\equiv&  
                 {\displaystyle 2\sum_{n=0}^\infty (-1)^n q^{(n+1/2)^2/2} }\nonumber\\
    \vartheta_2(q)&\equiv&  2 q^{1/8} \displaystyle\prod_{n=1}^\infty (1+q^n)^2 (1-q^n)~=~
                 2\sum_{n=0}^\infty q^{(n+1/2)^2/2} \nonumber\\
    \vartheta_3(q)&\equiv&  \displaystyle\prod_{n=1}^\infty (1+q^{n-1/2})^2 (1-q^n) ~=~
                1+ 2\sum_{n=1}^\infty q^{n^2/2} \nonumber\\
    \vartheta_4(q) &\equiv& \displaystyle\prod_{n=1}^\infty (1-q^{n-1/2})^2 (1-q^n) ~=~
                1+ 2\sum_{n=1}^\infty (-1)^n q^{n^2/2} ~.
\label{etathetadefs}
\eeqn
These functions satisfy the identities $\tthree^4 =\ttwo^4+\tfour^4$
and $\ttwo\tthree\tfour=2\eta^3$.
Note that $\vartheta_1(q)$ has a vanishing $q$-expansion and is modular invariant;  
its infinite-product representation has a vanishing coefficient 
and is thus not shown.
This function is nevertheless included here because it plays a role 
within string partition functions as the indicator
of the chirality of fermionic states, as discussed below.

The partition function of $n$ free bosons is given by 
\beq
         Z^{(n)}_{\rm boson} ~\equiv ~ {\tau_2}^{-n/2}\, (\overline{\eta}\eta)^{-n}~.
\label{bosons}
\eeq
By contrast, the characters of the level-one $SO(2n)$ affine Lie algebras
are defined in terms of both the $\eta$- and the $\vartheta$-functions.
Recall that at affine level one, the $SO(2n)$ algebra for each $n\in\IZ$ has
four distinct representations:  the identity ($I$),
the vector ($V$), the spinor ($S$), and the conjugate spinor ($C$).   
In general,
these representations have conformal dimensions $\lbrace h_I,h_V,h_S,h_C\rbrace=
\lbrace 0,1/2,n/8,n/8\rbrace$, and their characters are given by
\beqn
 \chi_I &=&  \half\,(\tthree^n + \tfour^n)/\eta^n ~=~ q^{h_I-c/24} \,(1 + n(2n-1)\,q + ...)\nonumber\\
 \chi_V &=&  \half\,(\tthree^n - \tfour^n)/\eta^n ~=~ q^{h_V-c/24} \,(2n + ...)\nonumber\\
 \chi_S &=&  \half\,(\ttwo^n + i^{-n} {\vartheta_1}^n)/\eta^n ~=~ q^{h_S-c/24} \,(2^{n-1} + ...)\nonumber\\
 \chi_C &=&  \half\,(\ttwo^n - i^{-n} {\vartheta_1}^n)/\eta^n ~=~ q^{h_C-c/24} \,(2^{n-1} + ...)~
\label{chis}
\eeqn
where the central charge is $c=n$ at affine level one.
The vanishing of $\vartheta_1$ implies that 
$\chi_S$ and $\chi_C$ have identical $q$-expansions;
this is a reflection of the conjugation symmetry between the spinor
and conjugate spinor representations. 
When $SO(2n)$ represents a transverse spacetime Lorentz group,
the distinction between $S$ and $C$ is equivalent to relative
spacetime chirality;  the choice of which spacetime chirality is to 
be associated with $S$ or $C$ is a matter of convention.
Note that the special case $SO(8)$ 
has a further triality symmetry under which
the vector and spinor representations are
indistinguishable.  Thus, for $SO(8)$, we find that $\chi_V=\chi_S$,
an identity already given 
below Eq.~(\ref{etathetadefs}) 
in terms of $\vartheta$-functions.

The above results are primarily of relevance for closed strings, 
where $\tau$ is the complex torus modular parameter
and where $q\equiv (2\pi i\tau)$.
However, with only small modifications, these functions can
also be used to describe the partition-function contributions in
Type~I strings.
In general, 
the one-loop Type~I partition function includes not only
a closed sector with contributions from a torus and a Klein bottle,
but also an open sector with
contributions from a cylinder and a M\"obius strip.
It turns out that all four of these contributions can be
written in terms of the above CFT characters $\chi$, which
are strictly defined as functions of $q\equiv (2\pi i\tau)$.
Indeed, all that changes is the definition of $\tau$:
for the torus, $\tau$ will
continue to represent the complex modulus, while for the Klein bottle, cylinder, and M\"obius strip,
$\tau$ will instead represent the modulus of the double-covering torus:
\beq
        \tau~\equiv ~\cases{
        \tau_1 + i\tau_2  & for torus \cr
        2 i \taut         & for Klein bottle \cr
        \half i \taut               & for cylinder \cr
        \half i \taut + \half & for M\"obius strip~.\cr}
\label{tauCases}
\eeq
Likewise, the contribution from 
worldsheet bosons will also change from that in Eq.~(\ref{bosons}) to
\beq
       Z_{\rm boson}^{\prime (n)} ~=~ \tau_2^{-n/2} \eta^{-n}~.
\eeq
Finally, in this paper we shall also define 
the hatted characters
$\chihat_i \equiv \exp(-i \pi h_i)\chi_i$,
where $h_i$ are the conformal weights of the corresponding primary fields $\Phi_i$ in the underlying CFT.
Thus, the hatted characters are 
explicitly real.  These characters
are particularly useful for expressing
the contributions from the M\"obius sectors.



\begin{thebibliography}{99}




\bibitem{Polbook}
      For an introduction, see:\\
        M.~B.~Green, J.~A.~Schwarz and E.~Witten,
        {\it Superstring Theory, Vols.~I and II}\/
        (Cambridge University Press, 1987);\\
      J. Polchinski, {\it String Theory, Vols.~I and II}\/
      (Cambridge University Press, 1998).  


\bibitem{Pol86}   J.~Polchinski,
     Commun.\ Math.\ Phys.\  {\bf 104}, 37 (1986).


\bibitem{McClainRoth}
  B.~McClain and B.~D.~B.~Roth,
  Commun.\ Math.\ Phys.\  {\bf 111}, 539 (1987);\\
  K.~H.~O'Brien and C.~I.~Tan,
  Phys.\ Rev.\ D {\bf 36}, 1184 (1987);\\
  M.~Trapletti,
  JHEP {\bf 0302}, 012 (2003)
  [arXiv:hep-th/0211281].



\bibitem{ft}  See, {\it e.g.}\/:
     A.~Roberge and N.~Weiss,
     Nucl.\ Phys.\ B {\bf 275}, 734 (1986);\\
     M.~G.~Alford, A.~Kapustin and F.~Wilczek,
  Phys.\ Rev.\ D {\bf 59}, 054502 (1999)
  [hep-lat/9807039];\\
  A.~Hart, M.~Laine and O.~Philipsen,
  Phys.\ Lett.\ B {\bf 505}, 141 (2001)
  [hep-lat/0010008];\\
  P.~de Forcrand and O.~Philipsen,
  Nucl.\ Phys.\ B {\bf 642}, 290 (2002)
  [hep-lat/0205016];\\
  M.~D'Elia and M.~-P.~Lombardo,
  Phys.\ Rev.\ D {\bf 67}, 014505 (2003)
  [hep-lat/0209146];\\
  P.~H.~Damgaard, U.~M.~Heller, K.~Splittorff, B.~Svetitsky and D.~Toublan,
  Phys.\ Rev.\ D {\bf 73}, 105016 (2006)
  [hep-th/0604054];\\
   M.~D'Elia, F.~Di Renzo and M.~P.~Lombardo,
  Phys.\ Rev.\ D {\bf 76}, 114509 (2007)
  [arXiv:0705.3814 [hep-lat]];\\
  Y.~Sakai, K.~Kashiwa, H.~Kouno and M.~Yahiro,
  Phys.\ Rev.\ D {\bf 77}, 051901 (2008)
  [arXiv:0801.0034 [hep-ph]];
  C.~Lehner, M.~Ohtani, J.~J.~M.~Verbaarschot and T.~Wettig,
  Phys.\ Rev.\ D {\bf 79}, 074016 (2009)
  [arXiv:0902.2640 [hep-th]].



\bibitem{Rohm}
      R.~Rohm,
      Nucl.\ Phys.\ B {\bf 237}, 553 (1984).


\bibitem{AlvOso}
       E.~Alvarez and M.~A.~R.~Osorio,
       Nucl.\ Phys.\ B {\bf 304}, 327 (1988)
       [Erratum-ibid.\ B {\bf 309}, 220 (1988)];
       Phys.\ Rev.\ D {\bf 40}, 1150 (1989); \\
        M.~A.~R.~Osorio,
         Int.\ J.\ Mod.\ Phys.\ A {\bf 7}, 4275 (1992).


\bibitem{AtickWitten}
     J.~J.~Atick and E.~Witten,
     Nucl.\ Phys.\ B {\bf 310}, 291 (1988).



\bibitem{KounnasRostand}
     M.~McGuigan,
     Phys.\ Rev.\ D {\bf 38}, 552 (1988); \\
      C.~Kounnas and B.~Rostand,
      Nucl.\ Phys.\ B {\bf 341}, 641 (1990);\\
  I.~Antoniadis and C.~Kounnas,
  Phys.\ Lett.\ B {\bf 261}, 369 (1991);\\
  I.~Antoniadis, J.~P.~Derendinger and C.~Kounnas,
  Nucl.\ Phys.\ B {\bf 551}, 41 (1999)
  [arXiv:hep-th/9902032].


\bibitem{DolanJackiw}
  L.~Dolan and R.~Jackiw,
  Phys.\ Rev.\  D {\bf 9}, 3320 (1974).



\bibitem{kounn}
     C.~Angelantonj, M.~Cardella and N.~Irges,
     Phys.\ Lett.\ B {\bf 641}, 474 (2006)
     [arXiv:hep-th/0608022];\\
     C.~Angelantonj, C.~Kounnas, H.~Partouche and N.~Toumbas,
     Nucl.\ Phys.\ B {\bf 809}, 291 (2009)
     [arXiv:0808.1357 [hep-th]].


\bibitem{HetGauge}
  K.~S.~Narain, M.~H.~Sarmadi and E.~Witten,
  Nucl.\ Phys.\  B {\bf 279}, 369 (1987);\\
  P.~H.~Ginsparg,
  Phys.\ Rev.\  D {\bf 35}, 648 (1987).



\bibitem{KLTclassification}
  H.~Kawai, D.~C.~Lewellen and S.~H.~H.~Tye,
  Phys.\ Rev.\ D {\bf 34}, 3794 (1986).



\bibitem{DH}
    L.~J.~Dixon and J.~A.~Harvey,
  Nucl.\ Phys.\ B {\bf 274}, 93 (1986).


\bibitem{SW}
  N.~Seiberg and E.~Witten,
  Nucl.\ Phys.\ B {\bf 276}, 272 (1986).




\bibitem{IT}
     H.~Itoyama and T.~R.~Taylor,
     Phys.\ Lett.\ B {\bf 186}, 129 (1987).


\bibitem{julie}
     J.~D.~Blum and K.~R.~Dienes,
     Phys.\ Lett.\ B {\bf 414}, 260 (1997)
     [arXiv:hep-th/9707148];
     Nucl.\ Phys.\ B {\bf 516}, 83 (1998)
     [arXiv:hep-th/9707160].


\bibitem{ADS1998}
  I.~Antoniadis, E.~Dudas and A.~Sagnotti,
  Nucl.\ Phys.\  B {\bf 544}, 469 (1999)
  [arXiv:hep-th/9807011].

\bibitem{AS2002}
  C.~Angelantonj and A.~Sagnotti,
  Phys.\ Rept.\  {\bf 371}, 1 (2002)
  [Erratum-ibid.\  {\bf 376}, 339 (2003)]
  [arXiv:hep-th/0204089].

\bibitem{earlyOrientifold}
  J.~A.~Harvey and J.~A.~Minahan,
  Phys.\ Lett.\  B {\bf 188}, 44 (1987);\\
  P.~Horava,
  Nucl.\ Phys.\  B {\bf 327}, 461 (1989);\\
  G.~Pradisi and A.~Sagnotti,
  Phys.\ Lett.\  B {\bf 216}, 59 (1989);\\
  A.~Sagnotti,
  arXiv:hep-th/0208020.


\bibitem{Hagedorn}
  R.~Hagedorn,
  Nuovo Cim.\ Suppl.\  {\bf 3}, 147 (1965).

\bibitem{Huang}
  S.~Fubini and G.~Veneziano,
  Nuovo Cim.\ A {\bf 64}, 811 (1969);\\
  K.~Huang and S.~Weinberg,
  Phys.\ Rev.\ Lett.\  {\bf 25}, 895 (1970);\\
  S.~Frautschi,
  Phys.\ Rev.\ D {\bf 3}, 2821 (1971);\\
  R.~D.~Carlitz,
  Phys.\ Rev.\ D {\bf 5}, 3231 (1972);\\
  N.~Cabibbo and G.~Parisi,
  Phys.\ Lett.\ B {\bf 59}, 67 (1975);\\
  L.~Susskind,
  Phys.\ Rev.\ D {\bf 20}, 2610 (1979).


\bibitem{cudell}
     P.~G.~O.~Freund and J.~L.~Rosner,
  Phys.\ Rev.\ Lett.\  {\bf 68}, 765 (1992);\\
  J.~R.~Cudell and K.~R.~Dienes,
  Phys.\ Rev.\ Lett.\  {\bf 69}, 1324 (1992)
  [Erratum-ibid.\  {\bf 69}, 2311 (1992)]
  [arXiv:hep-ph/9207242];
  Phys.\ Rev.\ Lett.\  {\bf 72}, 187 (1994)
  [arXiv:hep-th/9309126].



\bibitem{earlystringpapers}
   See, {\it e.g.}\/:\\
  M.~J.~Bowick and L.~C.~R.~Wijewardhana,
  Phys.\ Rev.\ Lett.\  {\bf 54}, 2485 (1985);\\
  S.~H.~H.~Tye,
  Phys.\ Lett.\ B {\bf 158}, 388 (1985);\\
  B.~Sundborg,
  Nucl.\ Phys.\ B {\bf 254}, 583 (1985);\\
  E.~Alvarez,
  Nucl.\ Phys.\ B {\bf 269}, 596 (1986);\\
  E.~Alvarez and M.~A.~R.~Osorio,
  Phys.\ Rev.\ D {\bf 36}, 1175 (1987);
  Physica {\bf A 158}, 449 (1989)
  [Erratum-ibid.\ A {\bf 160}, 119 (1989)];
  Phys.\ Lett.\ B {\bf 220}, 121 (1989);\\
  M.~Axenides, S.~D.~Ellis and C.~Kounnas,
  Phys.\ Rev.\ D {\bf 37}, 2964 (1988);\\
  Y.~Leblanc,
  Phys.\ Rev.\ D {\bf 38}, 3087 (1988);\\
  B.~A.~Campbell, J.~R.~Ellis, S.~Kalara, D.~V.~Nanopoulos and K.~A.~Olive,
  Phys.\ Lett.\ B {\bf 255}, 420 (1991).


\bibitem{vortices}
  B.~Sathiapalan,
  Phys.\ Rev.\ D {\bf 35}, 3277 (1987);\\
  Y.~I.~Kogan,
  JETP Lett.\  {\bf 45}, 709 (1987)
  [Pisma Zh.\ Eksp.\ Teor.\ Fiz.\  {\bf 45}, 556 (1987)];\\
  I.~I.~Kogan,
  Phys.\ Lett.\ B {\bf 255}, 31 (1991).



\bibitem{longstrings}
  M.~J.~Bowick and S.~B.~Giddings,
  Nucl.\ Phys.\ B {\bf 325}, 631 (1989);\\
  S.~B.~Giddings,
  Phys.\ Lett.\ B {\bf 226}, 55 (1989);\\
  N.~Deo, S.~Jain and C.~I.~Tan,
  Phys.\ Rev.\ D {\bf 40}, 2626 (1989);\\
  D.~A.~Lowe and L.~Thorlacius,
  Phys.\ Rev.\ D {\bf 51}, 665 (1995)
  [arXiv:hep-th/9408134].


\bibitem{general}
  F.~Englert and J.~Orloff,
  Nucl.\ Phys.\ B {\bf 334}, 472 (1990);\\
  M.~Hellmund and J.~Kripfganz,
  Phys.\ Lett.\ B {\bf 241}, 211 (1990);\\
  M.~A.~R.~Osorio and M.~A.~Vazquez-Mozo,
  Phys.\ Lett.\ B {\bf 280}, 21 (1992)
  [arXiv:hep-th/9201044];
  Phys.\ Rev.\ D {\bf 47}, 3411 (1993)
  [arXiv:hep-th/9207002];\\
  M.~Laucelli Meana, M.~A.~R.~Osorio and J.~Puente Penalba,
  Phys.\ Lett.\ B {\bf 400}, 275 (1997)
  [arXiv:hep-th/9701122].



\bibitem{Dbranes}
  M.~A.~Vazquez-Mozo,
  Phys.\ Lett.\ B {\bf 388}, 494 (1996)
  [arXiv:hep-th/9607052];\\
  J.~L.~F.~Barbon and M.~A.~Vazquez-Mozo,
  Nucl.\ Phys.\ B {\bf 497}, 236 (1997)
  [arXiv:hep-th/9701142];\\
  S.~Chaudhuri and D.~Minic,
  Phys.\ Lett.\ B {\bf 433}, 301 (1998)
  [arXiv:hep-th/9803120];\\
  K.~R.~Dienes, E.~Dudas, T.~Gherghetta and A.~Riotto,
  Nucl.\ Phys.\ B {\bf 543}, 387 (1999)
  [arXiv:hep-ph/9809406];\\
  S.~A.~Abel, J.~L.~F.~Barbon, I.~I.~Kogan and E.~Rabinovici,
  JHEP {\bf 9904}, 015 (1999)
  [arXiv:hep-th/9902058];\\
  I.~I.~Kogan, A.~Kovner and M.~Schvellinger,
  JHEP {\bf 0107}, 019 (2001)
  [arXiv:hep-th/0103235];\\
  S.~Chaudhuri,
  arXiv:hep-th/0502141.




\bibitem{geometries}
  M.~McGuigan,
  Phys.\ Rev.\ D {\bf 42}, 2040 (1990);\\
  D.~A.~Lowe and A.~Strominger,
  Phys.\ Rev.\ D {\bf 51}, 1793 (1995)
  [arXiv:hep-th/9410215];\\
  G.~Grignani, M.~Orselli and G.~W.~Semenoff,
  JHEP {\bf 0111}, 058 (2001)
  [arXiv:hep-th/0110152];\\
  J.~L.~F.~Barbon and E.~Rabinovici,
  JHEP {\bf 0203}, 057 (2002)
  [arXiv:hep-th/0112173];\\
  H.~Liu,
  arXiv:hep-th/0408001.



\bibitem{ppwaves}
  L.~A.~Pando Zayas and D.~Vaman,
  Phys.\ Rev.\ D {\bf 67}, 106006 (2003)
  [arXiv:hep-th/0208066];\\
  B.~R.~Greene, K.~Schalm and G.~Shiu,
  Nucl.\ Phys.\ B {\bf 652}, 105 (2003)
  [arXiv:hep-th/0208163];\\
  R.~C.~Brower, D.~A.~Lowe and C.~I.~Tan,
  Nucl.\ Phys.\ B {\bf 652}, 127 (2003)
  [arXiv:hep-th/0211201];\\
  G.~Grignani, M.~Orselli, G.~W.~Semenoff and D.~Trancanelli,
  JHEP {\bf 0306}, 006 (2003)
  [arXiv:hep-th/0301186];\\
  S.~j.~Hyun, J.~D.~Park and S.~H.~Yi,
  JHEP {\bf 0311}, 006 (2003)
  [arXiv:hep-th/0304239];\\
  F.~Bigazzi and A.~L.~Cotrone,
  JHEP {\bf 0308}, 052 (2003)
  [arXiv:hep-th/0306102].



\bibitem{magnetic}
  J.~G.~Russo,
  Phys.\ Lett.\ B {\bf 335}, 168 (1994)
  [arXiv:hep-th/9405118];\\
  J.~G.~Russo and A.~A.~Tseytlin,
  Nucl.\ Phys.\ B {\bf 454}, 164 (1995)
  [arXiv:hep-th/9506071];\\
  J.~G.~Russo,
  Nucl.\ Phys.\ B {\bf 602}, 109 (2001)
  [arXiv:hep-th/0101132].


\bibitem{tensionless}
  F.~Lizzi and G.~Sparano,
  Phys.\ Lett.\ B {\bf 232}, 311 (1989);\\
  F.~Lizzi and I.~Senda,
  Phys.\ Lett.\ B {\bf 244}, 27 (1990).




\bibitem{noncritical}
  S.~D.~Odintsov,
  Phys.\ Lett.\ B {\bf 237}, 63 (1990);
  Phys.\ Lett.\ B {\bf 274}, 338 (1992)
  [Sov.\ J.\ Nucl.\ Phys.\  {\bf 55}, 440 (1992)]
  [Yad.\ Fiz.\  {\bf 55}, 795 (1992)];\\
  A.~A.~Bytsenko and S.~D.~Odintsov,
  Phys.\ Lett.\ B {\bf 243}, 63 (1990)
  [Yad.\ Fiz.\  {\bf 52}, 1495 (1990\ SJNCA,52,945-948.1990)].




\bibitem{little}
  T.~Harmark and N.~A.~Obers,
  Phys.\ Lett.\ B {\bf 485}, 285 (2000)
  [arXiv:hep-th/0005021];\\
  M.~Berkooz and M.~Rozali,
  JHEP {\bf 0005}, 040 (2000)
  [arXiv:hep-th/0005047];\\
  D.~Kutasov and D.~A.~Sahakyan,
  JHEP {\bf 0102}, 021 (2001)
  [arXiv:hep-th/0012258].



\bibitem{matrix}
  B.~Sathiapalan,
  Mod.\ Phys.\ Lett.\ A {\bf 13}, 2085 (1998)
  [arXiv:hep-th/9805126].


\bibitem{NCOS}
  S.~S.~Gubser, S.~Gukov, I.~R.~Klebanov, M.~Rangamani and E.~Witten,
  J.\ Math.\ Phys.\  {\bf 42}, 2749 (2001)
  [arXiv:hep-th/0009140];\\
  J.~L.~F.~Barbon and E.~Rabinovici,
  JHEP {\bf 0106}, 029 (2001)
  [arXiv:hep-th/0104169].


\bibitem{cosmology}
  S.~A.~Abel,
  Nucl.\ Phys.\ B {\bf 372}, 189 (1992);\\
  S.~A.~Abel, K.~Freese and I.~I.~Kogan,
  JHEP {\bf 0101}, 039 (2001)
  [arXiv:hep-th/0005028];\\
  S.~Abel, K.~Freese and I.~I.~Kogan,
  Phys.\ Lett.\ B {\bf 561}, 1 (2003)
  [arXiv:hep-th/0205317].


\bibitem{ridge}
  J.~L.~F.~Barbon and E.~Rabinovici,
  arXiv:hep-th/0407236.


\bibitem{2Dhet}
  J.~L.~Davis, F.~Larsen and N.~Seiberg,
  JHEP {\bf 0508}, 035 (2005)
  [arXiv:hep-th/0505081];\\
  N.~Seiberg,
  JHEP {\bf 0601}, 057 (2006)
  [arXiv:hep-th/0511220].



\bibitem{HR}
  G.~H.~Hardy and S.~Ramanujan,
  Proc. London Math. Soc. {\bf 17}, 75 (1918).

\bibitem{Kani}
  I.~Kani and C.~Vafa,
  Commun.\ Math.\ Phys.\  {\bf 130}, 529 (1990).


\bibitem{missusy}
  K.~R.~Dienes,
  Nucl.\ Phys.\ B {\bf 429}, 533 (1994)
  [arXiv:hep-th/9402006];
  Nucl.\ Phys.\ B {\bf 611}, 146 (2001)
  [arXiv:hep-ph/0104274];\\
  K.~R.~Dienes, M.~Moshe and R.~C.~Myers,
  Phys.\ Rev.\ Lett.\  {\bf 74}, 4767 (1995)
  [arXiv:hep-th/9503055].


\bibitem{kutasov}
  D.~Kutasov and N.~Seiberg,
  Nucl.\ Phys.\ B {\bf 358}, 600 (1991).


\bibitem{intprojs}
  P.~C.~Argyres and K.~R.~Dienes,
  Phys.\ Rev.\ Lett.\  {\bf 71}, 819 (1993)
  [arXiv:hep-th/9305093].



\bibitem{dualityus}
  K.~R.~Dienes and M.~Lennek,
  arXiv:hep-th/0312173;
  Phys.\ Rev.\ D {\bf 70}, 126005 (2004)
  [arXiv:hep-th/0312216];
  Phys.\ Rev.\ D {\bf 70}, 126006 (2004)
  [arXiv:hep-th/0312217].

\bibitem{shyamoli}
  S.~Chaudhuri,
  Phys.\ Rev.\ D {\bf 65}, 066008 (2002)
  [arXiv:hep-th/0105110];
  arXiv:hep-th/0409301.

\bibitem{old}
  K.~R.~Dienes and M.~Lennek,
  hep-th/0505233;  
  hep-th/0507201.




\bibitem{nonSUSYgauge}
  V.~P.~Nair, A.~D.~Shapere, A.~Strominger and F.~Wilczek,
  Nucl.\ Phys.\  B {\bf 287}, 402 (1987);\\
  P.~H.~Ginsparg and C.~Vafa,
  Nucl.\ Phys.\  B {\bf 289}, 414 (1987).


\bibitem{EBstrings}
  A.~A.~Tseytlin,
  Nucl.\ Phys.\  B {\bf 524}, 41 (1998)
  [arXiv:hep-th/9802133];\\
  J.~Ambjorn, Yu.~Makeenko, G.~W.~Semenoff and R.~J.~Szabo,
  Phys.\ Rev.\  D {\bf 60}, 106009 (1999)
  [arXiv:hep-th/9906134];\\
  C.~Angelantonj, C.~Kounnas, H.~Partouche and N.~Toumbas,
  Nucl.\ Phys.\  B {\bf 809}, 291 (2009)
  [arXiv:0808.1357 [hep-th]].


\bibitem{AHI2003}
  P.~Anastasopoulos, A.~B.~Hammou and N.~Irges,
  Phys.\ Lett.\  B {\bf 581}, 248 (2004)
  [arXiv:hep-th/0310277].



\bibitem{DS}
  K.~R.~Dienes and M.~Sharma, ``S-Duality at Finite Temperature'', to appear.

\bibitem{D}
  K.~R.~Dienes, to appear.

\bibitem{DL}
  K.~R.~Dienes and M.~Lennek, to appear.


\bibitem{D-BranesT}
  M.~A.~Vazquez-Mozo,
  Phys.\ Lett.\  B {\bf 388}, 494 (1996)
  [arXiv:hep-th/9607052];\\
  S.~Lee and L.~Thorlacius,
  Phys.\ Lett.\  B {\bf 413}, 303 (1997)
  [arXiv:hep-th/9707167];\\
  S.~A.~Abel, J.~L.~F.~Barbon, I.~I.~Kogan and E.~Rabinovici,
  JHEP {\bf 9904}, 015 (1999)
  [arXiv:hep-th/9902058].



\bibitem{D-BranesPT}
  J.~L.~F.~Barbon and M.~A.~Vazquez-Mozo,
  Nucl.\ Phys.\  B {\bf 497}, 236 (1997)
  [arXiv:hep-th/9701142];\\
  J.~Ambjorn, Y.~M.~Makeenko and G.~W.~Semenoff,
  Phys.\ Lett.\  B {\bf 445}, 307 (1999)
  [arXiv:hep-th/9810170].



\bibitem{D-branesC}
  K.~R.~Dienes, E.~Dudas, T.~Gherghetta and A.~Riotto,
  Nucl.\ Phys.\  B {\bf 543}, 387 (1999)
  [arXiv:hep-ph/9809406];\\
  S.~Alexander, R.~H.~Brandenberger and D.~Easson,
  Phys.\ Rev.\  D {\bf 62}, 103509 (2000)
  [arXiv:hep-th/0005212];\\
  S.~A.~Abel, K.~Freese and I.~I.~Kogan,
  JHEP {\bf 0101}, 039 (2001)
  [arXiv:hep-th/0005028];\\
  M.~Majumdar and A.~Christine-Davis,
  JHEP {\bf 0203}, 056 (2002)
  [arXiv:hep-th/0202148];\\
  R.~Brandenberger, D.~A.~Easson and A.~Mazumdar,
  Phys.\ Rev.\  D {\bf 69}, 083502 (2004)
  [arXiv:hep-th/0307043].


  

\bibitem{NPTypeI}
  J.~Polchinski and E.~Witten,
  Nucl.\ Phys.\  B {\bf 460}, 525 (1996)
  [arXiv:hep-th/9510169];\\
  A.~Sen,
  JHEP {\bf 9809}, 023 (1998)
  [arXiv:hep-th/9808141].





\end{thebibliography}
\end{document}